\documentclass[12pt]{article}
\pdfoutput=1

\usepackage{amsmath}
\usepackage{amsfonts}
\usepackage{amssymb}
 
\usepackage{hyperref}
\newlength{\abstractwidth}
\usepackage{color}
\usepackage{graphicx}

\renewcommand{\thefootnote}{\fnsymbol{footnote}}
\renewcommand{\thanks}[1]{\footnote{#1}}
\newcommand{\starttext}{
\setcounter{footnote}{0}
\renewcommand{\thefootnote}{\arabic{footnote}}}

\newcommand{\bea}{\begin{eqnarray}}
\newcommand{\eea}{\end{eqnarray}}
\newcommand{\ee}{\end{equation}}
\newcommand{\be}{\begin{equation}}

\newcommand{\no}{\nonumber}

\def\cB{{\cal B}}

\def\Re{{\rm Re}}
\def\Im{{\rm Im}}

\def\half{ {1\over 2}}
\def\p{\partial}

\def\a{\alpha}
\def\b{\beta}

\def\ep{\varepsilon}

\def\g{\gamma}

\def\G{\Gamma}

\def\no{\nonumber}

\long\def\symbolfootnote[#1]#2{\begingroup%
\def\thefootnote{\fnsymbol{footnote}}\footnote[#1]{#2}\endgroup}

\begin{document}
\starttext
\setcounter{footnote}{0}

\begin{flushright}
KUL-TF-09/21
\end{flushright}

\bigskip

\begin{center}

{\Large \bf  Half-BPS  Solutions   locally asymptotic to $AdS_3\times S^3$ and interface conformal field theories}

\medskip

\vskip .4in

{\large  Marco Chiodaroli$^{a}$, Michael Gutperle$^{a}$,
and  Darya Krym$^{b}$}

\vskip .2in

{$\ ^{a}$ \sl Department of Physics and Astronomy }\\
{\sl University of California, Los Angeles, CA 90095, USA}\\
{\tt \small mchiodar@ucla.edu; gutperle@physics.ucla.edu}

\vskip .2in

{$\ ^{b}$\sl Instituut voor Theoretische Fysica, Katholieke Universiteit Leuven,\\
Celestijnenlaan 200D B-3001 Leuven, Belgium }\\
{\tt  \small daryakrym@gmail.com}
\end{center}

\vskip .2in

\begin{abstract}

\vskip 0.1in

Type IIB superstring theory has $AdS_3\times S^3 \times M_4$ (where the manifold  $M_4$ is either $K_3$ or $T^4$)  solutions which preserve sixteen supersymmetries. 
In this paper we consider half-BPS solutions which are locally asymptotic to  $AdS_3\times S^3 \times M_4$  and preserve eight of the sixteen supersymmetries. 
We reduce the BPS equations and the Bianchi identity for the self-dual five-form field  to a set of four  differential equations. The complete local solution can be parameterized 
in terms of two harmonic and two holomorphic functions and all bosonic fields have explicit expressions in terms of these 
functions.

We analyze the conditions for global regularity  and  construct  new half-BPS Janus-solutions which have two asymptotic $AdS_3$ 
regions. In addition, our analysis proves the global regularity of a class of solutions with more than two asymptotic $AdS_3$ regions.

 Finally, we discuss the  dual interpretation of the half-BPS Janus solutions carrying only Ramond-Ramond three-form  charge as supersymmetric interface theories.

\end{abstract}

\baselineskip=16pt
\setcounter{equation}{0}
\setcounter{footnote}{0}

\newpage

\tableofcontents

\newpage

\baselineskip 16pt 

\section{Introduction}
\setcounter{equation}{0}

The AdS/CFT correspondence \cite{Maldacena:1997re,Gubser:1998bc,Witten:1998qj} (for reviews see e.g. \cite{Aharony:1999ti,D'Hoker:2002aw}) relates a  gravity  theory in the bulk of a $d+1$-dimensional Anti-de Sitter (AdS) space to a $d$-dimensional conformal field theory (CFT) on the boundary of the space. One of the best understood examples is the $AdS_3/CFT_2$ correspondence,
 which is of central importance for the study of black holes in string theory.

The particular realization  of the $AdS_3/CFT_2$ correspondence which will be discussed in the present paper is the duality
 between type IIB string theory compactified on $AdS_3\times S^3 \times M_4$ and  a ${\cal N}=(4,4)$ two-dimensional superconformal
 field theory. The compactification manifold $M_4$ can be either the four torus $T^4$ or a $K_3$ manifold and either one  leads to  a theory with sixteen unbroken supersymmetries 
\footnote{Note that the $AdS_3 \times S^3$ vacuum of six-dimensional maximal supergravity preserves only half the supersymmetries of the 
six-dimensional Minkowski vacuum due to the non-zero self-dual or anti self-dual fluxes \cite{David:2002wn}.}.

This type IIB background can be obtained by taking the near-horizon limit of  a bound state of $Q_1$  D1-branes and $Q_5$ D5-branes  wrapped on $M_4$. This system was instrumental in the counting of black hole microstates for supersymmetric black holes \cite{Strominger:1996sh,Maldacena:1996ky}.  A complementary point of view is to consider the six-dimensional supergravity which is obtained by compactifying type IIB on $M_4$ \cite{Romans:1986er,Tanii:1984zk}  and study the near-horizon limit of a self-dual string soliton. 

The D1/D5 bound state is defined by the Higgs-branch of the two-dimensional $U(Q_1)\times U(Q_5)$ 
gauge theory living on the intersection of the branes. In the infrared limit the theory flows to a ${\cal N}=(4,4)$ two-dimensional
 superconformal theory \cite{Witten:1997yu}. The CFT  can also be understood as a hyperk\"ahler sigma model whose  target  space is $(M_4)^n/S_n$, where $S_n$ is the n-dimensional symmetric group \cite{Seiberg:1999xz,Vafa:1995bm,Dijkgraaf:1998gf}.  
 
In general, the AdS/CFT correspondence  maps local as well as non-local gauge 
invariant operators on the CFT side to supergravity solutions on the AdS side. In the limit of large 't Hooft coupling and large $N$ the classical gravity description becomes a reliable approximation. One goal for obtaining new supergravity solutions is to better understand the CFT side of the correspondence.

A particular example of such solutions is the  so-called Janus solution, which is  dual to interface configurations in the CFT. The original Janus solution \cite{Bak:2003jk} is a dilatonic deformation of the $AdS_5\times S^5$ vacuum of type IIB. The solution is constructed using $AdS_4$ slices and making the dilaton dependent on the slicing coordinate. 
 The dilaton approaches different constant values on the two boundary components. The solution has a $SO(2,3)\times SO(6)$ isometry 
but breaks all thirty-two supersymmetries.  On the ${\cal N}=4$ super Yang-Mills side, this solution corresponds to an interface theory,
 where the Yang-Mills coupling constant jumps across a $2+1$-dimensional interface.  On the field theory side, the $SO(3,2)$ isometry corresponds to the $2+1$-dimensional conformal symmetry preserved by the interface.

 In \cite{D'Hoker:2006uv} it was shown that up to half the broken supersymmetry can be restored by adding counterterms
 localized on the interface. The counterterms break the R-symmetry from $SO(6)$ to $SU(2)\times SU(2)$. Consequently, the ansatz
 for a  dual supergravity solution is constructed by a warped product of $AdS_4\times S^2\times S^2$ over a two-dimensional  
Riemann surface $\Sigma$. In \cite{D'Hoker:2007xy} a supersymmetric generalization of the Janus solution was found. Furthermore, it was shown that the conditions for the existence of sixteen preserved supersymmetries are related to solutions of a particular integrable system. Locally the integrable system can be solved in terms of harmonic functions.  
  
An important ingredient in the construction of  the solution is the fact that  the two-dimensional  surface $\Sigma$ has a boundary.
 At generic points on the boundary, one two-sphere shrinks to zero size, closing off the space.
 At special isolated points associated with poles of one of the harmonic functions, the $AdS_4$ metric factor goes to infinity.  The holographic map relates such points to boundaries of the space where the dual gauge theory lives. The supersymmetric Janus solution has two such
 boundary points corresponding to two 3+1-dimensional half spaces  glued together at a 2+1-dimensional defect.
 In \cite{D'Hoker:2007xz} the global regularity of the local solution was analyzed and apart from the supersymmetric Janus solution, an
 infinite class of "multi-Janus" solutions was found. These solutions display more than two asymptotic boundary regions.

Similar techniques were used to obtain supergravity duals to half-BPS Wilson 
loops in $AdS_5\times S^5$  \cite{D'Hoker:2007fq} and analogues of the Janus solution in M-theory \cite{D'Hoker:2008wc,D'Hoker:2008qm,D'Hoker:2009my}. For related work by other authors see, e.g. \cite{Gomis:2006cu,Clark:2005te,Lunin:2006xr,Lunin:2007ab,Yamaguchi:2006te,Gomis:2006sb}.

The primary goal of the present paper is to find half-BPS solutions that preserve eight of the sixteen supersymmetries of the vacuum and are  locally asymptotic to  $AdS_3\times S^3$. 
We use techniques developed in  \cite{Gomis:2006cu,D'Hoker:2007xy,D'Hoker:2007fq} for a specific 
 ansatz  which is a product of $AdS_2\times S^2\times M_4$ spaces warped over a two-dimensional Riemann surface $\Sigma$ with boundary. 
For simplicity we do not turn on the  moduli of  the compactification manifold $M_4$. The ansatz preserves a $SO(2,1)\times SO(3)$ subgroup of the $SO(2,2)\times SO(4)$ isometry of the $AdS_3\times S^3\times M_4$ 
vacuum.  \\
We derive the most general local solutions and find that all the fields can be expressed in terms of two harmonic and two holomorphic functions which are
 defined on $\Sigma$.  The requirement that the solutions are locally asymptotic to $AdS_3\times S^3$ and everywhere regular relates the harmonic and holomorphic functions and determines  the boundary conditions.
In particular, we present a half-BPS solution which is a supersymmetric generalization of the solution found in \cite{Bak:2007jm} and we give explicit expressions
for solutions having $n$ $AdS_3 \times S^3$ regions as long as $\Sigma$ has genus zero and one boundary component. 
 
Our solutions are dual to one-dimensional defects in the two-dimensional CFT\footnote{For earlier work in this direction see \cite{Kumar:2002wc,Kumar:2003xi,Kumar:2004me}.}. 
Defects, domain walls and interfaces in  two-dimensional conformal field theories can have applications in condensed matter physics- for example, in the description of impurities at critical points or in the study of the Kondo effect.  
 Domain walls and interfaces   have been discussed in the context of the AdS/CFT correspondence in \cite{Bachas:2001vj}. 
In $AdS_{d+1}/CFT_d$ $p$-dimensional defects can be realized as probe branes inside the bulk AdS space which have a lower dimensional $AdS_{p+1}$
 submanifold as worldvolume \cite{Karch:2000gx,Aharony:2003qf}. Probe branes in the context of $AdS_3\times S^3$ have been discussed in 
\cite{Bachas:2002nz,Bachas:2008jd,Raeymaekers:2006np,Yamaguchi:2003ay,Raju:2007uj,Mandal:2007ug}.

\begin{table}[htdp]
\begin{center}
\begin{tabular}{|c|c|c|c|}
\hline
probe brane&$AdS_{3}$&$S^{3}$& $M_{4}$\\
\hline
D1&$ AdS_{2}$ &$\cdot$&$\cdot$\\
\hline
D3&$ AdS_{2}$ &$S^{2}$&$\cdot$\\
\hline
D5&$ AdS_{2}$ &$\cdot$&$M_{4}$\\
\hline
D7&$ AdS_{2}$ &$S^{2}$&$M_{4}$\\
\hline
\end{tabular}
\caption{Half-BPS probe branes in $AdS_3\times S^3\times M_4$}
\end{center}
\label{default}
\end{table}%

As we can see in Table \ref{default}, there are four cases of probe D-branes, namely a probe D1 brane with $AdS_2$ worldvolume,  a probe D3-brane with $AdS_2\times S^2$ worldvolume, 
a probe D5-brane with $AdS_2\times M_4$ worldvolume and a probe $D7$ brane with $AdS_2\times S^2\times M_4$ worldvolume which are not wrapped on two cycles in the four-dimensional manifold. These brane configurations preserve the same $SO(2,1)\times SO(3)$ symmetries as our ansatz.
In
this paper, we obtain exact, fully back-reacted, solutions that
correspond to a configuration of D1 and D5-branes, as well as NS5-branes and fundamental strings.
However, the  regular solutions we find  have vanishing D7 and D3 brane charge.

  There also have  been some interesting recent developments in the description of interfaces on the CFT side, see for example \cite{Brunner:2008fa,Bachas:2007td,Fuchs:2007tx}.

The structure of the paper is as follows: in section \ref{sectwo}, we present  the  ansatz  for the bosonic fields in type IIB supergravity. The  dilatino and gravitino supersymmetry  transformations are reduced on $AdS_2\times S^2\times K_3\times \Sigma$ to express the BPS equations in terms of a two-dimensional spinor on $\Sigma$.
 In section \ref{secthree}, we find the complete
local solution of the BPS equations in terms of two harmonic and two holomorphic functions. Note that one harmonic function is obtained 
by solving the Bianchi identity for the self-dual five-form flux along the $M_4$ directions.  In section \ref{secfour}, we obtain the conditions for 
regularity on the boundary and bulk of $\Sigma$ and we present a family of half-BPS Janus solutions. 
In section \ref{secfive}, we review the dual two-dimensional CFT and the interpretation of the half-BPS Janus solution as an interface theory.  
In our concluding section, we discuss possible applications of the solutions, open questions and directions for further research.
Some technical details are relegated to the appendices.

\section{Ten-dimensional ansatz and reduction of BPS equations}\label{sectwo}
\setcounter{equation}{0}
In this  section, we present  the detailed ansatz for the bosonic fields and the reduction of the BPS conditions,
which need to be satisfied to have eight linearly independent unbroken supersymmetries.   
The reduced equations are  (\ref{BPSred2D1})-(\ref{BPSred2bzb}) and are solved in terms of two  harmonic functions and two holomorphic functions in section \ref{secthree}. 
Readers which are only interested in the solutions may wish to skip to section \ref{totsolution} after reading section \ref{tendimans}.

\subsection{Brief review of IIB supergravity fields}\label{sectwoone}
The IIB supergravity fields consist of the scalar fields $P$ and $Q$, composites of dilaton and axion:
 \be
 P= {1\over 2} \Big( d \phi + i e^{\phi}d\chi\Big), \quad Q= - {1\over 2} e^{\phi} d\chi 
 \ee 
The complex three-form $G$ is a composite of $H_{3}$, the NS-NS field strength, and $F_{3}$, the R-R field strength:
  \be
 G=e^{-\phi/2} H_{3}+ i e^{\phi/2}\Big(F_{3}-\chi H_{3}\Big) 
 \ee
The real self-dual five-form is:
\be
F_{(5)}  =  dC_{(4)} + { i \over 16} \left ( B_{(2)} \wedge \bar F_{(3)}
- \bar B_{(2)} \wedge  F_{(3)} \right ), \quad  F_{(3)} = d B_{(2)}
\ee
The fermionic fields are the dilatino $\lambda$ and the gravitino $\psi_\mu$,
both of which are complex Weyl spinors with opposite ten-dimensional
chiralities, given by $\Gamma_{11} \lambda =\lambda$, and $\Gamma_{11}
\psi_\mu  =-\psi_\mu$. The supersymmetry variations of the fermions  are
\bea
\delta\lambda &=& i (\G \cdot P) \cB^{-1} \ep^*
-{i\over 24} (\G \cdot G) \ep
\label{susy1main} \\
\delta \psi_\mu
&=& D _\mu  \ep
+ {i\over 480}(\G \cdot F_{(5)})  \Gamma_\mu \ep
-{1\over 96}\left ( \Gamma_\mu (\G \cdot G)
+ 2 (\G \cdot G) \G_\mu \right ) \cB^{-1} \ep^* \label{susy2main}
\eea
The complex conjugation matrix  $\cB$ satisfies $ \cB  \cB^{*}=1$ and $ \cB \Gamma_{\mu} \cB^{-1}=(\Gamma_{\mu})^{*}$.
For further review of supergravity definitions and equations, see Appendix A.

In obtaining IIB supergravity solutions with a large amount of supersymmetry, we solve the above BPS equations instead of the equations of motion. We will see that one of the Bianchi identities is not automatic for generic solutions of the BPS equations, but yields an extra condition. The solutions of the BPS equations and this Bianchi identity are shown to automatically satisfy the equations of  motion and all the remaining  Bianchi identities.

\subsection{The ten-dimensional ansatz}\label{tendimans}

The ten-dimensional metric ansatz is

\be
ds^{2} = f_{1}^{2 } ds^{2}_{AdS_{2}} + f^{2}_{2}ds^{2}_{S^{2}} + f^{2}_{3}ds^{2}_{K_{3}}  + \rho^{2 }dz  d\bar z
\ee
where $ds^{2}_{AdS_{2}}$,  $ds^{2}_{S^{2}}$, $ds^{2}_{K_{3}}$ are the unit radius metrics for $AdS_2$, $S^2$ and $K_3$ respectively. $\rho^2$ is an unspecified Riemannian 
metric on $\Sigma$, a two-dimensional surface with boundary. The metric factors $f_1^2$, $f_2^2$, $f_3^2$ and $\rho^2$ are real
 positive-definite functions on $\Sigma$ and will be determined by the BPS equations and Bianchi identities. 
The rest of the supergravity fields will be reduced on this ansatz. However, we specialize to the case where the supergravity 
fields are also independent of the $K_3$ coordinates. In other words, the three-form flux does not have any leg in the $K_3$ directions while
 the five-form flux must have four legs along the $K_3$ directions. 
For the sake of concreteness, we will consider the third metric factor to be a $K_3$ metric, but for the aforementioned restriction of the flux,
 the analysis would be completely analogous if the factor were to be a $T^4$ (see section \ref{t4vsk3}).

It is useful to introduce the orthonormal frame fields:
\bea
\label{frame1}
e^{i} & = & f_1 \, \hat e^{i} \hskip 1in i=0,1
\no \\
e^{j} & = & f_2 \, \hat e^{j} \hskip 1in j =2,3
\no \\
e^{k} & = & f_3 \, \hat e^{k} \hskip 1in k =4,5,6,7
\no \\
e^{a} \, &  & \hskip 1.33in a=8,9
\eea
where  $\hat e^{i}$, $\hat e^{j}$ and $\hat e^{k}$ refer to orthonormal
frames for the spaces $AdS_2$, $S^2$
and  $K_{3}$  which satisfy
\bea
 ds^{2}_{AdS_{2}}&=& \eta_{i_{1}i_{2}}\;\hat e^{i_{1}} \otimes \hat e^{i_{2}} \no\\
 ds^{2}_{S^{2}}&=& \delta_{j_{1}j_{2}}\;\hat e^{j_{1}} \otimes \hat e^{j_{2}} \no\\
  ds^{2}_{M_{4}}&=& \delta_{k_{1}k_{2}}\;\hat e^{k_{1}} \otimes \hat e^{k_{2}} \no\\
  \rho^{2 }dz \otimes d\bar z &=& \delta_{ab}\; e^{a} \otimes e^{b}
\eea
so that the unhatted frame fields contain the metric factor $f_{i}$, whereas the hatted ones do not.
The scalar one-form field strengths are given by
\be
Q =q_a e^a, \quad P = P_{a} e^{a}
\ee
The complex three-form is given by
\be
G= g^{(1)}_{a}e^{a01}+ g^{(2)}_{a}e^{a23}
\ee
The five-form flux is given by
\be
F_{5}= h_{a}  e^{a0123}+ \tilde h_{a} e^{a4567}
\ee
 Self-duality of the five-form field strength $F_{5}=*F_{5}$ imposes
 \be
 h_{a}=-\epsilon_{a}^{\; \;b} \tilde h_{b}
 \ee
 where $\epsilon_{8}^{\;\;9}=1$ and $\epsilon_{9}^{\;\;8}=-1$. See \ref{epconvention} for conventions regarding the $\epsilon_{\mu_{1} \cdots \mu_{10}}$ tensor.

\subsection{Reduction of the ten-dimensional spinor}
The supersymmetry parameter $\varepsilon$ must be globally well defined on the symmetric spaces $AdS_2$, $S^2$ and $K_3$. 
Therefore, the tensor products of Killing spinors on these symmetric spaces can be used as a basis for the spinors $\varepsilon$.
The Killing spinor equations on
$AdS_2 \times S^2   \times K_{3} $ are satisfied by a set of  basis spinors $\chi _{\eta _1, \eta _2, \eta _3}$ with $\{\eta_1, \eta_2, \eta_3\} \in \{ +1,-1\} $ 
\bea
\label{KS}
\left ( \hat \nabla _\mu  - {1 \over 2} \eta_1\, \gamma _\mu \otimes I_2 \otimes I_4 \right )
\chi _{\eta _1, \eta _2, \eta _3}
    & = & 0 \hskip 1in \mu =0,1
\no \\
\left ( \hat \nabla _i - {i \over 2} \eta_2 \, I_2 \otimes \gamma _i \otimes I_4 \right )
\chi_{\eta _1, \eta _2, \eta _3}
    & = & 0 \hskip 1in i=2,3
\no \\
 \hat \nabla _m \,
\chi _{\eta _1, \eta _2, \eta _3}
    & = & 0 \hskip 1in m=4,5,6,7
\eea
The covariant derivatives $\hat \nabla _\mu$ , $\hat \nabla _i$ , $\hat \nabla _m$ are taken with respect to the unit radius metrics of the corresponding spaces as explained in Appendix C.
We now expand the ten-dimensional spinor $\epsilon$ in terms of Killing spinors on $AdS_{2}\times S^{2}\times K_{3}$. 
\be
\epsilon= \sum_{\eta_{1},\eta_{2}}\chi_{\eta _1, \eta _2, \eta _3} \otimes \xi_{\eta_{1},\eta_{2},\eta_{3}  }
\ee
The Killing spinor equations are invariant under complex conjugation defined as
\be
\chi _{\eta _1, \eta _2, \eta _3} \rightarrow B_{(1)}\otimes B_{(2)}\otimes B_{(3)} \chi _{\eta _1, \eta _2, \eta _3}^{*}
\ee
Therefore, it is consistent to impose the following reality condition on $ \chi _{\eta _1, \eta _2, \eta _3}$
\be
 \big(B_{(1)}\otimes B_{(2)}\otimes B_{(3)}\big)\chi _{\eta _1, \eta _2, \eta _3}^{*}=\eta_{2}\chi _{\eta _1, \eta _2, \eta _3}\label{realconks}
\ee
where the presence of the $\eta_{2}$ prefactor comes from the fact that $B_{(2)}$ and $\gamma_{2}$ anticommute.
The chirality condition $\Gamma^{11}$ acts on $\epsilon$ in the following way (see \ref{gamma11}):
\bea
\Gamma^{11}\epsilon=
  \sum_{\eta_{1},\eta_{2}}  \chi^{(1)}_{\eta_{1}} \otimes \chi^{(2)}_{\eta_{2}}\otimes \chi^{(3)}_{\eta_{3}} \otimes \gamma_{(4)}\xi_{-\eta_{1},-\eta_{2},\eta_{3}  }
\eea
The chirality condition $\Gamma^{11}\epsilon=-\epsilon$ then implies the following condition on the two-dimensional spinors $\xi_{\eta_{1},\eta_{2},\eta_{3}  }$
\be
\gamma_{(4)}\xi_{-\eta_{1},-\eta_{2},\eta_{3}  }
=-\xi_{\eta_{1},\eta_{2},\eta_{3}  }\label{chiconab}
\ee
Ten-dimensional complex conjugation acts as follows (see \ref{bconj}):
 \bea
 B^{-1}\epsilon^{*}=
 \sum_{\eta_{1},\eta_{2}}  \chi^{(1)}_{\eta_{1}} \otimes \chi^{(2)}_{\eta_{2}}\otimes \chi^{(3)}_{\eta_{3}} \otimes (- i \;\eta_{2 }) B_{(4)}^{-1}(\xi_{\eta_{1},-\eta_{2},\eta_{3}  })^{* }
\eea
where we used the reality condition given in
(\ref{realconks}).

\subsection{Reduction of the ten-dimensional BPS equations}
There are five different sets of equations corresponding to the dilatino variation together with the variation of the gravitino in 
the $AdS_{2}$, $S^{2}$, $K_{3} $ and $\Sigma$ directions. In this section we will rely heavily on formulae from appendix C.

\subsubsection{Dilatino equation}
The terms in the dilatino variation can be reduced as follows:
\bea
i \;\Gamma^{M} P_{M} \mathcal{B}^{-1} \epsilon^{*}&=&   \gamma_{(1)}\otimes \gamma_{(2)}\otimes \gamma_{(3)}\otimes  P_{a}\gamma^{a}  \sum_{\eta_{1},\eta_{2}}  \chi^{(1)}_{\eta_{1}} \otimes \chi^{(2)}_{\eta_{2}}\otimes \chi^{(3)}_{\eta_{3}} \otimes  \eta_{2 } B^{-1}_{(4)}(\xi_{\eta_{1} -\eta_{2} \eta_{3}  })^{* }\no\\
&=& - \sum_{\eta_{1},\eta_{2}}  \chi^{(1)}_{\eta_{1}} \otimes \chi^{(2)}_{\eta_{2}}\otimes \chi^{(3)}_{\eta_{3}} \otimes  \eta_{2 }P_{a} \gamma^{a}B^{-1}_{(4)}(\xi_{-\eta_{1} \eta_{2} \eta_{3}  })^{* }
\eea
and
\bea
-{i\over 24} (\G \cdot G) \ep&=& {i\over 4} \Big( g^{(1)}_{a } 1_{2}\otimes \gamma_{(2)} \otimes \gamma_{(3)}\otimes \gamma^{a}+ i g^{(2)}_{a } \gamma_{(1)}\otimes 1_{2} \otimes \gamma_{(3)}\otimes \gamma^{a}\Big) \no\\
&& \;\; \times\;\sum_{\eta_{1},\eta_{2}} \chi^{(1)}_{\eta_{1}} \otimes \chi^{(2)}_{\eta_{2}}\otimes \chi^{(3)}_{\eta_{3}} \otimes \xi_{\eta_{1} \eta_{2} \eta_{3}  }\no\\
&=& {1\over 4}\sum_{\eta_{1},\eta_{2}}  \chi^{(1)}_{\eta_{1}} \otimes \chi^{(2)}_{\eta_{2}}\otimes \chi^{(3)}_{\eta_{3}}  \Big( i  g_{a}^{(1)}  \gamma^{a}  \xi_{\eta_{1} -\eta_{2} \eta_{3}  } -  g_{a}^{(2)}  \gamma^{a}  \xi_{-\eta_{1} \eta_{2} \eta_{3}  }\Big)
\eea
Putting the terms together we can rewrite condition (\ref{susy1main}) as:
\be
 - \eta_{2 }P_{a} \gamma^{a}B^{-1}_{(4)}(\xi_{-\eta_{1} \eta_{2} \eta_{3}  })^{* } + {1\over 4}  i  g_{a}^{(1)}  \gamma^{a}  \xi_{\eta_{1} -\eta_{2} \eta_{3}  } -  {1\over 4}  g_{a}^{(2)}  \gamma^{a}  \xi_{-\eta_{1} \eta_{2} \eta_{3}  }=0
\label{BPS-D}
\ee
\subsubsection{Gravitino  in $AdS_{2}$ direction}
The covariant derivative in the $AdS_{2}$ directions is given by
\be
\nabla_{\mu}\epsilon= {1\over f_{1}} \hat \nabla_{\mu} \epsilon
+{1\over 2} {D_{a}f_{1}\over f_{1}} \Gamma_{\mu}\Gamma^{a} \epsilon
\ee
where $ \hat \nabla_{\mu}$ is the covariant derivative on the unit radius $AdS_{2}$, $D_a\equiv e^M_a\p_M$ and $M$ is a spacetime (Einstein frame) index.
We use the same method of the previous section to extract an equation for the spinors $\xi$ and we obtain: 
\be
 {\eta_{1} \over f_{1}} \xi_{\eta_{1}  \eta_{2}\eta_{3}}+ {D_{a}f_{1}\over f_{1}} \gamma^{a} \xi_{-\eta_{1} - \eta_{2}\eta_{3}}-  h_{a} \gamma^{a}\xi_{\eta_{1}  \eta_{2}\eta_{3}}+i  {3 \eta_{2}\over 8} g_{a}^{(1)}\gamma^{a} B_{(4)}^{-1}\xi^{*}_{\eta_{1}  \eta_{2}\eta_{3}}-  {\eta_{2}\over 8} g_{a}^{(2)}\gamma^{a} B_{(4)}^{-1}\xi^{*}_{-\eta_{1}  -\eta_{2}\eta_{3}}=0
\label{BPS-AdS}
\ee

\subsubsection{Gravitino  in $S^{2}$ direction}
The covariant derivative in the $S^{2}$ directions is given by
\be
\nabla_{i}\epsilon= {1\over f_{2}} \hat \nabla_{i} \epsilon
+{1\over 2} {D_{a}f_{2}\over f_{2}} \Gamma_{i}\Gamma^{a} \epsilon
\ee
where $ \hat \nabla_{\mu}$ is the covariant derivative on the unit radius $S^{2}$.
The BPS condition is given by
\be
{i \eta_{2} \over f_{2}} \xi_{\eta_{1}  \eta_{2}\eta_{3}}+ {D_{a}f_{2}\over f_{2}} \gamma^{a} \xi_{\eta_{1} - \eta_{2}\eta_{3}}-  h_{a} \gamma^{a}\xi_{-\eta_{1}  \eta_{2}\eta_{3}}-i  {\eta_{2}\over 8} g_{a}^{(1)}\gamma^{a} B_{(4)}^{-1}\xi^{*}_{-\eta_{1}  \eta_{2}\eta_{3}}+ {3 \eta_{2}\over 8} g_{a}^{(2)}\gamma^{a} B_{(4)}^{-1}\xi^{*}_{\eta_{1}  -\eta_{2}\eta_{3}}=0
\label{BPS-S} \ee
\subsubsection{Gravitino  in $K_{3}$ direction}
The covariant derivative in the $K_{3}$ directions is given by
\be
\nabla_{l}\epsilon= {1\over f_{3}} \hat \nabla_{l} \epsilon
+{1\over 2} {D_{a}f_{3}\over f_{3}} \Gamma_{l}\Gamma^{a} \epsilon
\ee
where $ \hat \nabla_{l} \epsilon=0$ thanks to our basis of spinors.
The BPS condition is given by

\be
  {D_{a}f_{3}\over f_{3}} \gamma^{a} \xi_{\eta_{1} \eta_{2}\eta_{3}}+ h_{a} \gamma^{a}\xi_{-\eta_{1} -\eta_{2}\eta_{3}}+i  {\eta_{2}\over 8} g_{a}^{(1)}\gamma^{a} B_{(4)}^{-1}\xi^{*}_{-\eta_{1}  -\eta_{2}\eta_{3}}+ { \eta_{2}\over 8} g_{a}^{(2)}\gamma^{a} B_{(4)}^{-1}\xi^{*}_{\eta_{1}  \eta_{2}\eta_{3}}=0
\label{BPS-K}
\ee

\subsubsection{Gravitino  in $\Sigma$ direction}
The covariant derivative in the $\Sigma$ directions is given by
\be
\nabla_{a}= D_{a}+ {1\over 4} \omega_{a}^{\;\; bc} \Gamma_{bc} -{i\over 2} q_{a}
\ee
The gravitino variation in the $\Sigma$ direction is given by
\bea
&&D_{a} \xi_{\eta_{1}  \eta_{2}\eta_{3}}+{i\over 2}\omega_{a} \gamma_{(4)}  \xi_{\eta_{1}  \eta_{2}\eta_{3}}-{i\over 2}q_{a} \xi_{\eta_{1}  \eta_{2}\eta_{3}}-{1\over 2} h_{b}\big(\delta ^{b}_{a} + i \epsilon_{\;\;a}^{ b} \gamma_{(4)}\big) \xi_{-\eta_{1}  -\eta_{2}\eta_{3}}\no\\
&&-i{\eta_{2}\over 16} g^{(1)}_{b} (3 \delta_{a}^{b}+ i \epsilon_{\;\;a}^{ b} \gamma_{(4)}) B_{(4)}^{-1}\xi^{*}_{-\eta_{1}  -\eta_{2}\eta_{3}}-{\eta_{2}\over 16} g^{(2)}_{b} (3 \delta_{a}^{b}+ i \epsilon_{\;\;a}^{b} \gamma_{(4)}) B_{(4)}^{-1}\xi^{*}_{\eta_{1}  \eta_{2}\eta_{3}}=0
\label{BPS-Sig}
\eea

\subsection{$\tau$ matrix notation}

At this stage we can introduce a matrix notation which will be useful to express the BPS equations in a compact form
\bea \xi_{-\eta} &=& (\tau^1 \xi)_{\eta} \no \\
\eta \xi_{-\eta} &=& i(\tau^2 \xi)_{\eta} \no \\
\eta \xi_{\eta} &=& (\tau^3 \xi)_{\eta} \label{actxi} \eea
where $\tau^{1,2,3}$ are the standard Pauli matrices. We define the matrices $\tau^{i,j}$ as:
\be \tau^{i j} = \tau^{i} \otimes \tau^j \ee
The conditions (\ref{BPS-D}),(\ref{BPS-AdS}),(\ref{BPS-S}),(\ref{BPS-K}) and (\ref{BPS-Sig}) can then be rewritten as follows:
\bea (D) &&  {1 \over 4} \Big( i g^{(1)}_a \g^a \tau^{(01)} - g^{(2)}_a \g^a \tau^{(10)} \Big) \xi - P_a \g^a \tau^{(13)} B^{-1}_{(4)} \xi^*  =0  \label{BPSredD} \\
(GA) &&  \Big( {\tau^{(30)}\over f_1} + {D_a f_1 \over f_1} \g^a \tau^{(11)} - h_a \g^a \Big) \xi + i \Big( {3 \over 8} g^{(1)}_a \g^a  \tau^{(03)} - {1 \over 8} g^{(2)}_a \g^a  \tau^{(12)}  \Big) B^{-1}_{(4)} \xi^* =0  \label{BPSredGA}\;\;\;\;\;\;\; \\
(GS) &&  \Big( {i \tau^{(03)} \over f_2} + {D_a f_2 \over f_2} \g^a \tau^{(01)} - h_a \g^a \tau^{(10)} \Big) \xi \no\\ &&  - i \Big( {1 \over 8} g^{(1)}_a \g^a  \tau^{(13)}
 - {3 \over 8}  g^{(2)}_a \g^a \tau^{(02)}  \Big) B^{-1}_{(4)} \xi^* =0  \label{BPSredGS}\\
(GK) &&  \Big( {D_a f_3 \over f_3} \g^a + h_a \g^a \tau^{(11)} \Big) \xi - \Big( {1 \over 8} g^{(1)}_a \g^a \tau^{(12)} - {1 \over 8}  g^{(2)}_a \g^a \tau^{(03)}  \Big) B^{-1}_{(4)} \xi^* =0 \label{BPSredGK } \\
(G \Sigma) &&  \Big( {D_a } - {i \over 2} \omega_a \tau^{(11)}- i {q_a \over 2} - {1 \over 2}  h_a \tau^{(11)} - {i \over 2} \epsilon^b_{\; a} h_b  \tau^{(11)} \sigma^{3} \Big) \xi + \no \\
&& {1 \over 16} \Big(  3 g^{(1)}_a \tau^{(12)} +i  \epsilon^b_{\; a} g^{(1)}_b \tau^{(12)} \sigma^{3}  - 3 g^{(2)}_a \tau^{(03)} - i \epsilon^b_{\; a} g^{(2)}_b \tau^{(03)} \sigma^{3}  \Big) B^{-1}_{(4)} \xi^* =0 \label{BPSredGSig} \eea
This set of equations is diagonal with respect to the index $\eta_3$ which is not displayed.

\subsection{Reduction of BPS equations by discrete symmetries}
\subsubsection{Discrete symmetries}

The reduced BPS equations have several useful symmetries. First, ten-dimensional chirality induces the discrete symmetry:
\be \mathcal{I} \xi = - \tau^{(11)} \sigma^3 \xi \ee
It is easy to show that $\xi$ is invariant under the symmetry since Killing spinors in type IIB supergravity have the same chirality:
\be \mathcal{I} \xi = \xi \label{chiralcond} \ee
Moreover, it is possible to introduce another discrete symmetry:
\be \mathcal{J} \xi = \tau^{(32)} \xi \ee
$\{ \mathcal{I},\mathcal{J}\} $ form a maximal set of commuting generators. We will see in the next section that the unbroken supersymmetries 
are in eigenspaces of these discrete symmetries.

\subsubsection{Vanishing  of bilinear constraints}

We obtain a first set of constraints from the chirality condition (\ref{chiralcond}). We get that:
\be \xi^+ M \sigma^a \xi \qquad \mbox{if }  [M,\tau^{(11)}]=0  \ee
that is, the spinor bilinear vanishes if:
\be M \in \{ \tau^{(10)},\tau^{(01)},\tau^{(11)},\tau^{(00)},\tau^{(22)},\tau^{(23)},\tau^{(32)},\tau^{(33)} \} \label{constr1} \ee
To get a second set of constraints we consider  a matrix $T$ such that:
\be (T \tau^{(10)})^T = - T \tau^{(10)}, \qquad (T\tau^{(01)})^T=-T\tau^{(01)} \label{Tac} \ee
The above condition is satisfied if:
\be T \in \{ \tau^{(22)}, \tau^{(33)} \} \ee
Because of (\ref{Tac}) we obtain the relations:
\bea \xi^T T  \tau^{(01)} \sigma^a \sigma^b \xi = \xi^T T  \tau^{(10)} \sigma^a \sigma^b \xi &=& 0  \no \\
\xi^+ T  \tau^{(01)} \sigma^a  \sigma^b \xi^* = \xi^+ T  \tau^{(10)} \sigma^a \sigma^b \xi^* &=& 0  \label{BLvanish} \eea
with $a,b=1,2$. We then multiply equation (\ref{BPSredD}) by $\xi^T \sigma^{b} $ with $b=1,2$. Because of the relations (\ref{BLvanish}), the first two terms of the equation vanish. We are left with:
\be P_a \xi^T T \sigma^b \sigma^a \tau^{(13)} \sigma^2 \xi^* = 0 \ee
In case $P^2 \neq 0 $ the above condition can be rewritten as:
\be \xi^+ T \tau^{(13)}  \sigma^a \xi = 0, \qquad a=1,2  \ee
This gives the extra constraints:
\be \xi^+ M \sigma^a \xi =0, \qquad M \in \{ \tau^{(31)}, \tau^{(20)} \} \label{BLvanish2} \ee
A third set of constraints can be obtained from the gravitino equations (\ref{BPSredGA}) and (\ref{BPSredGS}). If we multiply the 
equation (\ref{BPSredGA}) by $\xi^+ T \sigma^p \tau^{(02)}$ and  equation (\ref{BPSredGS}) by $\xi^+ T \sigma^p \tau^{(03)}$ with $p=0,3$ we get that the terms proportional to $g^{(1,2)}_a$ vanish and we are left with:
\bea {1 \over f_1} \xi^+ T \sigma^p \tau^{(32)} \xi - i {D_a f_1 \over f_1} \xi^+ T \sigma^p \tau^{(13)} \sigma^a \xi - h_a \xi^+ T \sigma^p \tau^{(02)} \sigma^a \xi &=& 0 \no \\
{i \over f_2} \xi^+ T \sigma^p \xi + i {D_a f_2 \over f_2} \xi^+ T \sigma^p \tau^{(02)} \sigma^a \xi - h_a \xi^+ T \sigma^p \tau^{(13)} \sigma^a \xi &=& 0   \eea
The second and third term of each equation vanishes because of the conditions (\ref{BLvanish2}). We are left with the constraints:
\be \xi^+ M \sigma^p \xi =0, \qquad M \in \{ \tau^{(2,2)}, \tau^{(3,3)}, \tau^{(0,1)}, \tau^{(1,0)} \} \ee
It is easy to verify that all the constraints can be satisfied provided that:
\be \xi= \nu \tau^{(3,2)} \xi, \qquad \nu=\pm 1 \label{xicond}\ee

\subsubsection{Projection of the BPS equations}


Projection of the BPS equations on spinors which satisfy ${\cal I}\xi=\xi$ and ${\cal J}\xi= \nu \xi$ can be achieved by writing the 
 spinor $\xi$ as: 
\be
\xi_{\eta_{1}\eta_{2}}=\left(
\begin{array}{c}
\alpha_{\eta_{1}\eta_{2}}\\
\beta_{\eta_{1}\eta_{2}}
\end{array}
\right)
\ee
with $\eta_{1,2}=\pm 1$. The solution to the projection conditions (\ref{chiralcond}) and (\ref{xicond}) can be expressed in terms of two independent components $\alpha$ and $\beta$.
\be\label{defxip}
\xi=\left(
\begin{array}{c}
\alpha_{++}   \\
  \beta_{++}  \\
  \alpha_{+-}   \\
  \beta_{+-}  \\
  \alpha_{-+}   \\
  \beta_{-+}  \\
  \alpha_{--}   \\
  \beta_{--}  \\
\end{array}
\right)=\left(
\begin{array}{c}
\;\;\;\;\;\;\,\alpha\\
\;\;\;\;\;\;\,\beta\\
+i \nu \, \alpha\\
+i \nu \,\beta\\
-i \nu \, \alpha\\
+i \nu \,  \beta\\
\;\;\;\;-\alpha\\
\;\;\;\;\;\;\beta
\end{array}
\right)
\ee

At this point we are left with four linearly independent complex Killing spinors which can be labeled by the eigenvalue $\nu=\pm 1$ and by the
$K_3$ index $\eta_3$. \\
It is convenient to introduce complex coordinates on the two-dimensional surface $\Sigma$.
The conventions for the complex coordinates are  $ z= x_{8}  + i x_{9} $, which implies for tensor indices $v_{z}={1\over 2}(v_{8}- i v_{9})$, $v_{\bar z}={1\over 2}(v_{8}+ i v_{9})$.
We also rewrite the zweibein, the spacetime derivative and the connection on $\Sigma$ in the complex coordinates. $w$, $\bar{w}$ are the spacetime coordinates so that:
\be e^z = \rho dw, \qquad D_z = \rho^{-1} \partial_w, \qquad \omega_z = i \rho^{-2} \partial_w \rho  \label {omegadef} \ee
The projected equations for the dilatino and gravitino along $AdS_{2}$ , $S^{2}$ and $K_3$ are now given by:
\bea
D: &&4 P_{z}\alpha^{*} - \Big( g^{(1)}_{z}+i\,g^{(2)}_{z}\Big) \beta=0 \label{BPSred2D1} \\
&&  4 \bar P_{ z}\beta - \Big( \bar g^{(1)}_{z}+i\,\bar g^{(2)}_{z}\Big) \alpha^{*}=0 \label{BPSred2D2}   \\
\no  \\
GA:&& {1\over f_{1}} \alpha+{2 D_{z} f_{1}\over f_{1}}\beta-2 h_{z} \beta
-\Big({3\over 4} g_{z}^{(1)}  -{i\over 4} g_{z}^{(2)}\Big) \alpha^{* }=0 \label{BPSred2GA1}\\
&& {1\over f_{1}} \beta^{*}- {2 D_{ z} f_{1}\over f_{1}}\alpha^{*}-2 h_{  z} \alpha^{*}
+\Big({3\over 4} \bar g_{ z}^{(1)} -{i\over 4}\bar  g_{ z}^{(2)}\Big) \beta=0 \label{BPSred2GA2}\\
 \no\\
 GS:&& {\nu\over f_{2}} \alpha  +{2 D_{z} f_{2}\over f_{2}}\beta-2 h_{z} \beta
+\Big({1\over 4} g_{z}^{(1)}  -i{3\over 4} g_{z}^{(2)}\Big) \alpha^{* }=0 \label{BPSred2GS1}\\
&& {\nu\over f_{2}} \beta^{*}+ {2 D_{ z} f_{2}\over f_{2}}\alpha^{*}+2 h_{ z} \alpha^{*}
+\Big({1\over 4} \bar g_{z}^{(1)} -i {3\over 4} \bar g_{ z}^{(2)}\Big) \beta=0 \label{BPSred2GS2}\\
 \no\\
 GK:&&  {2 D_{z} f_{3}\over f_{3}}\beta+2 h_{z} \beta
+\Big({1\over 4} g_{z}^{(1)}  +i{1\over 4} g_{z}^{(2)}\Big) \alpha^{* }=0 \label{BPSred2GK1}\\
&& {2 D_{ z} f_{3}\over f_{3}}\alpha^{*}-2 h_{z} \alpha^{*}
+\Big({1\over 4} \bar g_{z}^{(1)} +i {1\over 4} \bar g_{ z}^{(2)}\Big) \beta=0 \label{BPSred2GK2} \eea
The projected equation along the base $\Sigma$ gives the conditions:
\bea
G\Sigma:&& \Big(D_{z}  +{i\over  2} \omega_{z}-{i \;q_{z} \over 2}+h_{z}\Big)\alpha   -{1\over 4} \Big(g^{(1)}_{z}- i g^{(2)}_{z}\Big) \beta^{*} =0 \label{BPSred2az}\\
&&  \Big(D_{z}   -{i\over  2} \omega_{z} -{i \;q_{z} \over 2}    \Big)\beta -{1\over 8} \Big(g^{(1)}_{z}+ i g^{(2)}_{z}\Big) \alpha^{*}=0 \label{BPSred2bz} \\
&&\Big(D_{ z}  -{i\over  2} \omega_{ z} +{i \;q_{ z} \over 2}\Big) \alpha^{*}   -{1\over 8} \Big(\bar g^{(1)}_{ z}+ i \bar g^{(2)}_{ z}\Big) \beta =0 \label{BPSred2azb} \\
&& \Big( D_{ z} +{i\over  2} \omega_{z} +{i \;q_{ z} \over 2} - h_{z}\Big)\beta^{*}  -{1\over 4} \Big(\bar g^{(1)}_{ z}- i \bar g^{(2)}_{ z}\Big) \alpha=0 \label{BPSred2bzb} \eea
Note that a $g^{(1)}$ and $g^{(2)}$ are both complex, so that $\bar g^{(i)}_{z}$ and $\bar g^{(i)}_z$ are independent fields,
The presence of two anti-symmetric complex fluxes is a major difference from the  analysis of  \cite{D'Hoker:2007fq} which deals with supergravity solutions dual to half-BPS 
Wilson loops in $AdS_5\times S^5$, but has only real fluxes.


\subsection{Replacing the $K_3$ with a four torus }\label{t4vsk3}
 Note that both the $AdS_3\times S^3\times K_3$ as well as the $AdS_3\times S^3\times T^4$ vacuum solutions of type IIB supergravity
 preserve sixteen real supersymmetries. Since our ansatz is  independent of the four-dimensional manifold $K_3$ or $T^4$, we expect
 the solution to be unaffected by which manifold we choose. However, a covariantly constant spinor on $K_3$ has a fixed four-dimensional chirality while a covariantly constant spinor on $T^4$ has  both chiralities. 
 
One can repeat the reduction of the BPS equations for the opposite chirality. It is easy to see that  
the reduction of the spinor (\ref{defxip}) is the only change. The new spinor is denoted by hatted components
\be\label{defxipnew}
\hat \xi=\left(
\begin{array}{c}
\hat\alpha_{++}   \\
 \hat \beta_{++}  \\
 \hat \alpha_{+-}   \\
 \hat \beta_{+-}  \\
  \hat\alpha_{-+}   \\
 \hat \beta_{-+}  \\
 \hat \alpha_{--}   \\
\hat  \beta_{--}  \\
\end{array}
\right)=\left(
\begin{array}{c}
\;\;\;\;\;\;\,\hat\alpha\\
\;\;\;\;\;\;\,\hat\beta\\
+i \nu \, \hat\alpha\\
+i \nu \,\hat\beta\\
+i \nu \, \hat\alpha\\
-i \nu \, \hat \beta\\
\;\;\;\;\hat\alpha\\
\;\;\;-\hat\beta
\end{array}
\right)
\ee
The reduction of the dilatino equation (\ref{BPSredD}) for the new spinor (\ref{defxipnew}) gives 
\bea
0&=& 4 P_{z}\hat\alpha^{*} - \Big( -g^{(1)}_{z}+i\,g^{(2)}_{z}\Big)\hat  \beta  \label{otherchia} \\
0&=& 4 \bar P_{ z}\hat \beta - \Big( -\bar g^{(1)}_{z}+i\,\bar g^{(2)}_{z}\Big) \hat \alpha^{*} \label{otherchib}
\eea 
 In the asymptotic $AdS_3\times S^3$ region  the axion and dilaton approach constant values and one can set $P_z=0$. 
 In this limit, demanding that 
 the original dilatino equations (\ref{BPSredD}), (\ref{BPSred2D2}) are  satisfied for non-vanishing spinors $\alpha,\beta$,  implies that   the components of the three-form tensor fields  are self-dual.
 On the other hand,   demanding that 
 the new dilatino equations  (\ref{otherchia}),(\ref{otherchib}) are  satisfied for non-vanishing spinors $\hat \alpha, \hat \beta$,  implies that   the components of the three-form tensor fields  are anti self-dual.
 The two  conditions can only be satisfied at the same time if $g^{(1)}=g^{(2)}=0$,  since  a three-form tensor field which is both self-dual  and anti self-dual is automatically zero in six dimensions.  The resulting solution corresponds to a six-dimensional Minkowski vacuum, not an asymptotically $AdS_3\times S^3$ spacetime.
 
 Hence, for non-trivial anti-symmetric tensor fields one of the two sets of spinors must be zero. We choose the unbroken supersymmetries to be associated with $\alpha,\beta$ and  hence have asymptotically self-dual anti-symmetric tensor fields. The other choice would correspond to flipping the sign of the D1-brane charges.  

 In conclusion, the BPS equations (\ref{BPSredD}-\ref{BPSredGSig}) are valid for both a $K_3$ manifold and a four torus. In both cases the solution preserves eight of the sixteen unbroken supersymmetries.  Note that there 
would be a difference if we were to turn on the internal moduli, since $K_3$ and $T^4$ have different Hodge numbers.

\section{Local solution of BPS equations}\label{secthree}
\setcounter{equation}{0}
\subsection{Expressions for the metric factors }\label{metricsec}
It is possible to obtain an expression for the metric factor $f_1$ taking a linear combination of equation (\ref{BPSred2GA1}) and (\ref{BPSred2GA2}): 
\be 2 { D_{z} f_{1}\over f_{1}}(\alpha \alpha^* + \beta \beta^*) + 2 h_{z} (\alpha \alpha^* - \beta \beta^*) -\Big({3\over 4} g_{z}^{(1)}  -{i\over 4} g_{z}^{(2)}\Big) \alpha^{* } \beta^* - \Big({3\over 4} \bar g_{z}^{(1)}  -{i\over 4} \bar g_{z}^{(2)}\Big) \alpha \beta=0 \ee  
We can then use equations (\ref{BPSred2az}-\ref{BPSred2bzb}) to eliminate the fluxes. We are left with the differential equation:
\be { D_{z} f_{1}\over f_{1}}(\alpha \alpha^* + \beta \beta^*) -  D_{z} (\alpha \alpha^* + \beta \beta^*)=0 \ee 
It is possible to obtain similar equations for $f_2$ and $f_3$. The metric factors are then found to be: 
\bea
f_{1}&=& c_{1}\big( \alpha^{*} \alpha+\beta^{*}\beta\big) \label{def-f1}\\
f_{2}&=& c_{2}\big( \alpha^{*} \alpha-\beta^{*}\beta\big) \label{def-f2} \\
f_{3}&=&  {1\over \sqrt {\rho \bar{ \sigma} \alpha^{*} \beta}} \label{def-f3}
\eea
where $c_1, c_2$ are real constants and $\sigma(w)$ is a holomorphic function such that:
\be \bar{\sigma} \alpha^* \beta = \sigma \alpha \beta^* \label{Realf3} \ee
Furthermore, is possible to eliminate the terms with $D_z$ in (\ref{BPSred2D1}-\ref{BPSred2GK2}) obtaining:
\bea {1\over c_1} - 4 h_z \a^* \b - {1 \over 4} (3 g^{(1)}_z - i g^{(2)}_z ) \a^{*2}+ {1 \over 4} (3 \bar{g}^{(1)}_z - i \bar{g}^{(2)}_z) \b^2 &=&  0 \\
{\nu\over  c_2} - 4 h_z \a^* \b + {1 \over 4} ( g^{(1)}_z - 3 i g^{(2)}_z ) \a^{*2}- {1 \over 4} ( \bar{g}^{(1)}_z - 3 i \bar{g}^{(2)}_z) \b^2 & = & 0 \\
4 h_z \a^* \b + {1 \over 4} (g^{(1)}_z + i g^{(2)}_z ) \a^{*2}- {1 \over 4} ( \bar{g}^{(1)}_z + i \bar{g}^{(2)}_z) \b^2 &=& 0  \eea
Combining the above equations we get the condition:
\be c_2 + \nu c_1 = 0 \ee
With a simple rescaling of the metric we can then set:
\be c_1 = 1 \qquad c_2 = -\nu \ee

\subsection{Spinor components in terms of two holomorphic functions}\label{spinconsec}

It is possible to use equations (\ref{BPSred2D1}-\ref{BPSred2GS2}) to solve for the fields $g^{(1,2)}$ and $\bar{g}^{(1,2)}$:
\bea g^{(1)}_z - i g^{(2)}_z &=& {2 \a^2 \over ( \a \a^*)^2 - (\b \b^*)^2 } + {2 \b \over \a^*} D_z \ln \Big( {  \a \a^* + \b \b^* \over   \a \a^* - \b \b^*} \Big)\label{gdefine1} \\
 \bar{g}^{(1)}_z - i \bar{g}^{(2)}_z &=& { 2 \b^{*2} \over ( \a \a^*)^2 - (\b \b^*)^2 } + {2 \a^* \over \b} D_z \ln \Big( {  \a \a^* + \b \b^* \over   \a \a^* - \b \b^*} \Big) \\
g^{(1)}_z + i g^{(2)}_z &=& 4 {\a^* \over \b} P_z \\
\bar{g}^{(1)}_z + i \bar{g}^{(2)}_z &=& 4 {\b \over \a^* } \bar{P}_z \label{gdefine4} \eea
Substituting these expressions into (\ref{BPSred2bz}-\ref{BPSred2azb}) gives two equations:
\bea \Big( D_z - i \omega_z - i q_z \Big) \b^2 - \a^{*2} P_z &=& 0 \label{a-sys1} \\
 \Big( D_z - i \omega_z + i q_z \Big) \a^{*2} - \b^2 \bar{P}_z &=& 0 \label{a-sys2} \eea
It is now convenient to use the expressions (\ref{compform}):
\bea P_z &=& {1 \over 2}  \big(D_z \phi  + i e^{\phi} D_{z} \chi \big)  \label{defPz} \\
q_z &=& - {1 \over 2} e^{\phi} D_z \chi \label{defqz} \eea
 We also use the spin connection and spacetime coordinates (\ref{omegadef}).
Equation (\ref{a-sys1}) then becomes:
\bea \Big( \partial_w + {\partial_w \rho \over \rho}  + {i \over 2} e^{\phi} \partial_w \chi  \Big) \b^2 - {\a^{*2} \over 2} e^{ \phi} \partial_w \big( e^{- \phi} + i \chi \big) &=& 0  \\
\Big( \partial_w + {\partial_w \rho \over \rho}  - {i \over 2} e^{\phi} \partial_w \chi  \Big) \a^{*2} - {\b^2 \over 2} e^{ \phi} \partial_w \big( e^{- \phi} - i \chi \big) &=& 0 \eea
Taking appropriate linear combinations simplifies the system to:
\bea \partial_w \ln \big[ \rho e^{\phi /2 } \big( \b^2 - \a^{*2} \big) \big]  &=& 0  \\
\partial_w \ln \big[ \rho e^{-\phi /2} \big( \b^2 + \a^{*2} \big)\big] &=& -i {\b^2 - \a^{*2} \over \b^2 + \a^{*2}} e^{\phi} \partial_w \chi \eea
These equations are solved by:
\bea \a^{*2} & = & - {\bar{k} \over \rho} \sinh (\bar{\lambda} + \Phi) - i {\bar{k} \over 2 \rho} e^{\bar{\lambda}- \Phi} \chi \\
\b^{2} & = &  {\bar{k} \over \rho} \cosh (\bar{\lambda} + \Phi) - i {\bar{k} \over 2 \rho} e^{\bar{\lambda}- \Phi} \chi \label{solab} \eea
Note that the spinors $\beta$ and $\alpha^*$ transform with weight $(-\half,0)$ with respect to the $SO(2)$ frame rotations. Since $\rho$ has weight $(\half,\half)$ it follows that $\bar k$ has weight $(-\half, \half)\sim(0,1)$ and that $e^\lambda$ has weight $(0,0)$.
 We have redefined the dilaton as
\be \phi= -2 \Phi \ee

\subsection{Reduction to one equation}

To simplify (\ref{BPSred2az}) and (\ref{BPSred2bzb}) we eliminate $h_z$ using
 (\ref{BPSred2GK1}-\ref{BPSred2GK2}) in (\ref{a-sys1}-\ref{a-sys2}).
\be\label{hzdefine} 4 h_z + P_z  {\a^{*2} \over \b^2} - \bar{P}_z {\b^2 \over \a^{*2}} = 0  \Rightarrow 2 h_z = D_z \ln{\alpha^* \over \b} + i q_z \ee
Plugging the above expression for the fluxes into (\ref{BPSred2az}, \ref{BPSred2bzb}) gives another system of differential equations.
\bea
D_z \ln \Big( {\a^2 \a^* \over \b} \Big) + i  \omega_z - {\b \b^* \over  \a \a^*}  D_z \ln \Big( { \a \a^* + \b \b^* \over   \a \a^* - \b \b^*} \Big) - {\a \b^*  \over ( \a \a^*)^2 - (\b \b^*)^2 } &=& 0 \label{a+sys1}\\
D_z \ln \Big( {\b \b^{*2} \over \a^*} \Big) + i \omega_z - {\a \a^* \over  \b \b^*}  D_z \ln \Big( { \a \a^* + \b \b^* \over   \a \a^* - \b \b^*} \Big) - {\a \b^*  \over ( \a \a^*)^2 - (\b \b^*)^2 } &=& 0 \qquad \label{a+sys2} \eea
Note that the axion and dilaton have dropped out of this system. Furthermore, this system is actually linearly dependent since 
the difference of the two equations is automatically true. The sum of the two equations can be simplified to
\be
\p_w \ln \big({\alpha\over\beta^* \rho}\big)- {|\alpha|^4+|\beta|^4 \over |\alpha|^4-|\beta|^4}\p_w\ln{|\beta|^2\over|\alpha|^2}     -{\alpha\beta^*\rho \over |\alpha|^4-|\beta|^4}=0
\ee
Using the spinor expressions (\ref{solab}) and defining the following field
\be e^{\psi} =- i {\rho^3  \a^* \b \over |k|^2 \cosh (\lambda- \bar \lambda)}
 = i {\rho  \a^* \b \over |\a|^4-|\b|^4}
 \label{defPsi}\ee
we can further reduce the equation to:
\be\label{firstord}
 \partial_w \psi + i  e^{\bar{\psi}} = 0
 \ee
The general solution is shown to be:
\be\label{resultpsi}
e^{\psi}=-{1 \over i} { {\partial_{\bar w}H}\over  H}, \quad \quad e^{\bar \psi}={1 \over i} {  {\partial_{w}H}\over H }
\ee
Using the definition (\ref{defPsi}) we can solve the condition on the holomorphic function $\sigma$ that was found in section \ref{metricsec}
in terms of the arbitrary harmonic function $H$ :
\be e^{\psi-\bar \psi} = -{\alpha^* \b \over \a \b^*} =- {\sigma \over \bar \sigma} \quad  \Rightarrow \qquad \sigma = {\text{const} \over \partial_w H } \label{def-sigma}  \ee
From now on we will set the constant to one. $H$ is an arbitrary harmonic function which parameterizes the solution.
Plugging $\sigma$ into (\ref{def-f3}) gives the $K_3$ metric factor:
\bea
 f_3^4&=& {(\partial_{\bar w} H)^2\over \rho^2 \alpha^{*2} \beta^2} = {\rho^4 H^2\over |k|^4 \cosh^2(\lambda-\bar \lambda)} \no \\
 &=&{4(\partial_{\bar w} H)^2\over \bar k^2 \Big( e^{-2\bar \lambda-2\Phi}-e^{2\bar \lambda +2\Phi}-e^{2\bar \lambda -2\Phi}\chi^2-2i e^{-2\Phi}\chi\Big)}
\label{f3inspinors}
\eea
We used (\ref{resultpsi}) and (\ref{defPsi}) in obtaining the second equality. The third equality is obtained using the relation:
\begin{equation}
\alpha^{*2} \beta^2={\bar k^2\over 4\rho^2} \Big( e^{-2\bar \lambda-2\Phi}-e^{2\bar \lambda +2\Phi}-e^{2\bar \lambda -2\Phi}\chi^2-2i e^{-2\Phi}\chi\Big)
\end{equation}
which follows from (\ref{solab}). Equation (\ref{f3inspinors}) determines $\rho$ in terms of the two holomorphic functions, the axion and the dilaton.

At this point,  spinor components, fluxes  and all the metric factors are known in terms of dilaton, axion, two holomorphic functions and one harmonic function. 
The BPS equations on their own are underdetermined and  the Bianchi identities are necessary to constrain axion and dilaton.

The benefit of hindsight allows us  to say that the extra condition is obtained from the Bianchi identity for the five-form in the $K_3$ directions.

\subsection{Five-form Bianchi identity}

The reduction of  (\ref{bianchi4}) in components along the $K_3$ directions yields
\be \partial_{\bar z} \big( f^4_3 \rho \tilde h_z \big) - \partial_z \big( f^4_3 \rho \tilde h_{\bar z} \big) = 0 \label{Bianchi-h1} \ee
We can find a convenient expression for $ \tilde h_z = - i h_z $ using equations (\ref{BPSred2GK1}), (\ref{BPSred2D1}), (\ref{defPz}) and (\ref{def-f3}):
\be f^4_3 \rho \tilde h_z =  {i \over 4} \partial_w f^4_3 +{i \over 2} {\partial_{\bar w} H \over \rho^2 \b^4} \rho P_z  =  {i \over 4} \partial_w f^4_3 - {i \over 2} {\partial_{\bar w} H \over \rho^2 \b^4} \Big( \partial_w \Phi - {i \over 2} e^{- 2 \Phi} \partial_w \chi \Big) \ee
Plugging in (\ref{solab}) we can rewrite the above equation as a total derivative:
\be f^4_3 \rho \tilde h_z =   i \partial_w \left( {f^4_3 \over 4} + \vartheta_5 \right), \qquad \vartheta_5 = { ( \partial_{\bar w} H)^2 e^{- 2 \bar \lambda}/ {\bar k}^2 \over e ^{2 \Phi} + e^{- 2 \bar \lambda} - i \chi } = {e^{-\Phi - \bar \lambda} \a^{* 2} \rho f^4_3 \over 2  \bar k }  \ee
The Bianchi identity then leads to the condition:
\be \partial_w \partial_{\bar w} \left( {f^4_3 \over 4} e^{-2 \Phi}  \Re \big( e^{-2 \lambda} \big)  \right) = 0 \label{resultBianchi} \ee
This condition is not automatic and gives us a restriction on the dilaton and axion (hidden within the metric factor $f_3$) in terms of a new harmonic function $\hat h$
defined as
\be
{1 \over 4} f_3^4 e^{-2\Phi} (e^{-2\lambda}+ e^{-2\bar \lambda})=\hat h \label{hhatdef}
\ee

\subsection{Complete local solution}\label{totsolution}

We now have two equations for  $f_3$ involving  holomorphic functions, the dilaton and the axion only, (\ref{hhatdef}) and (\ref{f3inspinors}).
We can use them to obtain an equation for the dilaton and the axion, together with its complex conjugate:
\bea
 (e^{-2\bar \lambda} - i\chi)^2-e^{4\Phi}&=& {(\partial_{\bar w} H)^2\over e^{2\bar \lambda}\bar k^2 } {e^{-2\lambda} +e^{-2\bar \lambda} \over \hat h}\\
 (e^{-2 \lambda} + i\chi)^2-e^{4\Phi}&=& {(\partial_{ w} H)^2\over e^{2\lambda} k^2 } {e^{-2\lambda} +e^{-2\bar \lambda} \over \hat h}
\eea
To simplify notation, we redefine our  holomorphic functions as follows:
\be
B={\partial_w H\over e^{\lambda} k}, \qquad  A= e^{-2 \lambda}
\ee
From the assignments of  weights under $SO(2)$ frame rotation in section \ref{spinconsec}, it follows that $k$ and $\partial_w H$ are forms of the same weight and hence both $A$ and $B$ have vanishing weight.

The following combinations of the metric factors have simple expressions in terms of $\alpha$ and $\beta$:
\bea
f^2_1-f^2_2 &=& 4 |\a \b |^2 = {e^{- 2 \Phi} \over \rho^2} {A + \bar A \over \hat h } |\partial_w H|^2\\
f^2_1+f^2_2 &=& 2 \big( |\beta|^4+|\alpha|^4 \big) = {e^{-2\Phi} \over 2 \rho^2} {A + \bar A  \over  |B|^2 \hat h } \Big( (A + \bar A) \hat h - B^2 -\bar B^2 \Big)
\eea

At this point we are able to find convenient expressions for all bosonic fields. The solutions for $\chi$ and $\Phi$ are
\be
 \chi= {1\over 2i} \Big( {B^2 -\bar B^2 \over \hat h} -A + \bar A \Big)
\ee
and
\bea
 e^{4\Phi} 
& = & {1\over 4} \Big( A + \bar A  - {(B + \bar B)^2 \over \hat h} \Big) \Big( A + \bar A - {(B - \bar B)^2 \over \hat h} \Big)
\eea
The expression for the metric factor  $f_3$ becomes
\be
f_3^4=4 {e^{2\Phi} \hat h \over A + \bar A} \label{sol-f3}
\ee
Equation (\ref{sol-f3}) and the second expression of (\ref{f3inspinors}), rewritten in terms of the new holomorphic functions, give the new form of  $\rho$:
 \be
\rho^4= e^{2\Phi} \hat h {|\partial_w H|^4  \over H^2} { A + \bar A \over |B|^4 } \label{sol-rho}
\ee
We then obtain the following expressions for the metric factors:
\bea
f^2_1 &=&  {e^{- 2 \Phi} \over 2 f_3^2} {|H| \over \hat h}  \Big( (A + \bar A) \hat h  -  (B - \bar B)^2 \Big) \label{sol-f1} \\
f^2_2 &=&  {e^{- 2 \Phi} \over 2 f_3^2} {|H| \over \hat h} \Big(  (A+ \bar A) \hat h  -   (B  + \bar B)^2  \Big) \label{sol-f2} \eea
In appendices \ref{bianchiappendix} and \ref{EOMappendix} it is shown that for a solution of the BPS equations and
 Bianchi identity (\ref{resultBianchi}), the remaining Bianchi identities and equations of motion are automatically satisfied.
 Since this is the case, we can derive the two-form potentials  along the two-sphere by rewriting the three-form field strengths as total derivatives.
\bea 
 f_2^2 \rho e^{-\Phi} \Re(g^{(2)})_z& = & \partial_w b^{(2)} \label{potdef2}\\
f_2^2 \rho e^{\Phi} \Im(g^{(2)}) _z+ \chi f_2^2 \rho e^{-\Phi} \Re(g^{(2)})_z& = & \partial_w c^{(2)}\label{potdeftwo} \eea
 The potentials written in terms of our holomorphic and harmonic functions are
\bea 
 b^{(2)} &=& -i  {H (B - \bar B) \over (A + \bar A) \hat h - (B - \bar B)^2 } + \tilde h_1, \qquad  \tilde h_1={1 \over 2 i} \int {\partial_w H \over B} + c.c. \label{potharmonic2}\\
c^{(2)} & = & - {H (A \bar B +  \bar A B) \over (A + \bar A) \hat h - (B - \bar B)^2 } + h_2, \qquad  h_2={1 \over 2 } \int {A \over B}\partial_w H + c.c.  \label{potharmonic4}\eea

The Maxwell charges related to the RR and NS-NS three form are defined as
\bea
q_{NS} = \int_{S^3}  	e^{-\Phi} Re(G), \quad q_{RR}= \int_{S^3} e^{\Phi} Im(G)
\eea
and can be calculated from (\ref{potdef2}) and (\ref{potdeftwo}).

A four-form potential can also be defined for the five-form field strength. The components  along $AdS_2\times S^2$ and $K_3$ are related by self-duality and we give
 the one along $K_3$:
\be 
 f^4_3 \rho \tilde h_z = \partial_w C_K \quad 
C_K = -{i \over 2} {B^2 - \bar B^2 \over A + \bar A} -{1\over 2} \tilde {h}
 \label{eq-Ck} \ee
Here $\tilde { h} $ is the harmonic function conjugate to $\hat h$ so that  $ \partial_w \tilde { h} = - i \partial_w  \hat  h $. Note that the harmonic function $\tilde h$ should not be confused with ${\tilde h}_z$.
Some details of the derivations are provided in Appendix \ref{bianchiappendix}.

In summary, our solution is determined by two independent holomorphic functions $A$, $B$ and two independent harmonic
 functions $\hat h$ and $H$. Alternatively, since $A\pm \bar A$ are dual harmonic functions and $B \pm \bar B$ are dual harmonic
 functions , we can parameterize our solution in terms of four independent harmonic functions. The conditions guaranteeing
 the regularity of $f_1^2$, $f_2^2$, and $e^{\Phi}$ are discussed in section \ref{reg-cond-sec}.

\section{Regularity and half-BPS Janus solution}\label{secfour}
\setcounter{equation}{0}

In this section we discuss the conditions imposed by regularity on our solutions. In particular, we will restrict our analysis to the case in which $\Sigma$ is a genus zero Riemann 
surface with a single boundary and the functions $H, \hat h, A$ and $B$ admit only singularities of a certain class on $\Sigma$. We will present simple Janus deformations of the three parameter family of $AdS_3 \times S^3$ vacua and general expressions for regular solutions having three or 
more $AdS_3 \times S^3$ regions. 

\subsection{Symmetries of the solutions}

The analysis of the regular BPS solutions can be simplified by using the symmetries of the theory. 
First of all, we note that we can rescale the harmonic functions as:
\be  B \rightarrow c B,  \qquad  \hat h \rightarrow c^2 \hat h, \qquad H \rightarrow c H \label{sym-1} \ee
leaving all fields unchanged provided that we change the constant in (\ref{def-sigma}) by a factor of $c^{-1}$.
Similarly, the we can rescale the harmonic function $H$ as:
\be  H \rightarrow c H  \label{sym-2} \ee
and find that the metric factors and two-form potentials change only by an overall scale if we also multiply the constant in (\ref{def-sigma}) by 
a factor of $c$.\\

Moreover, the $SL(2,R)$ symmetry of type IIB supergravity maps regular supersymmetric solutions into different regular supersymmetric solutions and has a simple action on our harmonic functions.
$S$-duality, acting as $\tau \rightarrow -1/\tau $ on the axion-dilaton system, transforms the harmonic and holomorphic functions as:
\be A \rightarrow {1 \over A}, \qquad B \rightarrow i {B \over A}, \qquad \hat h \rightarrow \hat h - {B^2 \over A } - {\bar B ^2 \over \bar A} \label{Sduality} \ee
The scale symmetry $\tau \rightarrow a^2 \tau $ acts as:
\be A \rightarrow a^2 A, \qquad B \rightarrow a B \ee
and the shift symmetry $\tau \rightarrow \tau + b$ has the action:
\be A \rightarrow A - i b \label{sym-shift} \ee 
Our solutions also display several discrete symmetries. In particular the discrete transformation:
\be   B \rightarrow -B, \qquad H \rightarrow -H  \label{sym-disc1} \ee
leaves all the fields unchanged while the transformation:
\be H \rightarrow - H  \label{sym-disc2} \ee
simply flips the sign of all the two-form potentials while leaving dilaton, axion, metric factors and the four-form potential invariant.\\
Finally, the transformation:
\be A \rightarrow - A, \qquad \hat h \rightarrow - \hat h \label{sym-disc3} \ee
flips the sign of the R-R potentials while leaving the NS-NS potentials and the other fields and metric factors unchanged.

\subsection{$AdS_3\times S^3$ vacua}

The $AdS_3 \times S^3$ slicing into $AdS_2\times S^2$ spaces is given in \cite{Bak:2003jk} and corresponds to the metric factors:
\be
f^2_1= \cosh^2 x, \qquad f^2_2 = \sin^2 y, \qquad \rho= 1,  \qquad f_3= \text{const}
\ee
where $x$ and $y$ are the real and imaginary part of $w$ and the $AdS_3$ and $S^3$ spaces both have unit radius. With this parameterization, the surface $\Sigma$ corresponds to a strip in the complex plane:
\be x \in (- \infty, +\infty), \qquad y \in [0,\pi] \ee  
and the boundary $\partial \Sigma$ is given by the lines $y = 0$ and $y = \pi$.
The dilaton and axion assume constant values and our vacua solutions are charged under the three-form anti-symmetric tensor fields. \\

In order to derive expressions for the harmonic functions corresponding to the vacua solutions, we first use equation (\ref{solab}) and (\ref{sol-f3}) to  
get the relations:
\bea
A &=&e^{- 2 \lambda} = e^{2 \Phi} {\b^{*2}+\a^2 \over \b^{*2}-\a^2 }-i \chi \label{A-sl-int}\\
B &=& f^2_3 e^{ \Phi} {\a \b^* \over  \b^{*2}-\a^2 } \\
\hat h &=& {f^4_3 \over 4} e^{- 2 \Phi} (A + \bar A)  \eea
Moreover, $H$ has a simple expression in terms of the metric factors while the spinor components $\a$ and $\b$ are determined from the metric factors up to a constant phase:
\be H^2=f^2_1 f^2_2 f^4_3, \qquad |\a|^2={f_1 - \nu f_2 \over 2}, \qquad |\b|^2= {f_1+ \nu f_2 \over 2} \label{H-sl-int} \ee
Using the relations (\ref{A-sl-int}-\ref{H-sl-int}) it is possible to show that the following functions lead to $AdS_3 \times S^3$ vacua:
\bea H &=& - i \sinh w + c.c. \label{harm-vac1}\\
 A &=& - i \epsilon^2 {\sin \gamma + \cos \gamma  \sinh w \over \cos \gamma - \sin \gamma \sinh w } - i \delta  \label{harm-sl1}\\
 B &=& - i \epsilon { \cosh w \over \cos \gamma - \sin \gamma \sinh w } \\
\hat h &=& {A+ \bar A \over \epsilon^2 }  \label{harm-sl3} \eea
The real parameters $\delta$ and $\epsilon$ are related to the values of dilaton and axion:
\be e^{\Phi}= \epsilon, \qquad \chi = \delta  \ee
It is easy to see that the harmonic functions $H$, $\hat h$, $A + \bar A$ and $B+ \bar B$
all obey Dirichlet boundary conditions for $y=0$ and $y=\pi$. \\

The harmonic function $H$ is singular for $x \rightarrow \pm \infty $ while, for generic values of the parameters, $A+\bar A$, $B + \bar B$ and $\hat h$  will vanish for $x \rightarrow \pm \infty $ and have singularities for:
\be \sinh x = \cot \gamma \; , \; y=0 \qquad \text{and} \quad \sinh x = - \cot \gamma \; , \; y= \pi \ee
The functions $H$, $A + \bar A$ and $\hat h$ do not have zeros in the bulk of $\Sigma$ while the holomorphic function $B$ vanishes for $w= i \pi/2$. \\
The charges of the solutions can be obtained finding the flux of the three-form anti-symmetric tensor fields on the three-sphere, which corresponds to a curve on $\Sigma$ starting on the $y=0$ boundary and ending on the $y=\pi$ boundary (as shown in Figure \ref{fig-wu}).
The integral of each three-form field along the curve is equal to the change in potential between the two endpoints. In conclusion we get for the Maxwell charges
\bea q_{RR} &=& c^{(2)}(y=\pi)-c^{(2)}(y=0)-  \chi\big(  b^{(2)}(y=\pi)-b^{(2)}(y=0)\big)\nonumber \\
& = &\pi    \epsilon   \sin \gamma  \\
q_{NS} &=& b^{(2)}(y=\pi)-b^{(2)}(y=0) \nonumber \\
&=&  \pi { \cos \gamma \over \epsilon } \label{qNSvac}\eea
In particular, a pure R-R solution can be obtained with $\gamma = \pi /2$ and has poles on the imaginary axis for $y=0$ and $y=\pi$.  

The harmonic functions (\ref{harm-vac1})- (\ref{harm-sl3}) depend on three parameters, we have however set the volume of $K_3$ as well as an overall scale to one for simplicity. In addition the dual harmonic function to $\hat h$ contains a constant related to the value of $C_4$ on $K_3$. Hence the vacuum solutions depend on six independent parameters.
\subsection{Regularity analysis \label{reg-cond-sec}}

\begin{figure}
\centering
\includegraphics[angle=90,scale=0.28]{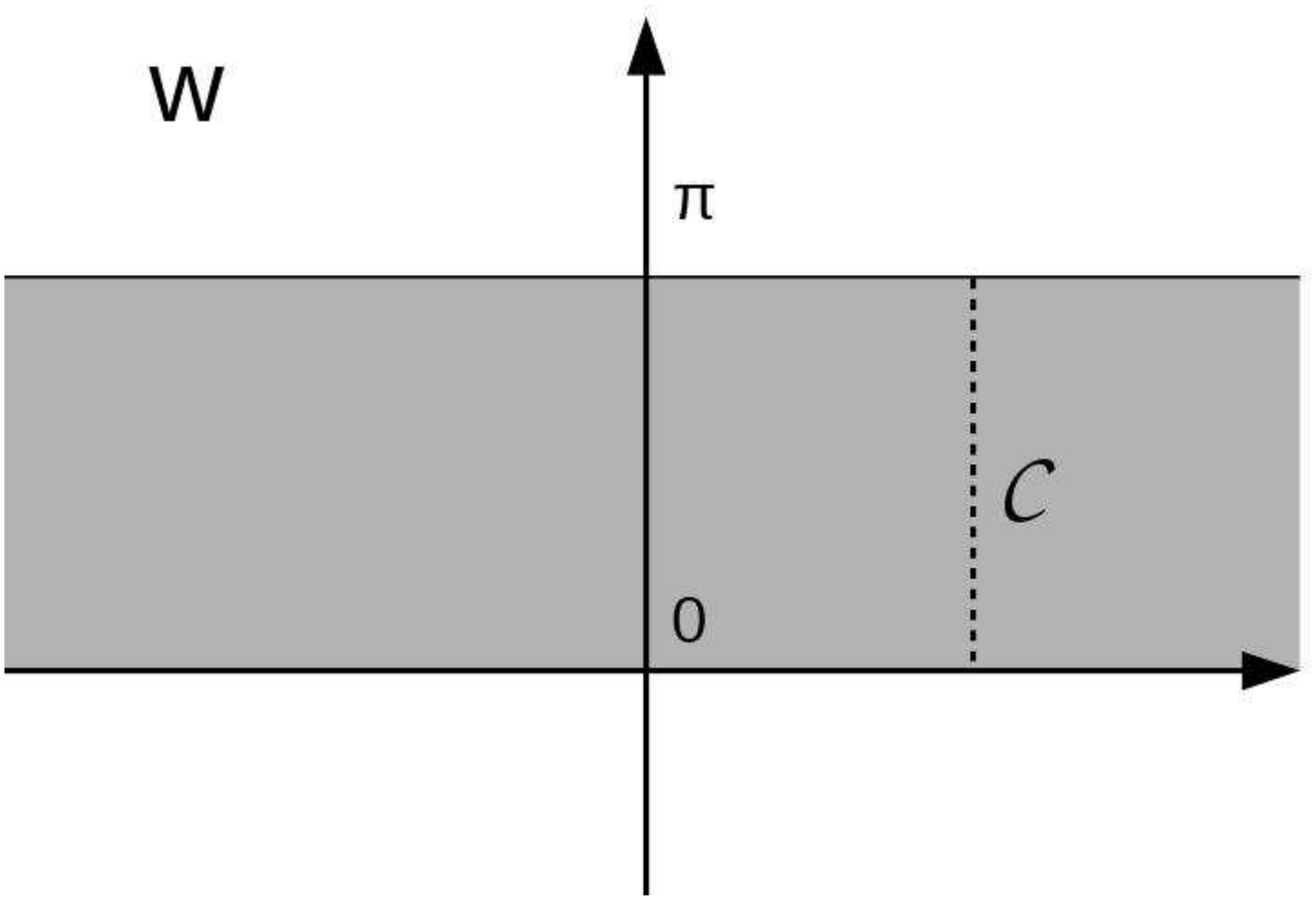}
\includegraphics[angle=90,scale=0.28]{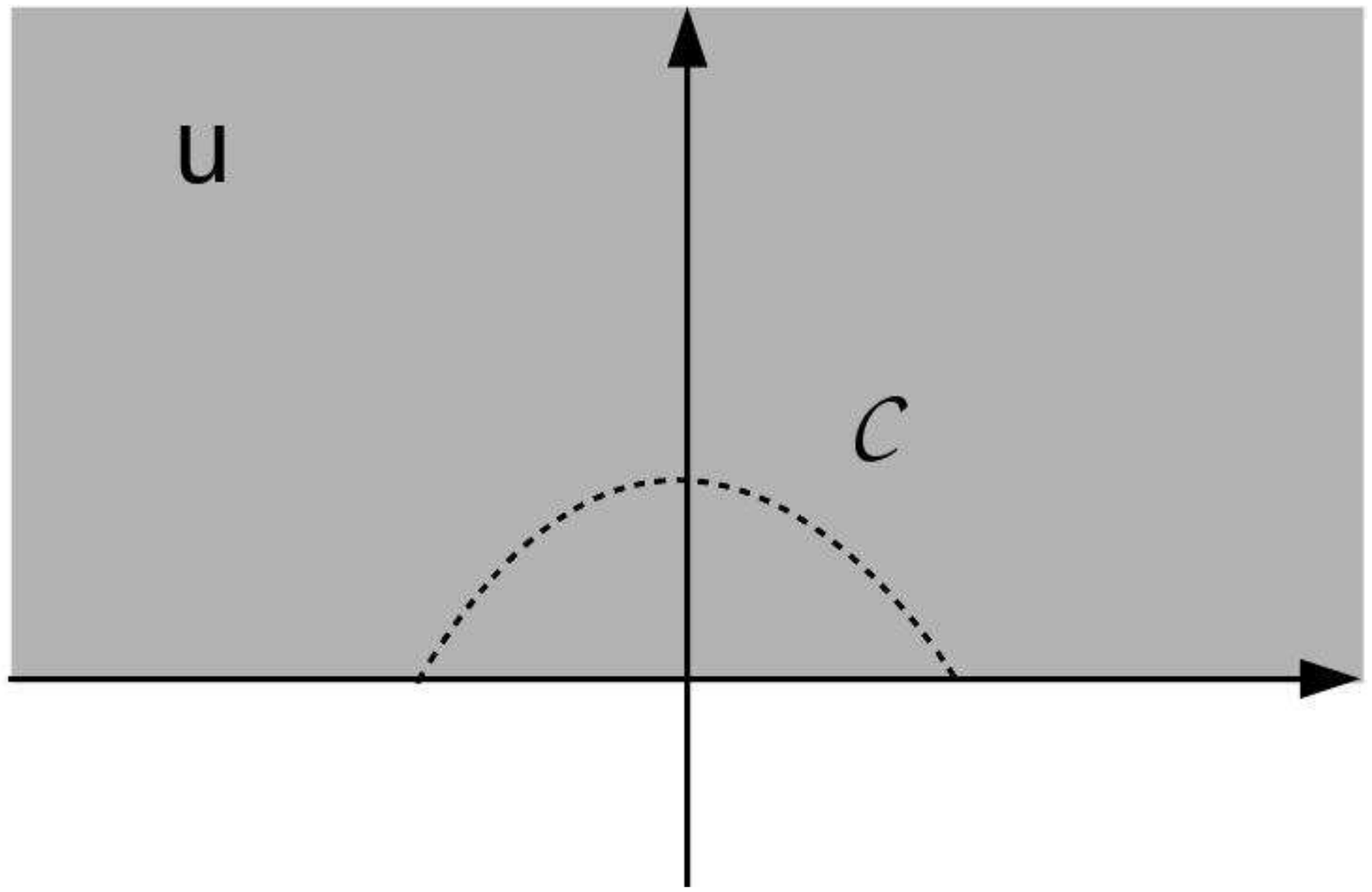}
\caption{Change of coordinates mapping the strip into the upper half plane.}
\label{fig-wu}
\end{figure}
At this stage, it is useful to introduce the new variable $u=e^{w}$. This change of coordinates maps the strip into the upper half-plane with the real axis $\Im u=0$ as the boundary of $\Sigma$ together with a point at infinity (as shown in Figure \ref{fig-wu}).\\
It is easy to see that the holomorphic functions in the previous section become rational functions with simple poles of order one when expressed in terms of $u$. \\
In this paper we will restrict our analysis to regular solutions for which the Riemann surface $\Sigma$ has genus zero and a single boundary component deferring the more general analysis to future work. 
Under this assumption we can always find a change of coordinates mapping $\Sigma$ into the upper half-plane. \\
Moreover, we will analyze only regular solutions with $AdS_3 \times S^3$ asymptotics. We will also restrict our analysis to the case in which the holomorphic functions $A$, $B$ and the holomorphic part of $H$ and $\hat h$ only admit simple poles of order one.\\
In order to avoid curvature singularities and in analogy with similar work in \cite{D'Hoker:2007fq} and \cite{D'Hoker:2007xz}, we will consider only solutions in which:
\begin{itemize}
\item the radius of the $AdS_2$ slice, given by the metric coefficient $f_1$, is non-zero and finite everywhere except at most isolated singular points. The singularities correspond to $AdS_3 \times S^3$ asymptotic regions. 
It is unclear whether there is a different class of regular solutions which do not respect this condition and hence have different asymptotics. 
\item the radius of the two-sphere, given by $f_2$, is finite on $\Sigma$ and zero on the boundary. The boundary is defined as the locus in which $f_2$ vanishes except at most a set of isolated points.
\item the volume of the $K_3$ manifold (given by $f_3$) and the dilaton are finite and non-zero everywhere.   
\end{itemize}
We can use the expressions (\ref{sol-f1}),(\ref{sol-f2}) and (\ref{sol-f3}) to prove that: 
\be f^2_1 f^2_2 f^4_3  = H^2  \label{reg1} \ee
Since $f_2=0 $ on $\partial \Sigma$, it follows that $H$ must vanish identically on the boundary. We can see from (\ref{sol-f1}) that, in order for $f_1$ to be 
finite on the boundary we need $\hat h$ to vanish identically as well. Moreover, given the relation (\ref{sol-f3}), we need $A + \bar A$ to vanish in order to have a non-zero value for
$f_3$. Finally, because of the factor of $(A+\bar A) \hat h - (B+\bar B)^2$ in the expression (\ref{sol-f2}), we need $B+\bar B$ to vanish as well to avoid a negative value for $f^2_2$ close to the boundary.   
In conclusion we get that:
\be \hat h = (A + \bar A) = (B + \bar B) = H = 0  \qquad \text{on} \quad \partial \Sigma \label{BC-harm} \ee
That is, the real  harmonic functions $H$, $\hat h$, $A+ \bar A$ and $B + \bar B$ all obey Dirichlet boundary conditions, where all harmonic function go to zero with the same rate as the argument approaches the boundary.
It follows that the conjugate harmonic functions $-i (A- \bar A)$ and $-i (B- \bar B)$ obey Neumann boundary conditions.
The conditions (\ref{BC-harm}) can be automatically satisfied if the harmonic functions with Dirichlet boundary conditions are taken in the form 
\be  i f(u) - i f(\bar u) \ee
where $f$ is a real analytic function so that $f(u)^* = f(\bar u)$.\\

Turning our analysis to singular points and zeros of the harmonic functions, it is possible to prove several necessary conditions to avoid a singularity in the solution. 
In all the solutions we construct in the
next sections these conditions, together with Dirichlet boundary
conditions, are also sufficient and regularity close to singularities
and zeros determines regularity everywhere in the bulk of $\Sigma$.

\medskip

{\bf R1: The harmonic functions $A+ \bar A$, $B + \bar B$ and $\hat h$ must have common singularities}\\

In order to have a finite value for $f_3$ we need $A + \bar A$ and $\hat h$ to have common singularities according to equation (\ref{sol-f3}). Moreover, we can see from expression (\ref{sol-f2}) that if $A + \bar A$ and $\hat h$ are singular, we need $B + \bar B$ to be singular as well so that:
\be A + \bar A - {(B+ \bar B)^2 \over \hat h} \rightarrow 0  \label{cond-poles} \ee   
If we expand our harmonic and holomorphic functions in the vicinity of a singularity as:
\bea A &=& i {c_A \over u - u_0} + i b_A + \dots \label{pole-expand1}\\
 B &=&i {c_B \over u - u_0} + i b_B + \dots \\
 \hat  h &=& i {\hat c \over u - u_0} + i \hat b + \dots + c.c. \label{pole-expand3} \eea
then equation (\ref{cond-poles}) gives a relation between the residues which needs to be satisfied:
\be  c_A \hat c = c^2_B \label{reg-poles} \ee 

{\bf R2: No singular points in the bulk of $\Sigma$ } \\

To prove this we can expand our harmonic functions in the vicinity of a common singularity as 
in equation (\ref{pole-expand1}-\ref{pole-expand3}). We then introduce the new coordinates $r e^{i \phi}=u-u_0$ in a 
neighborhood of the singular point and rewrite the harmonic functions as:
\be A + \bar A = 2 c_A {\sin \phi \over r} - 2 \Im b_a + \dots  \label{exp-A-rphi} \ee 
With similar expressions for the other functions. It follows from equation (\ref{exp-A-rphi}) that there exists a curve $\mathcal{C}$ on which $A + \bar A$ vanishes at least in a neighborhood of the singularity. In order to preserve the positivity of $f_3^4 $ and $e^{4 \Phi}$, 
the other two harmonic functions $B + \bar B$ and $\hat h$ must vanish on $\mathcal{C} $ as well. Since the ratio of metric factors can be expressed as:
\be { f^2_2 \over f^2_1 } = {(A + \bar A) \hat h - (B + \bar B)^2 \over (A+ \bar A) \hat h - (B - \bar B)^2 } \label{reg2} \ee
it follows that $f_2/f_1=0$ on $\mathcal{C}$, that is either $f_2=0$ or $f_1 \rightarrow \infty$ on the curve $\mathcal{C}$. The metric factor $f_1$ can have singularities only in isolated points and the boundary $\partial \Sigma$ is defined as the locus in which $f_2=0$, therefore the singular point must be on the boundary.
With a similar argument we can show that any singularity of $H$ must be on the boundary as well. \\
As we will see in the next sections, the absence of singularities in the bulk forces the $D3$ and $D7$ brane charges to be zero in case of a regular solution with a single boundary component. \\

{\bf R3: The functions $A + \bar A $, $ \hat h $ and $H$ cannot have any zero in the bulk of $\Sigma$}\\

The positivity of $f_3$ and $e^{4 \Phi}$ demands that if $A + \bar A $ or $\hat h$ have a zero then the zero is common to $A + \bar A$, $\hat h$ and $B + \bar B$.
The existence of a curve of zeros of $A+ \bar A$ or $\hat h$ can be excluded with an argument similar to the one of the previous paragraph: since $f_1$ can have singularities at most in isolated points, 
$f_2$ must vanish on the curve and the curve is just part of the boundary.
 With an analogous argument we can exclude the existence of a curve of zeros of $H$. \\
The existence of isolated zeros of the harmonic functions can be ruled out because the zero would be a global minimum in the bulk of $\Sigma$ and, due to the maximum principle for harmonic functions, the harmonic function would have to
be constant everywhere on $\Sigma$. \\
Similarly, it is possible to use the maximum principle to prove that an harmonic function does not change sign in the bulk only if all the residues of its holomorphic part have the same sign.
Moreover, since $\hat h$, $H$ and $A + \bar A$ cannot change sign we can use the transformations (\ref{sym-disc2}-\ref{sym-disc3}) to set them to be positive everywhere on $\Sigma$. 

\medskip

An expansion of the harmonic functions  $\hat h$, $H$ and $A + \bar A$ close to the boundary $Im(u)=0$   can be used to show that the absence of any zeros   in the bulk  implies that all three harmonic functions     vanish like $Im(u)$ as $Im(u)\to 0$.\\

{\bf R4: The holomorphic functions $B$ and $\partial_u H$ must have common zeros}\\

The two-dimensional curvature scalar can be expressed as:
\be R_{\Sigma} = - 4 {\partial_u \partial_{\bar u} \log \rho \over \rho^2 }  \ee
In order to avoid a curvature singularity we need $\rho$ to be strictly positive everywhere on $\Sigma$. 
In particular, since $H$, $\hat h$ and $A+\bar A$ vanish only on the boundary, equation (\ref{sol-rho}) implies that $B$ must have all the zeros of $\partial_u H$.\\
A separate analysis is required for the case of points which are zeros of $B$ but not zeros of $\partial_u H$: in these points the metric factor
$\rho$ is singular while $R_{\Sigma}$ vanishes. These zeros correspond to $AdS_2\times S^2 \times S^1\times R$ asymptotic regions and their analysis will be deferred to further work.\footnote{This solution is reminiscent of the ones found in \cite{Boonstra:1998yu} in a different context.}  \\

\subsection{Simple R-R Janus deformations}\label{simpljanusa}

\begin{figure}
\centering
\includegraphics[angle=90,scale=0.29]{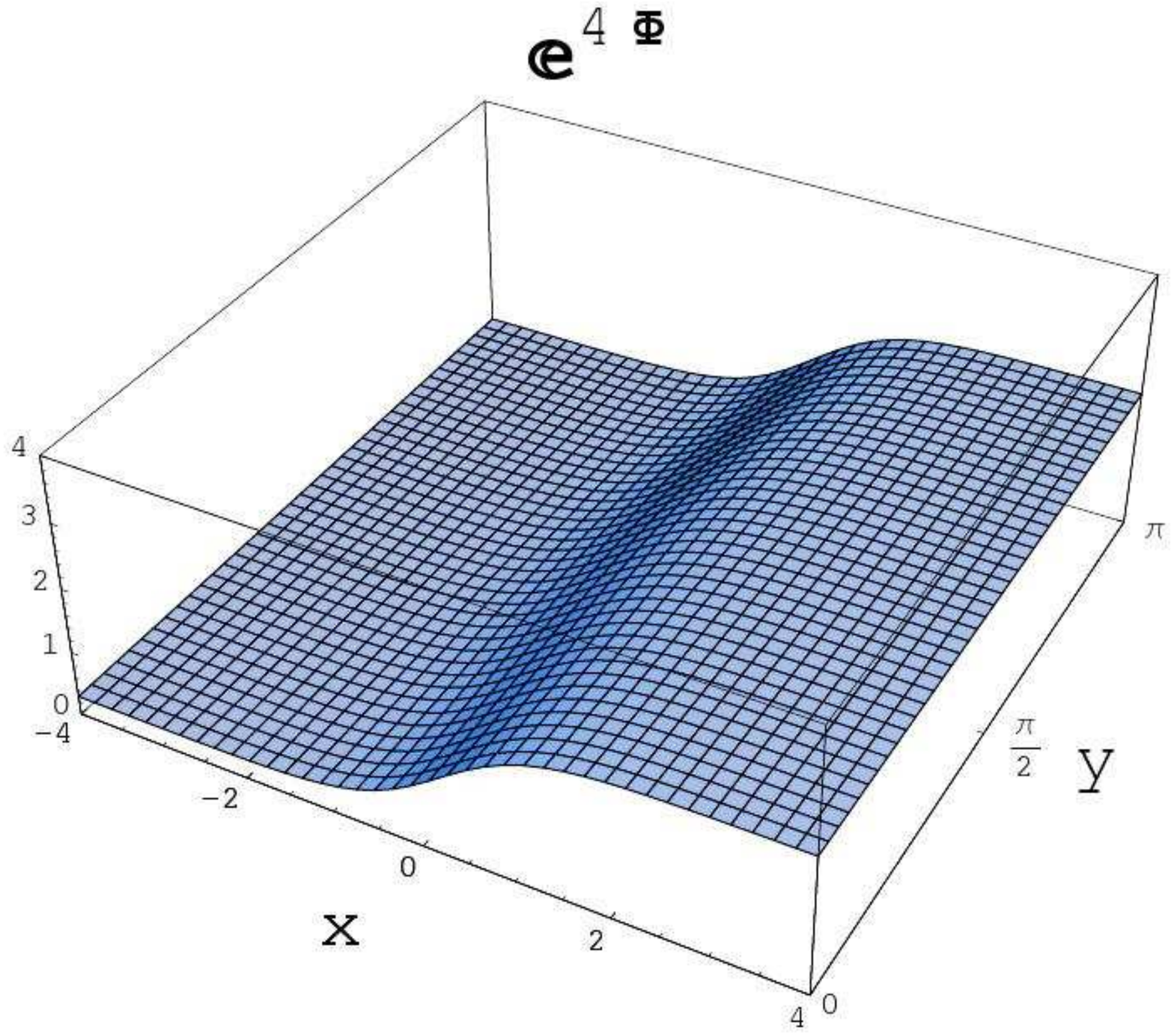}
\includegraphics[angle=90,scale=0.29]{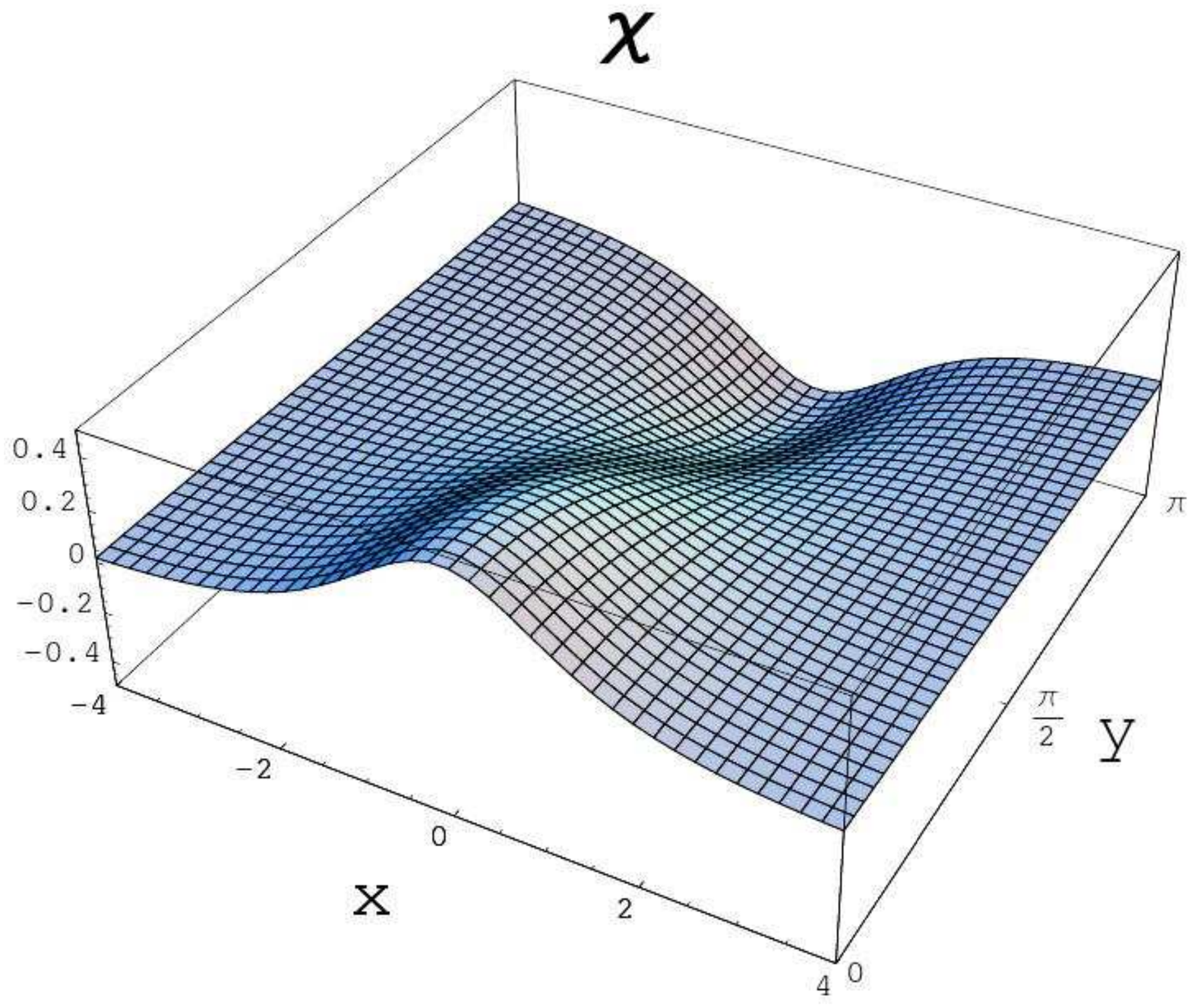}
\caption{ \small Dilaton and axion profiles for a Janus deformation with $\psi=1/2$ and $\theta=0$. The parameters $k$ and $L$ have been set to one. }
\label{fig-psi}
\end{figure}

In this section we look for a simple Janus deformation in which each of the harmonic functions has two singular points. We can see from equation (\ref{qNSvac}) that for the R-R $AdS_3 \times S^3$ vacuum the harmonic functions $A + \bar A, \; B+ \bar B$ and $\hat h$ have singularities
on the imaginary axis at $y=0 $ and $y = \pi$ while the harmonic function $H$ has singularities for $x \rightarrow \pm \infty$. We can look for a Janus deformation having poles in the same positions. This leads us to the ansatz:
\bea
H &=& -{i \over 2} \Big( c_1 e^w - c_2  e^{- w} \Big) + c.c. \\
A &=& i   { c_3  + c_4 \cosh w \over \sinh w} + i b_1 \\
B &=& i { c_5 + c_6 \cosh w \over \sinh w} + i b_2\\
\hat h &=& i  { c_7  + c_8 \cosh w  \over \sinh w} + c.c. \eea
Next, we can use the scale symmetry (\ref{sym-1}) to set $c_3=c_6 c_7$. We can also redefine the other coefficients as:
\be c_4 \rightarrow c_6 c_4, \qquad c_7 \rightarrow {c_7 \over c_6}, \qquad c_8 \rightarrow {c_8 \over c_6}  \ee
The regularity condition $R4$ implies that:
\be c_5 =0, \qquad b_2=c_6 {c_1-c_2 \over c_1 +c_2}  \ee
while equation (\ref{reg-poles}) gives:
\be c_8=-c_4, \qquad c_7^2-c_4^2 = c_6^2 \ee
Redefining $c_1=L e^{\psi}$, $c_2 =L e^{-\psi}$, $c_7 = k \cosh \theta$, $c_4 = k \sinh \theta$, $c_6 = k$ and $b_1=b$ we get a five parameters set of solutions\footnote{As was the case for the vacuum solution we have set the volume of $K_3$ to one and do not display the constant in the harmonic function  dual to $\hat h$.}. In conclusion, the harmonic functions are:
\bea
H &=& -i L \sinh (w + \psi) + c.c. \\
A &=& i k^2  { \cosh \theta + \sinh \theta \cosh w \over \sinh w} + i b  \\
B &=& i k { \cosh (w + \psi) \over \cosh \psi \sinh w} \\
\hat h &=& i  { \cosh \theta - \sinh \theta \cosh w \over \sinh w} + c.c. \eea

\begin{figure}
\centering
\includegraphics[angle=90,scale=0.29]{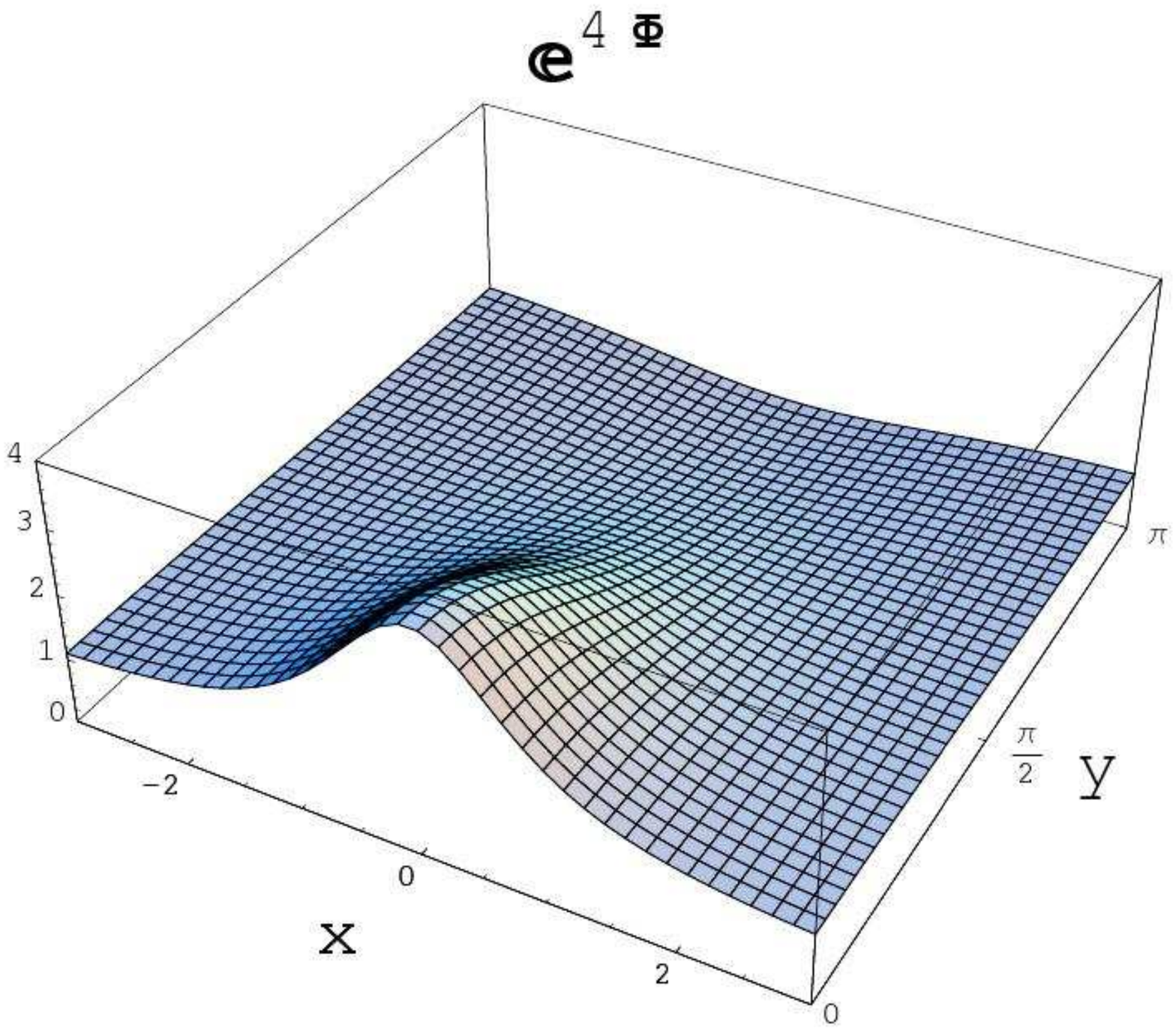}
\includegraphics[angle=90,scale=0.29]{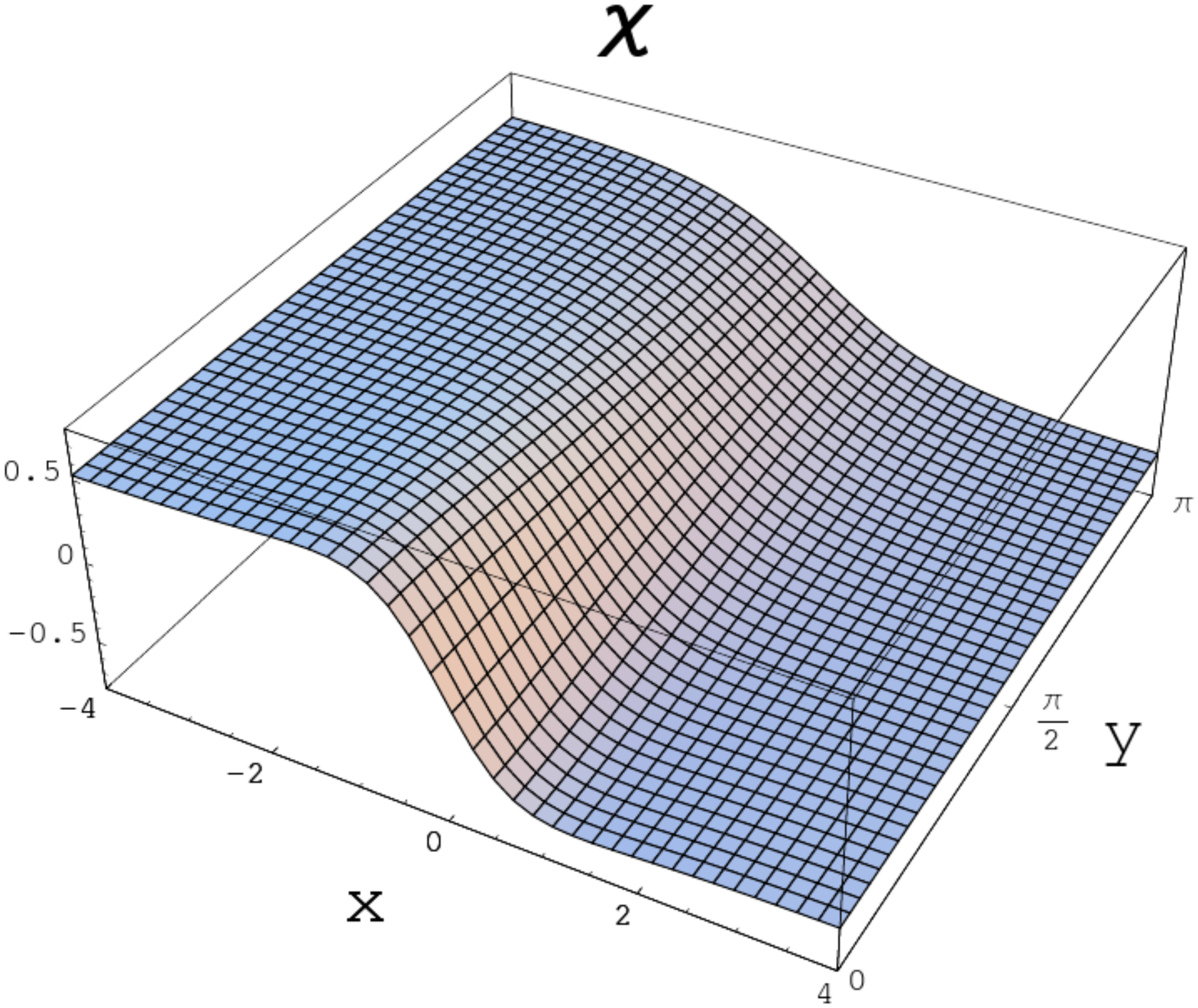}
\caption{ \small Dilaton and axion profiles for a Janus deformation with $\psi=0$ and $\theta=1/2$. The parameters $k$ and $L$ have been set to one. }
\label{fig-theta}
\end{figure}

\begin{figure}
\centering
\includegraphics[angle=90,scale=0.29]{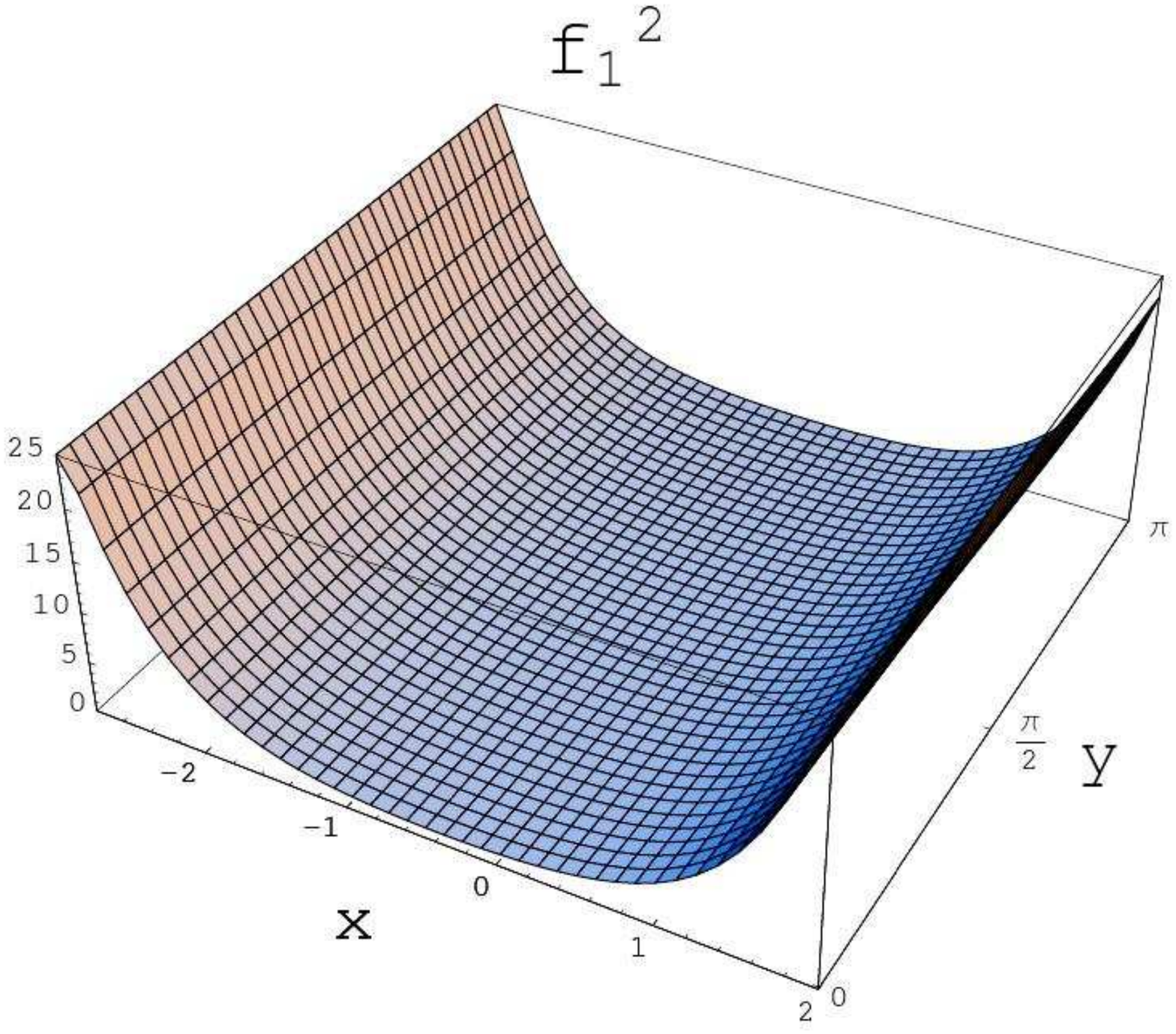}
\includegraphics[angle=90,scale=0.29]{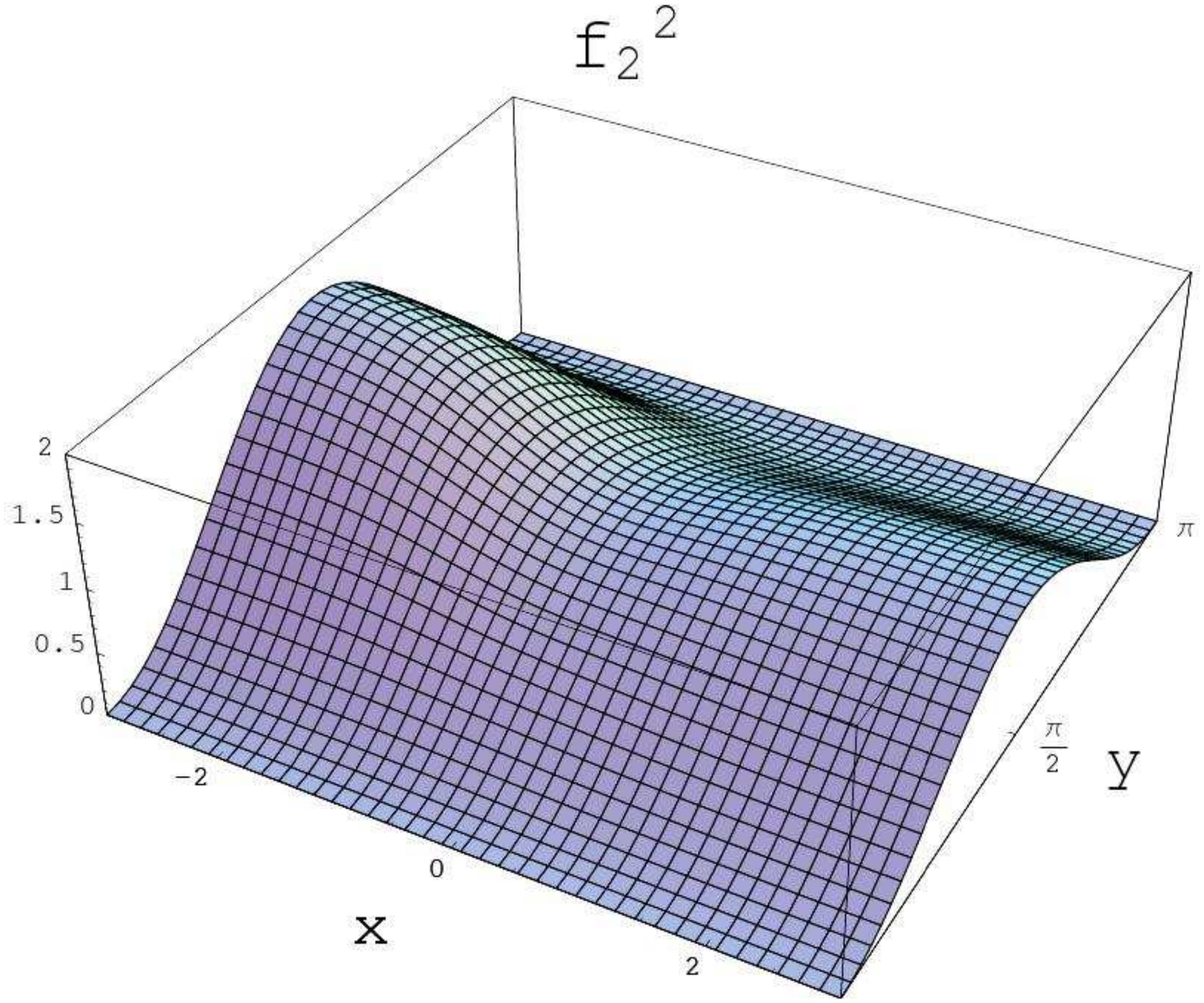}
\includegraphics[angle=90,scale=0.29]{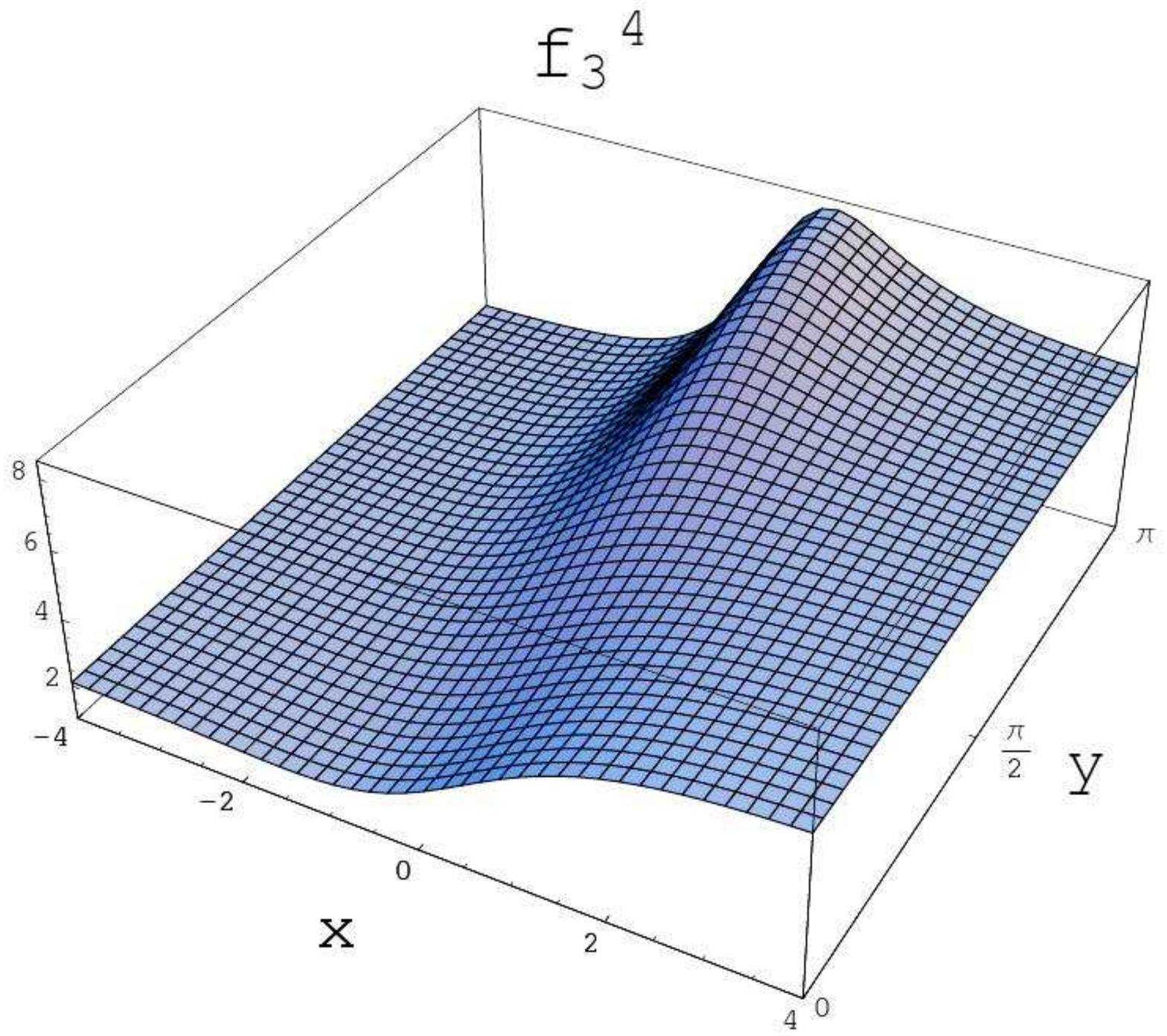}
\includegraphics[angle=90,scale=0.29]{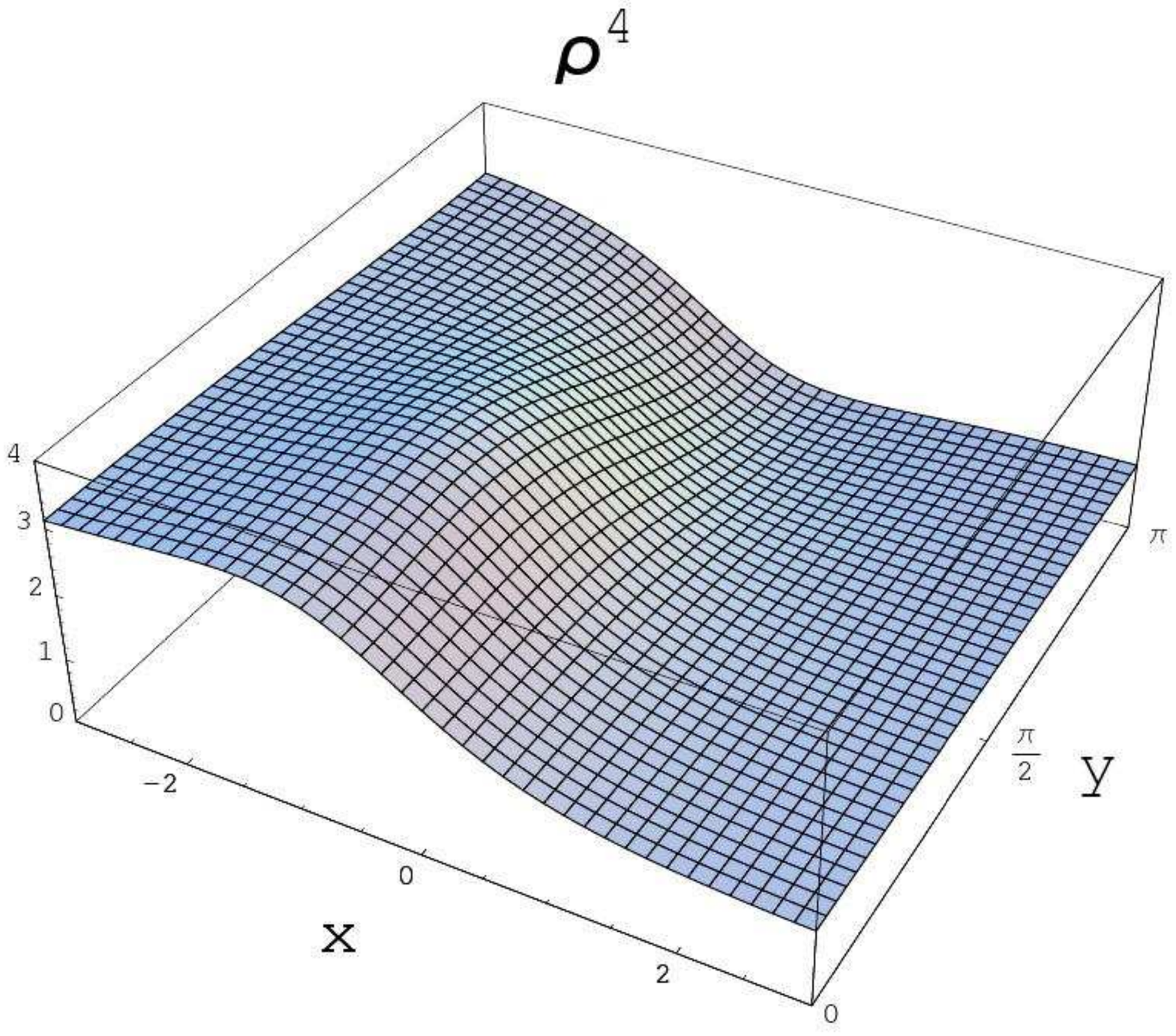}
\caption{ \small Metric factors for a Janus deformation with $\psi=\theta=1/2$, $L=k=1$ and $b=0$. }
\label{fig-metric}
\end{figure}

It is easy to see that the parameters $b$ and $k$ correspond to $SL(2,R)$ transformations keeping the NS-NS charge to zero. Specifically,
the scale transformation changes $k$ by an overall constant while the shift transformation acts as a shift of $b$. Similarly, the parameter $L$ corresponds 
to the transformation (\ref{sym-2}) and can be used to fix the radius of one of the two $AdS_3$ regions.\\
On the other side, the parameters $\theta$ and $\psi$ determine a different value for dilaton and axion in the two asymptotic regions. Dilaton and axion profiles for Janus deformations
with $\psi \neq 0, \theta =0$ and $\psi =0, \theta \neq 0$ are plotted in figure \ref{fig-psi} and figure \ref{fig-theta} respectively. \\
We have (relatively) simple expressions for the dilaton and axion:
\bea e^{4 \Phi} &=& k^4  {\cosh^2 (x + \psi)  {\rm sech}^2 \psi + \big(  \cosh^2 \theta -  {\rm sech}^2  \psi \big) \sin^2 y \over \big( \cosh  x  - \cos y \tanh \theta \big)^2} \\
\chi &=& - {k^2 \over 2}  {  \sinh 2 \theta  \sinh x - 2 \tanh \psi \cos y \over \cosh x \cosh \theta - \cos y \sinh \theta} - b  \eea
The metric factors are:
\bea \rho^4 &=& L^2 { e^{2 \Phi} \over k^2} { \cosh^2 x \cosh^2 \theta - \cos^2 y \sinh^2 \theta \over \cosh^2 (x + \psi) } \cosh^4 \psi \\
f^4_3 &=& 4 {e^{2 \Phi} \over k^2} {\cosh x \cosh \theta - \cos y \sinh \theta   \over \cosh x \cosh \theta + \cos y \sinh \theta} \eea
Note that the above expressions are manifestly regular. The profiles for the metric factors for a Janus deformation with $\psi \neq 0$ and $\theta \neq 0$ are plotted in Figure \ref{fig-metric}.
The value of the $K_3$ part of the four-form potential in the asymptotic regions is given by the function $\tilde{  h}$ since the first term of equation (\ref{eq-Ck}) vanishes for $x \rightarrow \pm \infty$:
\be C_k =  {1 \over 2}  {  \sinh 2 \theta  \sinh x + 2 \tanh \psi \cos y \over \cosh x \cosh \theta + \cos y \sinh \theta} \ee
Finally, the R-R charge of these solutions is equal to:
\be q_{RR} = \pi k L  \cosh \theta \cosh \psi \ee
While the NS-NS charge is equal to zero as expected for a pure R-R solution. \\

We can also compute the $D3$ brane charge calculating the flux of the five-form anti-symmetric tensor field over a five-dimensional closed surface of the form $K_3 \times \mathcal{C}$ where $\mathcal{C}$ is a closed curve
in the bulk of $\Sigma$. Since $\Sigma$ is simply connected and the harmonic functions do not admit any singularity in the bulk, 
$\mathcal{C}$ is contractible and the $D3$ brane charge must be zero. Similarly, the $D7$ brane charge 
must vanish as well.\\
\begin{table}[htdp]\begin{center}
\begin{tabular}{p{0.4in}|p{2.2in}|p{2.2in}|p{1in}}
& $\phi^0_-$ & $\phi^0_+$ & Jump
\\[5pt]
\hline
& & & \\
$e^{2 \Phi}$ & $k^2 (1 - \tanh \psi)$ & $k^2 (1 + \tanh \psi) $ & $2 k^2 \tanh \psi$ \\[5pt]
$\chi$ & $k^2 \sinh \theta - b $ & $ -k^2 \sinh \theta -b  $ & $- 2 k^2 \sinh \theta$  \\[5pt]
$f^4_3$ & $4 (1 - \tanh \psi) $ & $ 4 (1 + \tanh \psi)$  & $8 \tanh \psi$ \\[5pt]
$ C_K $ & $ -\sinh \theta  $ & $  \sinh \theta   $ & $  2  \sinh \theta $ \\
\end{tabular}
\label{tabletwo}
\caption{Different asymptotic values for scalars and metric factors.}
\end{center}
\end{table}
 
Note that the $\theta$ deformation leaves the dilaton and $f_3$ invariant while the axion has different values in the two $AdS_3 \times S^3$ regions. Similarly the $\psi$ deformation produces a jump in $\Phi$ and $f_3$ only.
The asymptotic values for the dilaton, axion and metric factors are given in Table 2.\\
We can see that the combinations:
\be  e^{-2 \Phi} f^4_3 \qquad \text{and} \quad \chi+k^2 C_K \ee
have the same values in the two asymptotic regions.
The fields approach their constant values in the two asymptotic regions as follows:
\bea \phi &=& \phi^0_- + \phi^1_-(y) e^{x} + \dots \qquad \text{for } x \rightarrow - \infty \\
\phi &=& \phi^0_+ + \phi^1_+(y) e^{-x} + \dots \qquad \text{for } x \rightarrow  \infty \eea
The profile functions $\phi^1_{\pm} (y)$ for the different fields are listed in Table 3.
\begin{table}[htdp]\begin{center}
\begin{tabular}{p{0.4in}|p{2.8in}|p{2.8in}}
& $\phi^1_-(y)$ & $\phi^1_+(y) $ \\[5pt]
\hline
& & \\
$e^{2 \Phi}$ & $ 2 k^2 (1 - \tanh \psi) \tanh \theta \cos y$ & $2 k^2 (1 + \tanh \psi)  \tanh \theta \cos y$  \\[5pt]
$\chi$ & $2 k^2    {\rm sech} \theta \cos y ( \tanh \psi  + \sinh^2 \theta )$ & $  2 k^2 {\rm sech} \theta \cos y ( \tanh \psi - \sinh^2 \theta ) $  \\[7pt]
$f^4_3$ & $- 8 (1 - \tanh \psi) \tanh \theta \cos y $ & $ -8 (1 + \tanh \psi) \tanh \theta \cos y $ \\[5pt]
$ C_K $ & $ 2  {\rm sech} \theta \cos y ( \tanh \psi + \sinh^2 \theta ) $ & 
$  2  {\rm sech} \theta \cos y ( \tanh \psi - \sinh^2 \theta ) $ \\
\end{tabular}
\label{tablethree}
\caption{$\phi^1_{\pm} (y)$ for scalars and metric factors.}
\end{center}
\end{table}

\subsection{NS-NS Janus Deformations}

The $SL(2,R)$ symmetry can be used to generate Janus solutions with non-zero NS-NS charge. In particular, all regular Janus solutions with two asymptotic regions can be obtained applying an $SL(2,R)$ transformation to the
pure R-R solution from the previous section. \\
The S-duality transformation maps the R-R solution to a solution charged only under the three-form NS-NS fields. Using the transformation (\ref{Sduality}) we get the following expressions for the harmonic functions:
\bea
H &=& -i L \sinh(w + \psi) + c.c. \\
A &=&   {- i  \sinh w \over k^2 \sinh \theta \cosh w + k^2 \cosh \theta + b \sinh w }  \\
B &=& { i   \cosh (w+ \psi) {\rm sech}  \psi \over k^2 \sinh \theta \cosh w + k^2 \cosh \theta + b \sinh w}\\
\hat h &=& {i  \over \sinh w}\left( \cosh \theta - \sinh \theta \cosh w - {( \tanh \psi \sinh w + \cosh w)^2 \over k^2 \sinh \theta \cosh w + k^2 \cosh \theta + b \sinh w} \right) + c.c. \qquad \quad \eea
These functions have singularities for:
\be e^w = - {\cosh \theta \pm \sqrt{b^2/k^4+1} \over \sinh \theta + b/k^2} \ee
The singular points are located on the $y=0$ and $y=\pi$ boundaries, but their positions now depend on the parameters. 
These solutions have NS-NS three-form charge equal to:
\be q_{NS} = - \pi k L \cosh \theta \cosh \psi \ee
and vanishing R-R charges. The expressions for the metric factors and the four-form potential are invariant under the S-duality transformation.
Using S-duality it is easy to see that the following fields combinations:
\be
(e^{-2 \Phi} + \chi^2 e^{2 \Phi})f^4_3, \qquad {\chi \over e^{4 \phi} + \chi^2} -k^2 C_K
\ee
have the same values in the two asymptotic regions.

\subsection{Multi-pole solutions}\label{multipolesol}

In this section we will use the conditions for regularity from section \ref{reg-cond-sec} to find a general ansatz for solutions having $n$ $AdS_3 \times S^3$ regions.
The relevant holomorphic functions will be rational functions in the variable $u$ and the position of poles and residues will parameterize the multi-pole solutions.
We start by taking the harmonic function $H$ to be in the form:
\be H = i  \sum^{n-1}_{i=1} { c_{H,i} \over u-x_{H, i} } - i {c_{H,n} u} + c.c. \label{multipoleH} \ee 
here $x_{H,1} \dots x_{H,n-1} $ are the poles of the holomorphic part of $H$, which also has a pole at infinity, and $c_{H,1} \dots c_{H,n}$ are the residues.
According to condition $R2$, the poles must be taken on the real axis while condition $R3$ determines the residues to be all positive.
With a change of coordinates on $\Sigma$ we can set the position of a pole at infinity, $x_{H,1}=0$ and $x_{H,2}=1$ bringing down the total number
of parameters to $2n-3$. \\
Similarly, we can take the function $A$ in the form: 
\be A = i  \sum^{2n-2}_{i=1} { c_{A,i} \over u-x_{A, i} } +i b \label{multipoleA} \ee 
As before, $x_{A,1} \dots x_{A,2n-2}$ and $x_{A,1} \dots x_{A,2n-2} $  are the poles and residues of $A$. The $x_{A,i}$ must be real since $A$ cannot have any zeros in the bulk and the residues must be all positive. 

At this point, the regularity conditions completely determines the other functions. Because of conditions $R4$ and $R1$, the function $B$ must have the same zeros of $\partial_u H$ and the same poles of $A$. These requirements fix its form up to an overall constant which can be set to one with the symmetry (\ref{sym-1}):
\be B=  { \prod^{n-1}_{i=1} (u-x_{H, i})^2 \over \prod^{2n-2}_{i=1} (u-x_{A, i})} \partial_u H \label{multipoleB} \ee
With this definition, the function $B + \bar B$ must have at least a curve of zeros in the bulk of $\Sigma$. To show this property, we first note that $\partial_u H$ can be expressed as:
\be \partial_u H = - i \Big( \sum^{n-1}_{i=1} {c_{H,i} \over (u- x_{H,i} )^2 } +c_{H,n} \Big) \label{multipoledH} \ee 
Since $\partial_u H$ is a rational function with a polynomial of degree $2n-2$ as numerator, the fundamental theorem of algebra
 guarantees that it must have $2n-2$ zeros. 
If we restrict $u$ to the real axis, the term in brackets in equation (\ref{multipoledH}) is strictly positive because the residues $c_{H,i}$ are all positive,
therefore $\partial_u H$ cannot have any zero on the real axis and all the zeros must be complex. In particular, since the numerator of $\partial_u H$ is a polynomial with real coefficients, half
of the zeros have positive imaginary part and are located in the bulk of $\Sigma$. Since $B$ has common zeros with $\partial_u H$, the harmonic function
$B+ \bar B$ must have the same $n-1$ zeros in the bulk of $\Sigma$. These zeros cannot be isolated due to the maximum principle for harmonic functions, therefore there must be at least
a curve of zeros in the bulk.  \\
Because of the presence of the curve of zeros, $B + \bar B$ changes sign in $\Sigma$ and the residues of $B$ cannot all have the same sign. 

The function $\hat h$ has the same singularities of $A + \bar A$ according to condition $R1$ and the residues are fixed by equation (\ref{reg-poles}):
\be \hat h = i \sum^{2n-2}_{i=1} {\hat c_i \over u - x_{A,i} } + c.c. \qquad \hat c_i = {c^2_{B,i} \over c_{A,i} } \label{multipoleh} \ee
The residues $c_{B,i}$ can be obtained from:
\be 
c_{B,i} = \lim_{u \rightarrow x_{A,i}} (u-x_{A,i}) B (u)
\ee
Note that our solution depends on a total of $6n-4$ parameters. Four of these parameters must still correspond to the $SL(2,R)$ transformations generated by (\ref{Sduality}-\ref{sym-shift}) and to the scale 
transformation (\ref{sym-2}). \\

We now need to prove that the harmonic functions (\ref{multipoleH}-\ref{multipoleh}) provide a solution which is regular everywhere on $\Sigma$.
Regularity on $\partial \Sigma$ is satisfied because the harmonic functions obey to Dirichlet boundary conditions and respect condition $R1$ together with equation (\ref{reg-poles}).  
 Note that away from singularities located at the boundary, the harmonic functions $A+\bar A, B+\bar B, \hat h, H$ all vanish  linearly in $y$.

From equation (\ref{sol-f1}) we see that $f_1$ is manifestly positive and non-vanishing in the bulk of $\Sigma$. Similarly,
according to equation (\ref{sol-rho}), $\rho$ is always finite and strictly positive in the bulk since condition $R4$ is satisfied. $f_3$ is finite and positive as well because $A + \bar A $ and $\hat h$ are finite and positive in the bulk.
The only non-trivial requirement is coming from the regularity of the dilaton and of the metric factor $f^2_2$. We must prove that:
\be (A + \bar A) \hat h - (B + \bar  B)^2 > 0 \label{cond-R5} \ee
everywhere on $\Sigma$. 
\begin{figure}
\centering
\includegraphics[angle=90,scale=0.29]{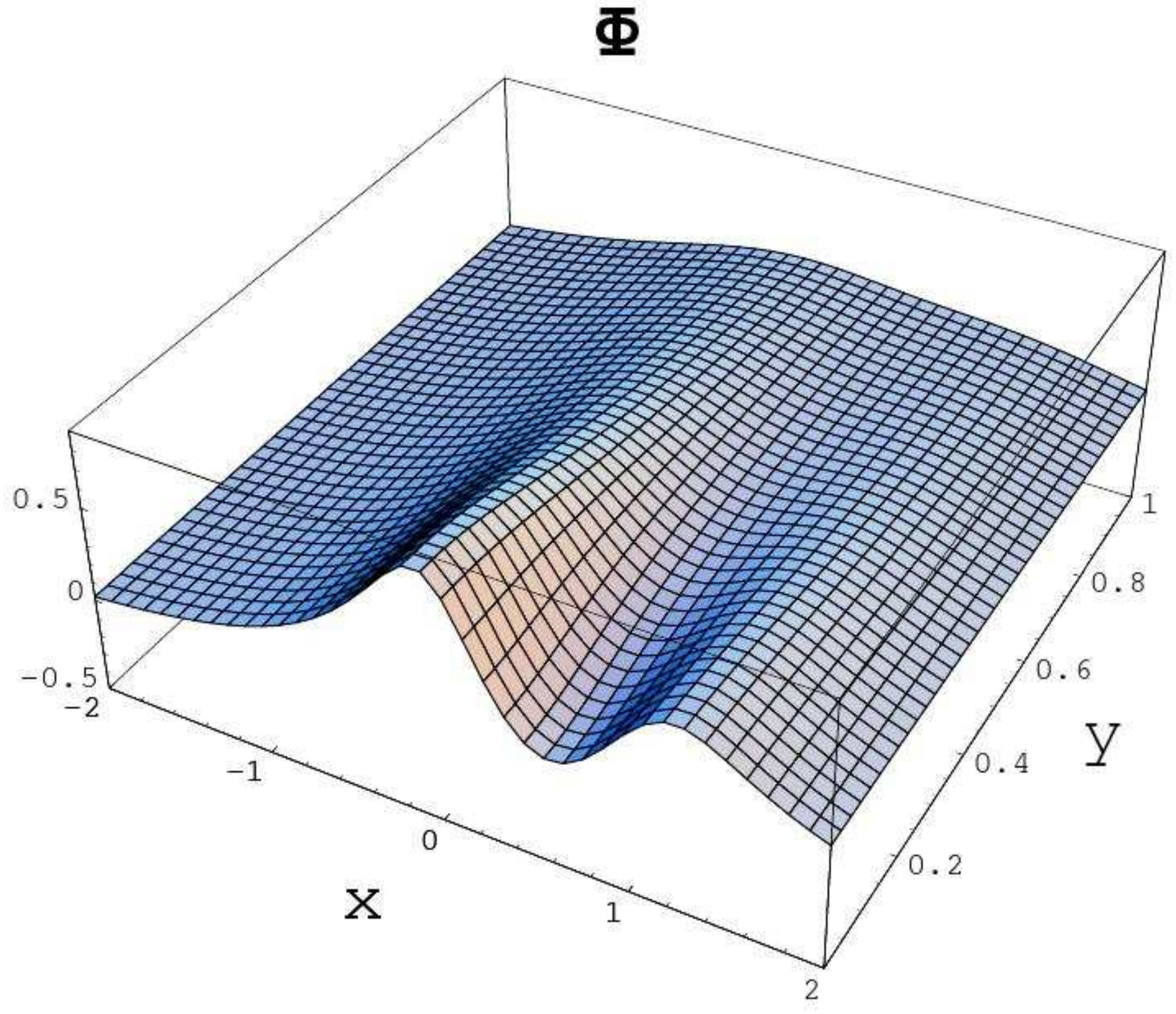}
\includegraphics[angle=90,scale=0.29]{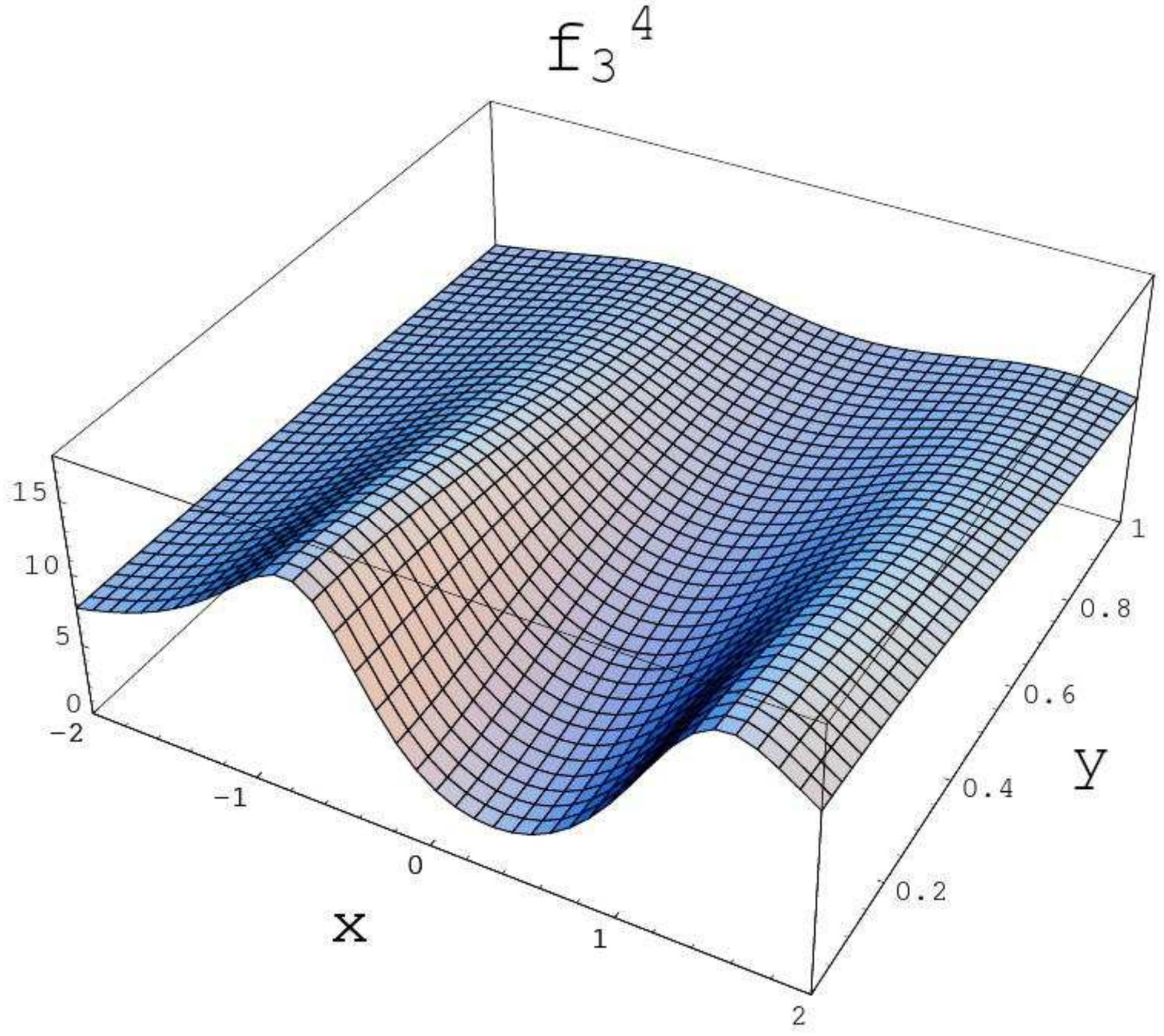}
\caption{\small Dilaton and metric factor $f_3$ for a multi-pole Janus deformation. $H$ is singular for $x=0,1, \infty$ with unit residues while $A$ has
poles in $x=0,-1,2$ with unit residues and in $x=1$ with residue $1/2$.}
\label{fig-multipole1}
\end{figure}
Using the expressions (\ref{multipoleA}-\ref{multipoleh}) we can show that:
\be (A+\bar A) \hat h - (B+\bar B)^2 = 4 y^2 \sum^{2n-2}_{i=1} \sum^{i-1}_{j=1} { { \textstyle c_{A,i}
c^2_{B,j} \over \textstyle c_{A,j}} + { \textstyle c_{A,j} c^2_{B,i} \over \textstyle c_{A,i}} - 2
c_{B,i} c_{B,j} \over
 \big((x-x_{A,i})^2 + y^2 \big) \big((x-x_{A,j})^2 + y^2 \big) } 
\label{multipole-sum}\ee
where the diagonal terms with $i=j$ have canceled due to equation (\ref{cond-poles}). We then note that the denominators of the terms in
the summation are manifestly positive while the numerators can be rewritten as squares:
\be  { \textstyle c_{A,i} c^2_{B,j} \over \textstyle c_{A,j}} + { \textstyle c_{A,j} c^2_{B,i} \over \textstyle
c_{A,i}} - 2 c_{B,i} c_{B,j} = \Big( \sqrt{  c_{A,i} \over c_{A,j} }  c_{B,j} -
\sqrt{ c_{A,j} \over c_{A,i} }  c_{B,i} \Big)^2  \ee 
Since $B + \bar B$ has a curve of zeros in the bulk of $\Sigma$, the residues $c_{B,i}$ cannot all have the same sign and at least one of 
the terms in the summation (\ref{multipole-sum}) will be non-zero. Hence, the left-hand side of equation (\ref{multipole-sum}) will be strictly
positive causing the metric factor $f^2_2$ and the dilaton to be positive everywhere in the bulk. \\

\begin{figure}
\centering
\includegraphics[angle=90,scale=0.29]{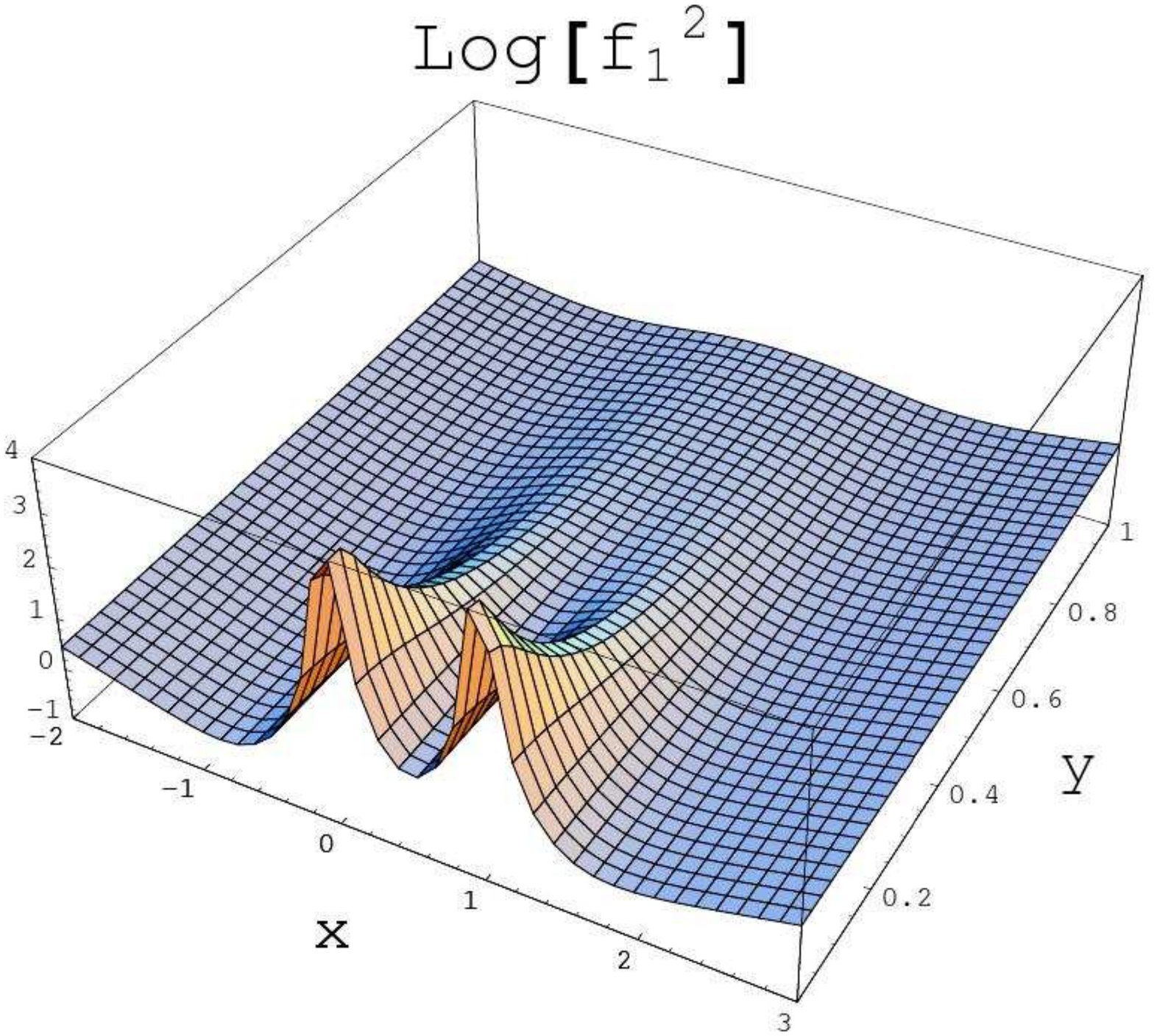}
\includegraphics[angle=90,scale=0.29]{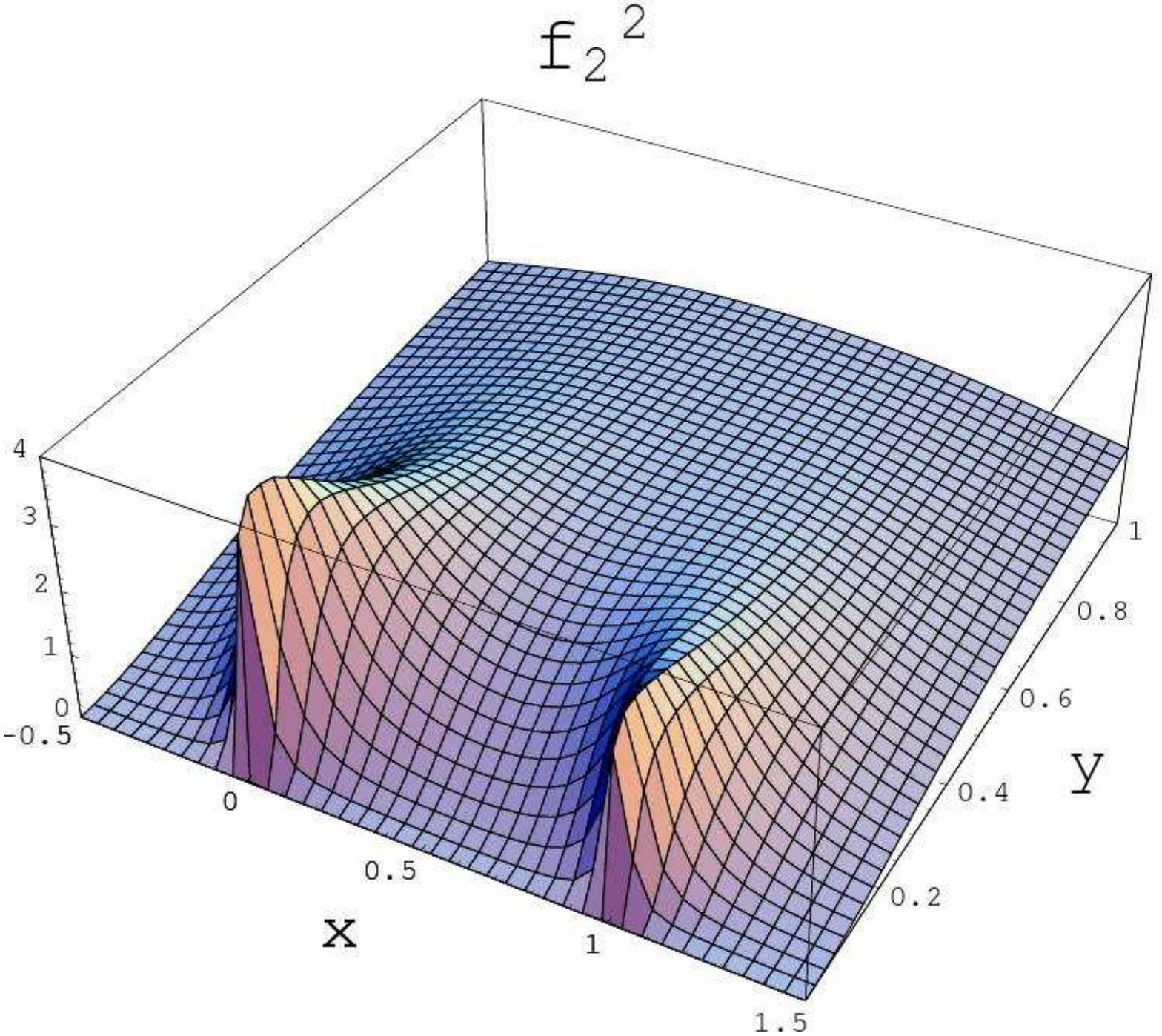}
\caption{\small Metric factors $f_1$ and $f_2$ for a multi-pole Janus deformation. $H$ is singular for $x=0,1, \infty$ while $A$ has
poles in $x=0,-1,1,2$.}
\label{fig-multipole2}
\end{figure}
The profiles of the various fields in case of a solution with three asymptotic regions are plotted in Figure \ref{fig-multipole1} and Figure \ref{fig-multipole2}. 

The solutions we have found carry in general $D1$,$D5$, as well as $NS5$ and fundamental string charges.  
Regular multi-pole solutions cannot have any $D3$ and $D7$-brane charge. The argument is similar to the one presented in the previous section:
the $D3$ and $D7$ charges are constructed integrating the five-form flux and $d\chi$ over closed surfaces of the form
$\mathcal C \times K_3$ and $\mathcal C$ respectively, where $\mathcal C$ is a closed curve in the bulk of $\Sigma$. 
Since $\Sigma$ is simply connected and the harmonic functions do not admit any singularity in the bulk, 
$\mathcal{C}$ is contractible and the charges are zero. 
In other words, the back-reacted solutions carrying $D3$ and $D7$  charges  are either singular or require a Riemann surface $\Sigma$ which is not 
simply connected.   

\newpage

\section{Janus solution and  interface CFT}\label{secfive}
\setcounter{equation}{0}
In this section we review the 
two-dimensional CFT dual
of type IIB string theory on $AdS_{3}\times S^{3}\times M_{4}$ with  self-dual R-R three-form  flux. The compactification manifold 
  $M_{4}$ is either $T^{4}$ or $K_{3}$. A comprehensive review can be found in \cite{David:2002wn} which we will follow to a large extent. 
\subsection{Review of the two-dimensional CFT}

The  $AdS_{3}\times S^{3}\times M_{4}$  vacuum  with  self-dual R-R three-form   flux  can be obtained by taking the near-horizon limit  
of $Q_{1}$ D1-branes and $Q_{5}$ D5-branes wrapping $M_{4}$.
The  theory living on the common $1+1$-dimensional
worldvolume of the D1/D5 bound state is a ${\cal N}=(4,4)$  supersymmetric field theory.
Taking the near-horizon limit of the D1-D5 system
\cite{Maldacena:1997re} corresponds to flowing to  the IR fixed point of the ${\cal N}=(4,4)$ theory. This fixed point  defines the dual CFT which can more explicitly be described as a $1+1$-dimensional
supersymmetric sigma-model with  target space given by 
the moduli space of $Q_{1}$ instantons in a two-dimensional
 $SU(Q_{5})$ gauge theory.  The moduli space is $4n$-dimensional where $n= Q_{1}Q_{5}$ (for $M_{4}=T_{4}$) or
$n=Q_{1 }Q_{5}+1$ (for $M_{4}=K_{3}$). This conformal field
 theory is given
by \cite{Witten:1997yu,Vafa:1995bm,Dijkgraaf:1998gf}
 the smooth resolution  of the orbifold CFT of the
symmetric product  $M^{n}/S_{n}$. For definiteness   we focus
 on the case where $M_{4}= T^{4}$ in this section.  The central charge
 of the CFT is then $c= 6n = 6 Q_{1}Q_{5}$.
The orbifold $T_{4}^{n}/S_{n}$ can be constructed by
starting with the free field CFT representing the tensor
product $T_{4}^{n}$.

\begin{equation}\label{orbcft}
S= {1\over 4 \pi }\int d^{2}z \sum_{i,a}
\Big( \partial X_{i,a} \bar \partial X_{i,a} + \psi_{i,a }\bar
\partial \psi_{i,a}+ \bar \psi_{i,a }\partial \bar \psi_{i,a}\Big)\ .
\end{equation}
 The indices $i=1,2,\cdots 4$, and $a=1,2,\cdots n$
parameterize  $n=Q_1 Q_5$ copies of the four torus $T^{4}$.

The global sub-superalgebra of the ${\cal N}=(4,4)$ superconformal algebra  is $SU(1,1|2)\times SU(1,1|2)$.   There are several $SU(2)$ symmetries which will be important in the following analysis.
First of all, the  $SU(2)_R\times SU(2)_{R'}$ R-symmetry is part of the global superalgebra and acts  on the holomorphic and anti-holomorphic part of the CFT respectively.
In addition, there exists  an $ SU(2)_{I}\times SU(2)_{I'}$   group of outer automorphisms of the superconformal algebra.

 For holomorphic (and anti-holomorphic) fields each state in the CFT is labelled by the conformal dimension $h$ (and $h'$) and its $SU(2)_R$  and  $SU(2)_{R'}$  quantum numbers $j$ and $j'$, respectively. In addition, 
 states related to the internal structure of the $M_4$ can be charged with respect to the   $SU(2)_I$  and  $SU(2)_{I'}$ symmetries.

We now describe the holomorphic side of the CFT with analogous expressions for the anti-holomorphic side.

The fermionic generators of  the $SU(1,1|2)$ superalgebra are $G^\a_{\pm 1/2}, G^{\a\dagger}_{\pm 1/2}$ with $ \a=1,2$.   A special class of states is formed by  chiral primaries which have $h=j$ and are in short multiplets of the superalgebra. Chiral primaries are annihilated by half the superconformal generators, namely
\be
G^1_{-1/2} \mid \phi\rangle=0 ,\quad \quad  G^{2\dagger}_{-1/2} \mid \phi\rangle=0
\ee
Chiral primaries  are protected from quantum corrections to their conformal dimension.

For an interface  CFT there are two kinds of possible operators which can be added to deform the theory preserving the interface conformal symmetry. First, exactly marginal operators of dimension $(h,\bar h)=(1,1)$ can be added to the bulk Lagrangian (possibly with a position dependent coupling constant). Second,  dimension $(h,\bar h)=(1/2,1/2)$ operators which are localized at the interface can be added.
We will now construct these operators from the chiral primaries in the untwisted and twisted sectors of the CFT.

In the untwisted sector, the chiral primaries with the lowest dimension have $(h,\bar h)=(1/2,1/2)$. They are constructed from  holomorphic and anti-holomorphic fermion bilinears $\psi^i \bar \psi^j$. These chiral primaries will transform as $({\bf 2,2 })$ under the $SU(2)_R\times SU(2)_{R'}$ R-symmetry. 
Acting with non-trivial set of superconformal generators produces descendants and the bottom components of the supermultiplet have  $(h,\bar h)=(1,1)$. This means that  they  are marginal operators.
 These operators all transform as $({\bf 1,1})$ under $SU(2)_R\times SU(2)_{R'}$.
 They can be further classified into representations of  $SU(2)_I\times SU(2)_{I'}$. The only operator which transforms as  a  ${\bf (1,1)}$ singlet under $SU(2)_I\times SU(2)_{I'}$ is of the following form:
\be
{\cal O}_0= \sum_{i,a} \partial X_{i,a} \bar \partial X_{i,a} +{\rm fermions}
\ee
 and is proportional to the Lagrangian density for the untwisted sector.

Other short multiplets come from the twisted sector of the symmetric orbifold. The twist fields in a symmetric orbifold are 
labeled by the conjugacy classes of the symmetric group $S(Q_1 Q_5)$. The simplest twist fields are associated with $Z_2$ twists
 which exchange a pair of bosons and fermions, for example  $x_{i,1}\leftrightarrow x_{i,2}, \psi_{i,1}\leftrightarrow \psi_{i,2}$
 while all other fields remain unchanged. A twist operator is constructed by summing over all elements in a given conjugacy class.  
By doing so, one obtains a chiral primary $\Sigma^{{1\over 2},{1\over 2}}$ with ${h,\bar h}=(1/2,1/2)$.   Descendants  with $(h,\bar h)=(1,1)$ are obtained by acting with 
superconformal generators on the chiral primary. Among the descendants there is one operator  $T_0$ transforming  as $({\bf 1,1})$ under $SU(2)_R\times SU(2)_{R'}$ and  
 $({\bf 1,1})$ under $SU(2)_I\times SU(2)_{I'}$. $T_0$ can be obtained from the following operator product expansion
\be
\lim_{z\to w  } \big( G^2(z) \tilde G^{1\dagger}(\bar z) - G^{1\dagger}(z) \tilde G^{2\dagger}(\bar z)\big) \Sigma^{{1\over 2},{1\over 2}}(w,\bar w) = {1\over (z-w) (\bar z-\bar w) }T^0(w,\bar w)+\cdots
\ee

There are higher order twist fields associated with other conjugacy classes  of $S_n$, but they will not be needed here. The interpretation of turning on $T_0$ is that the orbifold CFT  will be deformed by turning on a  particular blow-up mode. 

In summary,  there are five chiral primary states with dimension $({h,\bar h})=(1/2,1/2)$ which do not transform under  $SU(2)_I\times SU(2)_{I'}$: four from the untwisted sector and one from the twisted sector. Among the superconformal descendants of these states there are two operators with dimension $(h,\bar h)=(1,1)$  which do not transform   under  $SU(2)_I\times SU(2)_{I'}$ and under  $SU(2)_R\times SU(2)_{R'}$. We denote the operator from the untwisted sector as $O_0$ and the operator from the twisted sector as $T_0$.

\subsection{Holographic interpretation of Janus solution}

In this section we provide the holographic interpretation of the R-R charged Janus solutions obtained in the previous section. 
We recall the relation between the mass of a scalar field in $AdS_3$ and the conformal dimension of the dual operator in the $CFT_2$.
\be
\Delta=1+\sqrt{1+m^2}
\ee
Hence a massless field corresponds to a marginal operator with $\Delta=2$ and  fields which saturate the 
Breitenlohner-Freedman bound $m^2=-1$ correspond to operators with conformal dimensions $\Delta=1$. 
In the following analysis we briefly review the holographic map between asymptotic  behavior of fields near the boundary of $AdS_3$ and the presence of  sources and/or expectation values for the dual operators.
 For simplicity, we exhibit the map for Euclidean $AdS_3$:
\be
ds^2= {dz^2+dx_1^2+dx_2^2\over z^2}
\ee
where the boundary is approached as $z\to 0$.  A massless field with $m^2=0$ and $\Delta=2$ behaves near the boundary as
\be
\lim_{z\to 0}\phi_{\Delta=2}\sim \phi^0 (x)+ z^2  \phi^1(x)+\cdots 
\ee
If $\phi_0$ is non-vanishing, a source for the marginal operator dual to the scalar field is turned on.
A non-vanishing $\phi_1$ is interpreted as a non-trivial expectation value for the dual operator.

The (global part) of the $SU(2)_{R} \times SU(2)_{R'}$ R-symmetry is realized by the isometry of the $S^3$ in the $AdS_3\times S^3$ vacuum, hence scalar fields transform trivially under the R-symmetry.
Furthermore since the $K_3$ or $T^4$ are not touched in the ansatz, all fields are neutral with respect  to the $SU(2)_{I_1}\times SU(2)_{I_2}$.  Scalar fields which deform the $T^4$ or $K_3$
 will be charged under the  $SU(2)_{I}\times SU(2)_{I'}$.
 The four scalars which are present in our ansatz from the perspective
 of six-dimensional supergravity (i.e. the theory obtained by compactification of  type IIB on $T^4$ or $K_3$) are the ten-dimensional dilaton $\Phi$, the axion $\chi$ the volume of $T^4$  (related to $f_3^4$)
 and the four-form potential $C_4$ evaluated along $T^4$. For a $AdS_3\times S^3$ background  with self-dual R-R flux, two of these four scalars will become massive and two will remain massless \cite{de Boer:1998ip,Avramis:2006nb,Deger:1998nm}. It is the two massless fields which we can identify to be dual to the marginal operators $O_0$ and $T_0$.  One finds the relation
\be\label{masslessf}
O_0 \;\; \sim\;\; \phi_6, \quad \quad T_0 \;\; \sim  \chi-  C_4
\ee
where the six-dimensional dilaton is defined by $e^{-2\phi_6}= e^{2\Phi} f_3^4$.  In particular, turning on the operator $O_0$ corresponds to a non-zero value for the parameter $\psi$ in the R-R Janus solution in
 section \ref{simpljanusa} while the operator $T_0$ corresponds to a Janus deformation with $\theta \neq 0$.

In the strip coordinates the asymptotic behavior of the $AdS_3$ metric is given by
\be\label{adsmetricb}
\lim_{x\to  \pm \infty} ds^2 \sim dx^2+ e^{2|x|} \; { dz^2-dt^2 \over z^2} +o(e^{-2|x|})
\ee
The behavior of the massless fields (\ref{masslessf}) is given by:
\be
\lim_{x\to  \pm \infty} \phi_{\Delta=2} = \phi^0_{\pm;\Delta=2}+ \phi^{1}_{\pm; \Delta=2}(y) e^{-2|x|}+o(e^{-4|x|})
\ee
where we denote both  massless fields given in (\ref{masslessf}) by  $\phi_{\Delta=2}$; the detailed expressions for $\phi^0_{\pm;\Delta=2}$ can be obtained from table 2 and $ \phi^{1}_{\pm; \Delta=2}(y)$ can be read off from table 3.
The two massless scalar fields take two different values as $|x|\to \pm \infty$, this means that a different source  for the dual operators will be added in the two half spaces $x^\perp>0$ and $x^\perp<0$ respectively. We obtain the Lagrangian:
\be\label{jumpdil}
\mathcal{L}_1=\mathcal{L}_0 + \Theta(x^\perp)  c_1 O_0+ \Theta(x^\perp) c_2 T_0
\ee
where $\mathcal{L}_0$ is given by (\ref{orbcft}). In general the addition of terms  like (\ref{jumpdil}) will break the supersymmetry and it is necessary to add counterterms (\ref{counterterm}) to restore some fraction of it.
\be\label{counterterm}
\mathcal{L}_{total}=\mathcal{L}_1+   \delta(x^\perp ) {\cal O}_{\Delta=1}
\ee
 Since the defect is one-dimensional the appropriate operator has conformal dimension one and should correspond to an operator with $(h, \bar h)=(1/2,1/2)$.   The exact combination of operators which appears as a counterterm is determined by the preservation of the supersymmetry. The analysis of the counterterms on the CFT side  and the precise match to the supergravity solution are left for future work.
\medskip

\section{Discussion}
\setcounter{equation}{0}

In this paper we have found the general local solutions of type IIB supergravity which can be viewed as deformations of the $AdS_3\times S^3 \times K_3$ (or $T^4$) vacuum. 
The solutions  preserve eight of the sixteen supersymmetries as well as a $SO(2,1)\times SO(3)$ subgroup of the $SO(2,2)\times SO(4)$ 
global symmetry of the vacuum. Local solutions are parameterized in terms of two holomorphic and two harmonic functions. \\
There is one  interesting feature of our solutions that differs  from the half-BPS solutions discovered in other AdS/CFT contexts
 \cite{D'Hoker:2007xy,D'Hoker:2008wc}.  The BPS equations alone do not completely determine the solutions and the Bianchi identity of
 the self-dual five-form field strength is necessary to determine all fields in terms of harmonic functions.
 We showed that non-trivial solutions exist and presented  half-BPS Janus solutions carrying  R-R or NS-NS fluxes. The holographic interpretation of these solutions is given by a superconformal interface theory. 

There are quite a few  possible applications of these solutions as well as open questions and directions for future research. 
It would be interesting to perform an analysis of the structure of the counterterms similar to what has been done in the ${\cal N}=4$  SYM case in \cite{D'Hoker:2006uv,Clark:2004sb,Gaiotto:2008sd}.
The holographic solution suggests that a particular linear combination of dimension $(h,\bar h)=(1/2,1/2)$ chiral primary operators 
from the untwisted and twisted sectors is localized on the defect.  On the field theory side, the supersymmetry variation of the
 counterterm is expected to cancel the supersymmetry variation of the bulk which would become a total derivative and localize on
 the interface. 

Our ansatz for the solution did not turn on the moduli of the $K_3$ or $T^4$.  This can in principle be done using a ten-dimensional ansatz as well. It is however simpler to consider the six-dimensional supergravity theory which is obtained by  the compactification of type IIB on $K_3$ (or $T^4$) \cite{de Boer:1998ip}. The moduli of $K_3$ or $T^4$ are scalar fields in the six-dimensional theory  which, together  with the universal scalar fields discussed in the present paper, live in a coset manifold.  It would be interesting to analyze possible interface theories where the  moduli take the role of the six-dimensional dilaton and axion  and fluxes associated with cycles on $K_3$ (or $T^4$) are turned on.

In general it is also possible to have conformal defects  in a CFT which carry additional degrees of freedom localized on the boundary. In $AdS_3/CFT_2$ such defects can be realized by probe branes with $AdS_2$ worldvolume in $AdS_3$. The probe approximation neglects the back-reaction. As discussed at the end of  section \ref{multipolesol}, the presence of other  branes and the associated conserved charges makes it necessary drop some of the  requirements which we have been imposing on the solutions. It would be interesting to explore the  possibility of  half-BPS solutions which 
correspond to completely back-reacted  brane solutions.

In the two-dimensional conformal field theory there have been interesting recent developments (see  for example \cite{Brunner:2008fa,Bachas:2007td,Fuchs:2007tx}) concerning one-dimensional interfaces. It is an open question whether some of these developments have a counterpart  on the dual supergravity side. For example, it would be interesting to explore whether there are topological defects and whether the notion of fusion of defects has a gravitational analogue.

\bigskip\bigskip

\noindent{\Large \bf Acknowledgements}

\medskip

MG gratefully acknowledges the hospitality of   the Department of Physics and
Astronomy, Johns Hopkins University, while some of the work  in this paper was done. We thank Eric D'Hoker and Per Kraus for useful conversations.
The work of MG and MC was
supported in part by NSF grant PHY-07-57702. The work of MC was supported in part by the 2009-10 Siegfried W. Ulmer Dissertation Year Fellowship.
The work of DK was supported in part by the FWO - Vlaanderen, project G.0235.05
 and in part by the Federal Office for Scientific, Technical and Cultural Affairs
 through the Interuniversity Attraction Poles Programmeâ Belgian Science
 Policy, P6/11-P.

\appendix
\section{IIB supergravity in ten dimensions}
\label{convten}
\setcounter{equation}{0}

In this appendix we gather our conventions for ten-dimensional IIB supergravity. We use the SU(1,1) formalism of \cite{Schwarz:1983qr}.
The bosonic fields are: the metric $g_{\mu\nu}$;
the complex axion-dilaton scalar $B$;  the complex two-form $B^{(2)}_{\mu\nu}$
and the real four-form $C_{(4)}$. We introduce composite fields
in terms of which the field equations are expressed simply as follows,
\bea
\label{sugra1}
P_{\mu}  &=& {1\over 1-|B|^2} \partial_{\mu} B \no \\
Q_{\mu}  &= & {1\over 1-|B|^2}   \Im( B \partial_{\mu}  \bar B)
\eea
We use form notation for the field strengths: $F_{(3)} = d B_{(2)}$,
\bea
\label{GF5}
G & = &  {1\over \sqrt{1-|B|^2}} (F_{(3)} - B \bar F_{(3)} )
\no \\
F_{(5)} & = & dC_{(4)} + { i \over 16} \left ( B_{(2)} \wedge \bar F_{(3)}
- \bar B_{(2)} \wedge  F_{(3)} \right )
\eea
The scalar field $B$ is related to the complex string coupling $\tau$,
the axion $\chi$, and dilaton $\phi$  by
\bea
\label{Btau}
B = {1 +i \tau \over 1 - i \tau } \hskip 1in \tau =  \chi + i e^{- \phi}
\eea
 Note that these definitions do not give the standard $SL(2,R)$ invariant form of the fields. They are however related by a gauge transformation where:
 \be
 P\to e^{{2i \theta}} P, \quad  Q\to Q+ d\theta, \quad G\to e^{{i\over 2} \theta} G
 \ee
 with
 \be
 \theta= {1\over 2i} \log\left({1+ e^{-\phi}- i \chi\over 1+ e^{-\phi}+i\chi}\right)
 \ee
 After this transformation the bosonic fields read
 \be\label{compform}
 P= {1\over 2} \Big( d \phi + i e^{\phi}d\chi\Big), \quad Q= - {1\over 2} e^{\phi} d\chi, \quad G=e^{-\phi/2} H_{3}+ i e^{\phi/2}\Big(F_{3}-\chi H_{3}\Big)
 \ee
 where $H_{3}$ is the NS-NS three-form field strength and $F_{3}$ is the R-R three-form field strength. \\
 In general  the composite fields $P,Q$, and $G$ satisfy  Bianchi identities
given as follows,
\bea
0 &=& dP-2i Q\wedge P
\label{bianchi1} \\
0 &=& d Q + i P\wedge \bar P
\label{bianchi2} \\
0 &=& d G - i Q\wedge G +  P\wedge \bar G
\label{bianchi3} \\
0 &=& d F_{(5)} -  {i\over 8} G \wedge \bar G
\label{bianchi4}
\eea
The field strength $F_{(5)}$ is required to be self-dual,
\bea
\label{SDeq}
 F_{(5) } = * F_{(5) }
\eea
Note that the self-duality condition is related to the convention of the ten-dimensional alternating symbol which we choose:
\be \label{epconvention}
\epsilon^{0123456789}=+1, \quad \epsilon_{0123456789}=-1
\ee
The field equations are given by,
\bea
0 & = &
\nabla ^\mu P_\mu - 2i Q^\mu P_\mu
+ {1\over 24} G_{\mu\nu\rho }G^{\mu\nu\rho}
\label{Peq}
\\
0 & = &
\nabla ^\rho  G_{\mu\nu\rho } -i Q^\rho G_{\mu\nu\rho}
- P^\rho \bar G_{\mu\nu\rho }
+ {2\over 3} i F_{(5)\mu\nu\rho\sigma\tau }G^{\rho\sigma\tau}
\label{Geq}
\\
0 & = & R_{\mu\nu }
- P_\mu  \bar P_\nu  - \bar P_\mu  P_\nu
- {1\over 6} (F_{(5)}^2)_{\mu\nu }
\no \\ && \hskip .5in
- {1\over 8} (G_\mu {} ^{\rho\sigma } \bar G_{\nu \rho\sigma }
+ {\bar G_\mu} {} ^{ \rho\sigma } G_{\nu \rho\sigma })
+{1\over 48 } g_{\mu\nu } G^{\rho\sigma\tau } \bar G_{\rho\sigma\tau }
\label{Eeq}
\eea
The fermionic fields are the dilatino $\lambda$ and the gravitino $\psi_\mu$,
both of which are complex Weyl spinors with opposite ten-dimensional
chiralities, given by $\Gamma_{11} \lambda =\lambda$, and $\Gamma_{11}
\psi_\mu  =-\psi_\mu$. The supersymmetry variations of the fermions  are
\bea
\label{BPS}
\delta\lambda
&=& i (\G \cdot P) \cB^{-1} \ep^*
-{i\over 24} (\G \cdot G) \ep
\label{susy1} \\
\delta \psi_M
&=& D _\mu  \ep
+ {i\over 480}(\G \cdot F_{(5)})  \Gamma_\mu \ep
-{1\over 96}\left ( \Gamma_\mu (\G \cdot G)
+ 2 (\G \cdot G) \G_\mu \right ) \cB^{-1} \ep^*
\no
\eea
The complex conjugation matrix  $\cB$ satisfies $ \cB  \cB^{*}=1$ and $ \cB \Gamma_{\mu} \cB^{-1}=(\Gamma_{\mu})^{*}$.

\section{Basis of gamma matrices}
\setcounter{equation}{0}
Our conventions for the gamma matrices equal to the ones used in \cite{D'Hoker:2007fq}:
\bea
\Gamma^{\mu}&=&\gamma^{\mu}\otimes 1_{2}\otimes 1_{4}\otimes1_{2}, \quad \mu=0,1\no\\
\Gamma^{i}&=&\gamma_{(1)}\otimes \gamma^{i}\otimes 1_{4}\otimes1_{2}, \quad i=2,3\no\\
\Gamma^{l}&=&\gamma_{(1)}\otimes \gamma_{(2)}\otimes \gamma^{l}\otimes1_{2}, \quad l=4,5,6,7\no\\
\Gamma^{a}&=&\gamma_{(1)}\otimes \gamma_{(2)}\otimes \gamma_{(3)} \otimes \gamma^{a}, \quad a=8,9\label{gammatena}
\eea
Where we denote $\mu=0,1$ as the $AdS_{2}$ directions, $i=2,3$ as the $S^2$ directions, $l=4,5,6,7$ as the $K_{3}$ directions and $a=8,9$ as the $\Sigma$ directions.\\
 The sets of two-dimensional and four-dimensional gamma matrices are given by
\be
\gamma^{0}= -i \sigma^{2}, \quad \gamma^{1}=\sigma^{1}, \quad \gamma_{(1)}=\sigma^{3}\label{gammaads2}
\ee
\be
\gamma^{2}=  \sigma^{2}, \quad \gamma^{3}=\sigma^{1}, \quad \gamma_{(2)}=\sigma^{3}\label{gammas2}
\ee
\be
\gamma^{4}=  \sigma^{1}\otimes 1_{2},  \quad \gamma^{5}=\sigma^{2}\otimes 1_{2},\quad\gamma^{6}=  \sigma^{3}\otimes \sigma^{1} \ee
\be   \gamma^{5}=\sigma^{3}\otimes \sigma^{2} \quad \gamma_{(3)}=\sigma^{3}\otimes \sigma^{3}
\ee
\be
\gamma^{8}= \sigma^{1}, \quad \gamma^{9}=\sigma^{2}, \quad \gamma_{(4)}=\sigma^{3}
\ee
The explicit form of the ten-dimensional Gamma matrices (\ref{gammatena}) is then given by
\bea
\Gamma^{0}&=& -i \sigma^{2 }\otimes 1_{2}\otimes
1_{2}\otimes 1_{2}\otimes 1_{2}\no\\
\Gamma^{1}&=&\;\; \;\;\sigma^{1 }\otimes 1_{2}\otimes  1_{2}\otimes 1_{2}\otimes 1_{2}\no\\
\Gamma^{2}&=&\;\; \;\;\sigma^{3 }\otimes \sigma^{2}\otimes 1_{2}\otimes 1_{2}\otimes 1_{2}\no\\
\Gamma^{3}&=&\;\; \;\;\sigma^{3 }\otimes \sigma^{1}\otimes 1_{2}\otimes 1_{2}\otimes 1_{2}\no\\
\Gamma^{4}&=&\;\; \;\;\sigma^{3 }\otimes \sigma^{3}\otimes \sigma_{1}\otimes 1_{2}\otimes 1_{2}\no\\
\Gamma^{5}&=&\;\; \;\;\sigma^{3 }\otimes \sigma^{3}\otimes \sigma_{2}\otimes 1_{2}\otimes 1_{2}\no\\
\Gamma^{6}&=&\;\; \;\;\sigma^{3 }\otimes \sigma^{3}\otimes \sigma_{3}\otimes \sigma_{1}\otimes 1_{2}\no\\
\Gamma^{7}&=&\;\; \;\;\sigma^{3 }\otimes \sigma^{3}\otimes \sigma_{3}\otimes \sigma_{2}\otimes 1_{2}\no\\
\Gamma^{8}&=&\;\; \;\;\sigma^{3 }\otimes \sigma^{3}\otimes \sigma_{3}\otimes \sigma_{3}\otimes \sigma_{1}\no\\
\Gamma^{9}&=&\;\; \;\;\sigma^{3 }\otimes \sigma^{3}\otimes \sigma_{3}\otimes \sigma_{3}\otimes \sigma_{2}
\eea
the ten-dimensional chirality matrix is given by
\bea
\Gamma^{11}&=&\gamma_{(1)}\otimes\gamma_{(2)}\otimes\gamma_{(3)}\otimes\gamma_{(4)}
=\sigma_{3}\otimes\sigma_{3}\otimes\sigma_{3}\otimes\sigma_{3}\otimes\sigma_{3}
\eea
The supersymmetry transformation parameter $\epsilon$, dilatino and gravitino satisfy the following chirality condition:
\be
\Gamma^{11}\epsilon=-\epsilon, \quad \Gamma^{11}\lambda=+\lambda, \quad \Gamma^{11}\psi_{\mu}=-\psi_{\mu}
\ee

The complex conjugation matrices  and their properties for the gamma matrices $\gamma^{\mu}, \gamma^{i}, \gamma^{l }$ and $\gamma^{a}$ are the same as in  appendix A of \cite{D'Hoker:2007fq}
\bea
& B^{(1)}=1_{2}, & \qquad B^{(2)}=\sigma_{2}, \no \\
& B^{(3)}=\sigma^{2}\otimes \sigma^{1}, & \qquad B^{(4)}= \sigma^{2}
\eea
The ten-dimensional complex conjugation matrix is
\be
B= i B^{(1)}\otimes \gamma^{(2)}B^{(2)}\otimes B^{(3)}\otimes B^{(4)}= 1_{2}\otimes \sigma^{1}\otimes \sigma^{2}\otimes \sigma^{1}\otimes \sigma^{2}
\ee

\section{Useful formulae for the reduction}\label{fromulared}
\setcounter{equation}{0}

Here we gather some   formulas for products of gamma matrices which are useful for the reduction
\bea
\Gamma^{01}&=& -\gamma_{(1)}\otimes 1_{2}\otimes 1_{(4)} \otimes 1_{2}
\\
\Gamma^{23}&=&  -i \; 1_{2}\otimes \gamma_{(2)}\otimes 1_{(4)} \otimes 1_{2}
\\
\Gamma^{89}&=& i \; 1_{2} \otimes 1_{2}\otimes 1_{(4)} \otimes \gamma_{(4)}
\\
\Gamma^{a01}&=& -1_{2}\otimes \gamma_{(2)}\otimes \gamma_{(3)} \otimes \gamma^{a}
\\
\Gamma^{a23}&=&  -i \; \gamma_{(1)}\otimes 1_{2}\otimes \gamma_{(3)} \otimes \gamma^{a}
\\
\Gamma^{0123}&=&  i\; \gamma_{(1)}\otimes \gamma_{(2)}\otimes 1_{(4)} \otimes 1_{2}
\\
\Gamma^{4567}&=&  - 1_{2}\otimes 1_{(2)}\otimes \gamma_{(3)} \otimes 1_{2}
\\
\Gamma^{a0123}&=& i\; 1_{(2)}\otimes 1_{(2)}\otimes \gamma_{(3)} \otimes \gamma^{a}
\\
\Gamma^{a4567}&=&  - \gamma_{(1)}\otimes \gamma_{(2)}\otimes 1_{(4)} \otimes \gamma^{a}
\\
\Gamma^{a0123}\Gamma^{\mu}&=&  i\; \gamma^{\mu}\otimes 1_{(2)}\otimes \gamma_{(3)} \otimes \gamma^{a}
\\
\Gamma^{a4567}\Gamma^{\mu}&=& +\gamma^{\mu}\gamma_{(1)}\otimes \gamma_{(2)}\otimes 1_{(4)} \otimes \gamma^{a}
\\
\Gamma^{a0123}\Gamma^{i}&=&  i\; \gamma_{(1)} \otimes \gamma^{i} \otimes \gamma_{(3)} \otimes \gamma^{a}
\\
\Gamma^{a4567}\Gamma^{i}&=&  1_{2}\otimes \gamma^{i }\gamma_{(2)}\otimes 1_{(4)} \otimes \gamma^{a}
\\
\Gamma^{a0123}\Gamma^{l}&=&-  i\; \gamma_{(1)} \otimes \gamma_{(2)} \otimes \gamma^{l}\gamma_{(3)} \otimes \gamma^{a}
\\
\Gamma^{a4567}\Gamma^{l}&=&
 - 1_{2}\otimes 1_{2} \otimes \gamma^{l} \otimes \gamma^{a}
 \\
\Gamma^{a0123}\Gamma^{b}&=& i\; \gamma_{(1)} \otimes \gamma_{(2)} \otimes 1_{(4)}\otimes(\delta^{ab} 1_{2}+i \epsilon^{ab} \gamma_{(4)})
\\
\Gamma^{a4567}\Gamma^{b}&=&
 - 1_{2}\otimes 1_{2} \otimes \gamma_{(3)} \otimes (\delta^{ab} 1_{2}+i \epsilon^{ab} \gamma_{(4)})
\\
\Gamma^{\mu} \Gamma^{a01}+2 \Gamma^{a01}\Gamma^{\mu}&=&-3 \gamma^{\mu}\otimes \gamma_{(2)}\otimes \gamma_{(3)}\otimes \gamma^{a}
\\
\Gamma^{\mu} \Gamma^{a23}+2 \Gamma^{a23}\Gamma^{\mu}&=&i\; \gamma^{\mu} \gamma_{(1)}\otimes 1_{2}\otimes \gamma_{(3)}\otimes \gamma^{a}
\\
\Gamma^{i} \Gamma^{a01}+2 \Gamma^{a01}\Gamma^{i}&=& \gamma_{(1)}\otimes \gamma^{i}\gamma_{(2)}\otimes\gamma_{(3)}\otimes \gamma^{a}
\\
\Gamma^{i} \Gamma^{a23}+2 \Gamma^{a23}\Gamma^{i}&=& -3 i \;1_{2}\otimes \gamma^{i}\otimes\gamma_{(3)}\otimes \gamma^{a}
\\
\Gamma^{l} \Gamma^{a01}+2 \Gamma^{a01}\Gamma^{l}&=& \gamma_{(1)}\otimes 1_{2}\otimes \gamma^{l}\gamma_{(3)}\otimes\gamma^{a}
\\
\Gamma^{l} \Gamma^{a23}+2 \Gamma^{a23}\Gamma^{l}&=&i\;  1_{2}\otimes \gamma_{(2)}\otimes \gamma^{l}\gamma_{(3)}\otimes\gamma^{a}
\\
\Gamma^{b} \Gamma^{a01}+2 \Gamma^{a01}\Gamma^{b}&=& -\gamma_{(1)}\otimes 1_{2}\otimes 1_{(4)} \otimes ( 3 \delta^{ab} 1_{2} + i \epsilon^{ab} \gamma_{(4)} )
\\
\Gamma^{b} \Gamma^{a23}+2 \Gamma^{a23}\Gamma^{b}&=& -i \; 1_{2}\otimes \gamma_{(2)}\otimes 1_{(4)} \otimes ( 3 \delta^{ab} 1_{2} +i \epsilon^{ab} \gamma_{(4)} )
\eea
Where the following relation was used $\gamma^{a}\gamma^{b} = \delta^{ab} 1_{2}+ i \epsilon^{ab} \gamma_{(4)}$ with $ \epsilon^{89}=+1, \; \epsilon^{98}=-1$.

\section{Killing spinors on $AdS_{2}$ and $S^{2}$ and $K_3$}
\setcounter{equation}{0}
Following the general philosophy for the half-BPS Janus solutions found in \cite{D'Hoker:2007xy,D'Hoker:2007fq},  the supersymmetry parameters $\epsilon^{a}$ should be expanded in terms of Killing spinors on $AdS_{2}$ and $S^{2}$ since a less symmetric choice would break additional supersymmetries.

\subsection{Killing spinors  for $AdS_{2}$}

There are two possible equations for Killing spinors on $AdS_{2}$.
An $AdS_{2}$ space with unit radius satisfies $R_{\mu\nu}= - g_{\mu\nu}$
The Killing spinor equation is given by
\be\label{killingads}
\partial_{\mu} \chi_{\eta}^{(1)}+{1\over 4} \omega_{\mu}^{\; ab} \gamma_{ab}\chi_{\eta}^{(1)}-\eta {i\over 2} \gamma_{\mu } \gamma_{(1)} \chi^{(1)}_{\eta} =0
\ee
where $\omega_{\mu}^{ab}$ is the spin connection and $\eta=\pm 1$. $\gamma_{(1)}$ denotes the chirality matrix with the property $\{\gamma_{(1)}, \gamma_{\mu}\}=0$. Integrability demands that $\eta=\pm 1$ and the two solutions are linearly independent.
There is an alternative equation for the Killing spinor which is related to (\ref{killingads}) by an unitary rotation
\be\label{killingadsb}
\partial_{\mu}  \chi^{'(1)}_{\eta}+{1\over 4} \omega_{\mu}^{\; ab} \gamma_{ab}\chi^{'(1)}_{\eta}-\eta {1\over 2}  \gamma_{\mu }  \chi^{'(1)}_{\eta} =0
\ee
We use the  representation of the two-dimensional gamma matrices given in (\ref{gammaads2}).

\subsection{Killing spinors  for  $S^{2}$}

There are two possible equations for Killing spinors on $S^{2}$.
An $S^{2}$ space with unit radius satisfies $R_{\mu\nu}= + g_{\mu\nu}$
The Killing spinor equation is given by
\be
\partial_{i} \chi^{(2)}_{\eta}+{1\over 4} \omega_{i}^{\; ab} \gamma_{ab}\chi^{(2)}_{\eta} -\eta {1\over 2}  \gamma_{i } \gamma^{(2)} \chi_{\eta} =0
\ee
where $\omega_{\mu}^{ab}$ is the spin connection and by integrability $\eta=\pm 1$.
The alternative equation is
\be
\partial_{i} \chi^{'(2)}_{\eta}+{1\over 4} \omega_{i}^{\; ab} \gamma_{ab}\chi^{'(2)}_{\eta} -\eta {i\over 2} \gamma_{i }  \chi^{'(2)}_{\eta}=0
\ee

We use the  representation of the two-dimensional gamma matrices given in (\ref{gammas2}).

\subsection{Killing spinors on $K_{3}$}

A $K_{3}$ surface is Ricci flat, i.e. $R_{lm}=0$. The Killing spinor equation is simply
\be
D_{l} \chi^{(3)}= \partial_{l} \chi^{(3)}+{1\over 4} \omega_{l}^{\; ab} \gamma_{ab}\chi^{(3)}=0
\ee
The Killing spinor has a definite four-dimensional chirality, since the holonomy is restricted to $SU(2)$, and we choose
\be
\gamma^{(3)}\chi^{(3)}_{\eta}=+\chi^{(3)}_{\eta}
\ee
There are two linearly independent Killing spinors labelled by $\eta=\pm 1$.

\subsection{Expansion of the supersymmetry parameter $\epsilon$}

The chirality condition $\Gamma^{11}$ acts on $\epsilon$ in the following way:
\bea \label{gamma11}
\Gamma^{11}\epsilon&=& \gamma_{(1)}\otimes\gamma_{(2)}\otimes\gamma_{(3)}\otimes\gamma_{(4)} \sum_{\eta_{1},\eta_{2}}  \chi^{(1)}_{\eta_{1}} \otimes \chi^{(2)}_{\eta_{2}}\otimes \chi^{(3)}_{\eta_{3}} \otimes \xi_{\eta_{1},\eta_{2},\eta_{3}  }\no\\
&=& \sum_{\eta_{1},\eta_{2}}  \chi^{(1)}_{-\eta_{1}} \otimes \chi^{(2)}_{-\eta_{2}}\otimes \chi^{(3)}_{\eta_{3}} \otimes \gamma_{(4)}\xi_{\eta_{1},\eta_{2},\eta_{3}  } =   \sum_{\eta_{1},\eta_{2}}  \chi^{(1)}_{\eta_{1}} \otimes \chi^{(2)}_{\eta_{2}}\otimes \chi^{(3)}_{\eta_{3}} \otimes \gamma_{(4)}\xi_{-\eta_{1},-\eta_{2},\eta_{3}  } \qquad
\eea
Ten-dimensional complex conjugation acts as follows on the spinor
 \bea\label{bconj}
 B^{-1}\epsilon^{*} &=&  i \;B_{(1)}\otimes \gamma^{(2)}B_{(2)}\otimes B_{(3)}\otimes B_{(4)}  \Big(\sum_{\eta_{1},\eta_{2}}  \chi^{(1)}_{\eta_{1}} \otimes \chi^{(2)}_{\eta_{2}}\otimes \chi^{(3)}_{\eta_{3}} \otimes \xi_{\eta_{1},\eta_{2},\eta_{3}  }\Big)^{*}\no\\
 &=&   i  \sum_{\eta_{1},\eta_{2}}  \eta_{2 }\; \chi^{(1)}_{\eta_{1}} \otimes \chi^{(2)}_{-\eta_{2}}\otimes \chi^{(3)}_{\eta_{3}} \otimes B_{(4)}^{-1}(\xi_{\eta_{1},\eta_{2},\eta_{3}  })^{* } \no\\
  &=&  \sum_{\eta_{1},\eta_{2}}  \chi^{(1)}_{\eta_{1}} \otimes \chi^{(2)}_{\eta_{2}}\otimes \chi^{(3)}_{\eta_{3}} \otimes (- i \;\eta_{2 }) B_{(4)}^{-1}(\xi_{\eta_{1},-\eta_{2},\eta_{3}  })^{* }
\eea
where we used the reality condition  given in
(\ref{realconks}).
\section{Remaining   Bianchi identities and potentials}\label{bianchiappendix}
\setcounter{equation}{0}

\subsection{Reduction of the three-form Bianchi Identity} \label{bianchi3app}

The Bianchi identities for $P$ and $Q$, (\ref{bianchi1}) and (\ref{bianchi2}), are automatically satisfied using the definitions \ref{compform}.
The  Bianchi identity for the three-form $G$, (\ref{bianchi3}) is reduced using definition (\ref{compform})
to
\bea
0 = d H_3 = d(e^{-\Phi}Re(G)) \no \\
0= d F_3 = d(e^{\Phi}Im(G)+\chi e^{-\Phi}Re(G)) \label{bianchig}
\eea
When we use the ansatz in section \ref{sectwoone}, the identity reduces to:
\bea
\partial_{\bar z} \Big( e^{-\Phi } f_{1}^{2} \rho Re(g^{(1)})_{z}\Big)-\partial_{z} \Big( e^{-\Phi }  f_{1}^{2} \rho Re(g^{(1)})_{\bar z}\Big)&=&0\no\\
\partial_{\bar z} \Big( e^{-\Phi }  f_{2}^{2} \rho Re(g^{(2)})_{z}\Big)-\partial_{z} \Big( e^{-\Phi }  f_{2}^{2} \rho Re(g^{(2)})_{\bar z}\Big)&=&0\no\\
\partial_{\bar z} \Big( e^{\Phi }  f_{1}^{2} \rho Im(g^{(1)}) _{z}+ \chi e^{-\Phi } f_{1}^{2} \rho Re(g^{(1)})_{z} \Big)-\partial_{z} \Big( e^{\Phi }  f_{1}^{2} \rho Im(g^{(1)})_{\bar z} + \chi e^{-\Phi } f_{1}^{2} \rho Re(g^{(1)})_{\bar z} \Big)&=&0\no\\
\partial_{\bar z} \Big( e^{\Phi }  f_{1}^{2} \rho Im(g^{(2)}) _{z}+ \chi e^{-\Phi } f_{1}^{2} \rho Re(g^{(2)})_{z} \Big)-\partial_{z} \Big( e^{\Phi }  f_{1}^{2} \rho Im(g^{(2)})_{\bar z} + \chi e^{-\Phi } f_{1}^{2} \rho Re(g^{(2)}) _{\bar z}\Big)&=& 0 \no \qquad \\ 
  \label{Bianchi} \eea
We write the real and imaginary components of the fluxes in terms of the combinations displayed in (\ref{gdefine1}-\ref{gdefine4}). 
\bea
Re(g^{(1)})_{z} &=& {1\over 4} ( g^{(1)}_z - i g^{(2)}_z ) +  {1\over 4} ( g^{(1)}_z + i g^{(2)}_z )+ {1\over 4} ( \bar g^{(1)}_z - i \bar g^{(2)}_z)+  {1\over 4} ( \bar g^{(1)}_z + i \bar g^{(2)}_z )\no
\\
Im(g^{(1)})_{z}&=& {1\over 4 i} ( g^{(1)}_z - i g^{(2)}_z ) +  {1\over 4i} ( g^{(1)}_z + i g^{(2)}_z )-{1\over 4i} ( \bar g^{(1)}_z - i \bar g^{(2)}_z)- {1\over 4i} ( \bar g^{(1)}_z + i \bar g^{(2)}_z )\no \\
Re(g^{(2)})_{z}&=& -{1\over 4i} ( g^{(1)}_z - i g^{(2)}_z ) +  {1\over 4i} ( g^{(1)}_z + i g^{(2)}_z )- {1\over 4i} ( \bar g^{(1)}_z - i \bar g^{(2)}_z)+  {1\over 4i} ( \bar g^{(1)}_z + i \bar g^{(2)}_z )\no \\
Im(g^{(2)})_{z}&=& {1\over 4} ( g^{(1)}_z - i g^{(2)}_z ) -  {1\over 4} ( g^{(1)}_z + i g^{(2)}_z )- {1\over 4} ( \bar g^{(1)}_z - i \bar g^{(2)}_z)+  {1\over 4} ( \bar g^{(1)}_z + i \bar g^{(2)}_z ) \quad \eea
The expressions in terms of the spinor components introduced in (\ref{defxip}) are:
\bea
\rho \; \Re (g^{(1)}) _z &=& {1 \over 2} \left( - \nu {\a^2+\b^{*2} \over f_1 f_2 } \rho + {\a^{*2} + \b^2 \over \a^* \b} \partial_w \log \Big( {f_1 \over f_2} e^{-2 \Phi}   \Big) + i {\a^{*2}-\b^2 \over \a^* \b} e^{-2 \Phi} \partial_w \chi \right) \\
 \rho \; \Re (g^{(2)})_z &=& {i \over 2} \left( -\nu {\a^2+\b^{*2} \over f_1 f_2 } \rho + {\a^{*2} + \b^2 \over \a^* \b} \partial_w \log \Big( {f_1 \over f_2} e^{2 \Phi}   \Big) - i {\a^{*2}-\b^2 \over \a^* \b} e^{-2 \Phi} \partial_w \chi \right) \\
 \rho \;  \Im (g^{(1)})_z &=& -{i \over 2} \left( -\nu {\a^2-\b^{*2} \over f_1 f_2 } \rho - {\a^{*2} - \b^2 \over \a^* \b} \partial_w \log \Big( {f_1 \over f_2} e^{2 \Phi}   \Big) + i {\a^{*2}+\b^2 \over \a^* \b} e^{-2 \Phi} \partial_w \chi \right) \\
 \rho \; \Im (g^{(2)})_z&=& {1 \over 2} \left( -\nu {\a^2-\b^{*2} \over f_1 f_2 } \rho - {\a^{*2} - \b^2 \over \a^* \b} \partial_w \log \Big( {f_1 \over f_2} e^{-2 \Phi}   \Big) - i {\a^{*2}+\b^2 \over \a^* \b} e^{-2 \Phi} \partial_w \chi \right) \qquad  \eea
Using equation (\ref{defPsi}) and the expressions (\ref{def-f1}-\ref{def-f3}) we observe that:
\be \rho \a^* \b  = i \nu f_1 f_2 e^{\psi} = - \nu {\partial_{\bar w} H \over H} f_1 f_2 \ee
Then, using (\ref{resultpsi}) we can show that:
\bea f_1^2 \rho e^{-\Phi} \Re(g^{(1)})_z & = & -{1 \over 2} \Big( \vartheta_1 \partial_w H + H \partial_w \bar \vartheta_1 \Big) \label{gvartheta1} \\
 f_2^2 \rho e^{-\Phi} \Re(g^{(2)})_z& = & -{i \over 2} \Big( \vartheta_2 \partial_w H - H  \partial_w \bar \vartheta_2 \Big) \label{gvartheta2} \\
f_1^2 \rho e^{\Phi} \Im(g^{(1)})_z & = & - {i \over 2} \Big( \vartheta_3 \partial_w H - H \partial_w \bar \vartheta_3 - i H \bar \vartheta_1  \partial_w \chi \Big) \label{gvartheta3} \\
f_2^2 \rho e^{\Phi} \Im(g^{(2)})_z & = &  {1 \over 2} \Big( \vartheta_4 \partial_w H + H \partial_w \bar \vartheta_4 + i H \bar \vartheta_2  \partial_w \chi \Big) \label{gvartheta4} \eea
where:
\bea \vartheta_1 = k {f_1 \over f_2} e^{-2 \Phi}  { e^{-\lambda } + i e^{\lambda} \chi \over \partial_w H }, && 
\vartheta_3 = k {f_1 \over f_2} e^{+ 2 \Phi}  { e^{\lambda } \over \partial_w H } \label{def-vartheta1} \\
\vartheta_2 = k {f_2 \over f_1} e^{-2 \Phi}  { e^{-\lambda } + i e^{\lambda} \chi \over \partial_w H},  &&
\vartheta_4 = k {f_2 \over f_1} e^{+ 2 \Phi}  { e^{\lambda } \over \partial_w H} \label{def-vartheta4} \eea
The Bianchi identities (\ref{Bianchi}) reduce to:
\bea \partial_w \partial_{\bar w}  \Im \vartheta_1 = 0, && 
\partial_w \partial_{\bar w} \Big(  \Re \vartheta_3 + \chi \Im \vartheta_1 \Big) = 0  \label{BI-vartheta1} \\
\partial_w \partial_{\bar w} \Re \vartheta_2 = 0,&&
\partial_w \partial_{\bar w} \Big(  \Im \vartheta_4 - \chi \Re \vartheta_2 \Big) = 0  \label{BI-vartheta4}\eea

Using the expressions for the fields from section \ref{totsolution} it is possible to show that the Bianchi identity for the three-form anti-symmetric tensor field is automatically satisfied.
We obtain from (\ref{sol-f1}-\ref{sol-f2})
\bea {f_2 \over f_1} e^{2 \Phi} &=& {1 \over 2} \Big( (A + \bar A)  - { (B + \bar B)^2 \over \hat h}  \Big) \\
 {f_1 \over f_2} e^{2 \Phi} &=& {1 \over 2} \Big( (A + \bar A)  -  { (B - \bar B)^2 \over \hat h} \Big)  \eea
Using the above expressions in equations (\ref{def-vartheta1}-\ref{def-vartheta4}) we get:
\bea \Im \vartheta_1 &=& {f_1 \over f_2} e^{- 2 \Phi} \Im { A + i \chi \over B } = {1 \over 2 i} \Big( {1\over B}- {1 \over \bar B} \Big) \\
\Re \vartheta_2 &=& {f_2 \over f_1} e^{- 2 \Phi} \Re { A + i \chi \over B } = {1 \over 2 } \Big( {1\over B} + {1 \over \bar B} \Big) \\
\Re \vartheta_3 + \chi \Im \vartheta_1 &=& {f_1 \over f_2} e^{ 2 \Phi} \Re { 1 \over B } + { \chi \over 2 }  \Im {1\over B} = {1 \over 2} \Big( {A \over B} + {\bar A \over \bar B} \Big) \\
\Im \vartheta_4 - \chi \Re \vartheta_2 &=& {f_2 \over f_1} e^{ 2 \Phi} \Im { 1 \over B } - { \chi \over 2 } \Re {1\over B} = {1 \over 2 i} \Big( {A \over B}- {\bar A \over \bar B} \Big) 
\eea
The above combinations are manifestly harmonic functions and therefore tautologically  satisfy equations (\ref{BI-vartheta1}-\ref{BI-vartheta4}).

We can now obtain the potentials by rewriting the three-form field strengths as total derivatives:
\bea f_1^2 \rho e^{-\Phi} \Re(g^{(1)})_z & = & \partial_w b^{(1)} \label{potdef1}\\
 f_2^2 \rho e^{-\Phi} \Re(g^{(2)})_z & = & \partial_w b^{(2)} \label{potdef2b}\\
f_1^2 \rho e^{\Phi} \Im(g^{(1)})_z + \chi f_1^2 \rho e^{-\Phi} \Re(g^{(1)})_z  & = & \partial_w c^{(1)} \label{potdef3}\\
f_2^2 \rho e^{\Phi} \Im(g^{(2)}) _z+ \chi f_2^2 \rho e^{-\Phi} \Re(g^{(2)})_z & = & \partial_w c^{(2)} \label{potdef4}\eea
The two-form potentials are then shown to be:
\bea b^{(1)} = & - \Big( { \textstyle H \bar \vartheta_1 \over \textstyle 2} + i \mu_1 \Big)  \qquad & \text{with: } \partial_w \mu_1 =  \Im \vartheta_1 \partial_w H \label{def-b1} \\
 b^{(2)} = & i \Big( {\textstyle H \bar \vartheta_2 \over \textstyle 2} -  \mu_2 \Big)  \qquad & \text{with: } \partial_w \mu_2 =  \Re \vartheta_2 \partial_w H \label{def-b2}\\
  c^{(1)} = & i \Big( {\textstyle \bar \vartheta_3 + i \chi \bar \vartheta_1 \over \textstyle 2} H - \mu_3 \Big)  \qquad & \text{with: } \partial_w \mu_3 =  \big( \Re \vartheta_3 + \chi \Im \vartheta_1 \big) \label{def-c1} \partial_w H \\
  c^{(2)} = & \Big( {\textstyle \bar \vartheta_4 + i \chi \bar \vartheta_2 \over \textstyle 2} H +i \mu_4 \Big)  \qquad & \text{with: } \partial_w \mu_4 =  \big( \Im \vartheta_4 - \chi \Re \vartheta_2 \big)  \partial_w H  \label{def-c2}\eea
 The potentials written in terms of our holomorphic and harmonic functions are
\bea b^{(1)} &=& - {H (B + \bar B) \over (A + \bar A) \hat h - (B + \bar B)^2 } - h_1, \qquad h_1={1 \over 2} \int {\partial_w H \over B} + c.c. \label{potharmonic1app}\\
 b^{(2)} &=& -i  {H (B - \bar B) \over (A + \bar A) \hat h - (B - \bar B)^2 } + \tilde h_1, \qquad  \tilde h_1={1 \over 2 i} \int {\partial_w H \over B} + c.c. \label{potharmonic2app}\\
c^{(1)} & = & - i {H (A \bar B -  \bar A B) \over (A + \bar A) \hat h - (B + \bar B)^2 } + \tilde h_2, \qquad \tilde h_2={1 \over 2 i} \int {A \over B}\partial_w H + c.c.  \label{potharmonic3app}\\
c^{(2)} & = & - {H (A \bar B +  \bar A B) \over (A + \bar A) \hat h - (B - \bar B)^2 } + h_2, \qquad  h_2={1 \over 2 } \int {A \over B}\partial_w H + c.c.  \label{potharmonic4app}\eea
where one should note that the harmonic functions $\tilde h_i$ and $h_i$ are conjugate to each other.
A four-form potential can also be defined for the five-form field strength. By self-duality the two components are related and we give the one along $K_3$
\be f^4_3 \rho \tilde h_z = \partial_w C_K \quad 
C_K = -{i \over 2} {B^2 - \bar B^2 \over A + \bar A} - {1\over 2}\tilde { h}
 \label{eq-Ckb} \ee
Here $\tilde { h} $ is the harmonic function conjugate to $\hat h$ so that  $ \partial_w \tilde { h} = - i \partial_w  \hat  h $.

\subsection{$AdS_2\times S^2$ component of the five-form Bianchi-identity}

The five-form component along the $AdS_2$ and $S^2$ directions is given by
\begin{eqnarray}
F_5&=& h_z e^{z0123}+ h_{\bar z} e^{\bar z 0123}\\
&=& \rho f_1^2 f_2^2 h_z\;  \hat e^{z0123}+ \rho f_1^2 f_2^2 h_{\bar z} \;\hat e^{\bar z0123}
\end{eqnarray}
hence
\begin{equation}
dF_5 = \Big( \partial_w (\rho f_1^2 f_2^2 h_{\bar z})- \partial_{\bar w} (\rho f_1^2 f_2^2 h_{ z})\Big) \hat e^{z\bar z0123}
\end{equation}
The third rank anti-symmetric tensor forms are given by
\begin{eqnarray}
G=g_z^{(1)} \rho f_1^2 \hat e^{z01}+ g_{\bar z}^{(1)} \rho f_1^2 \hat e^{\bar z01}+g_z^{(2)} \rho f_2^2 \hat e^{z23}+ g_{\bar z}^{(2)} \rho f_2^2 \hat e^{\bar z23}\\
\bar G=\bar g_z^{(1)} \rho f_1^2 \hat e^{z01}+ \bar g_{\bar z}^{(1)} \rho f_1^2 \hat e^{\bar z01}+\bar g_z^{(2)} \rho f_2^2 \hat e^{z23}+\bar g_{\bar z}^{(2)} \rho f_2^2 \hat e^{\bar z23}
\end{eqnarray}
Hence we get
\begin{eqnarray}
G\wedge \bar G= \Big( g_z^{(1)} \bar g_{\bar z}^{(2)} -  g_{\bar z}^{(1)} \bar g_{ z}^{(2)}+g_z^{(2)} \bar g_{\bar z}^{(1)} -  g_{\bar z}^{(2)} \bar g_{ z}^{(1)}\Big)\rho^2 f_1^2 f_2^2 \; \hat e^{z\bar z0123}
\end{eqnarray}
Hence the second part of the Bianchi identity becomes
\begin{equation}\label{bianchifinal}
 \partial_w (\rho f_1^2 f_2^2 h_{\bar z})- \partial_{\bar w} (\rho f_1^2 f_2^2 h_{ z})-{i\over 8} \rho^2 f_1^2 f_2^2  \big( g_z^{(1)} \bar g_{\bar z}^{(2)} -  g_{\bar z}^{(1)} \bar g_{ z}^{(2)}+g_z^{(2)} \bar g_{\bar z}^{(1)} -  g_{\bar z}^{(2)} \bar g_{ z}^{(1)}\big)=0
\end{equation}
Using (\ref{hzdefine}) and the expressions from the previous sections one obtains
\be
\rho f_1^2 f_2^2 h_{\bar z}={1\over 8}{H^2\Big( -B^2 A' + \bar B^2 A'+2(A+\bar A)B B'-(A+\bar A)^2 \partial_w \hat h\Big)\over B^4+(\bar B^2-(A+\bar A)\hat h)^2-2 B^2 (\bar B^2+ (A+\bar A)\hat h) }
\ee
employing the relation
\bea\label{gcombine}
&& g_z^{(1)} \bar g_{\bar z}^{(2)} -  g_{\bar z}^{(1)} \bar g_{ z}^{(2)}+g_z^{(2)} \bar g_{\bar z}^{(1)} -  g_{\bar z}^{(2)} \bar g_{ z}^{(1)}= 
{1\over 2i}\Big\{ ( g_z^{(1)}+i g_z^{(2)})( \bar g_{\bar z}^{(1)}+ i  \bar g_{\bar z}^{(2)}) - \no \\
&& \quad ( \bar g_z^{(1)}+i \bar g_z^{(2)})(  g_{\bar z}^{(1)}+ i   g_{\bar z}^{(2)})
+( \bar g_z^{(1)}-i \bar g_z^{(2)})(  g_{\bar z}^{(1)}- i   g_{\bar z}^{(2)})
 -( g_z^{(1)}-i g_z^{(2)})( \bar g_{\bar z}^{(1)}- i  \bar g_{\bar z}^{(2)})\Big\} \qquad \; \; \;
\eea
Using (\ref{gdefine1})-(\ref{gdefine4}) one obtains the following expression:
\bea\label{gcombine2}
&& ( g_z^{(1)}+i g_z^{(2)})( \bar g_{\bar z}^{(1)}+ i  \bar g_{\bar z}^{(2)})-( \bar g_z^{(1)}+i \bar g_z^{(2)})(  g_{\bar z}^{(1)}+ i   g_{\bar z}^{(2)})=\no\\
&& \qquad= -{8\rho^3 H\over \partial_{\bar w} H} \Big( \bar P_w \beta^4- P_w \alpha^{*4}\Big) +{8\rho^3 H^2\over \partial_{\bar w} H \partial_w H} \Big(  P_w \alpha^{*2}\beta^{*2} - \bar P_w\alpha^2  \beta^2 \Big)
\eea
Note that the third and fourth term in (\ref{gcombine}) are the complex conjugate of (\ref{gcombine2}).
Plugging in the expressions in section \ref{totsolution} it can be shown that the Bianchi identity (\ref{bianchifinal}) is automatically satisfied.

\section{Equations of motion } \label{EOMappendix}
\setcounter{equation}{0}
In this section we  check that the local half-BPS solution which is parameterized by the two harmonic functions $\hat h$
 and $H$ as well as two holomorphic functions $A,B$ satisfies the bosonic equations of motion. In particular:

 $\bullet$ Einstein equation along $AdS_2$ is equivalent to
\be
R_{ \mu\nu}+ 8 g_{\mu\nu}  h_z h_{\bar z} + {3\over 8} g_{\mu\nu} \big( g^{(1)}_z \bar g^{(1)}_{\bar z}   +g^{(1)}_{\bar z} \bar g^{(1)}_{ z}\big)+{1\over 8} g_{\mu\nu} \big( g^{(2)}_z \bar g^{(2)}_{\bar z} +  g^{(2)}_{\bar z} \bar g^{(2)}_{ z}\big)=0, \quad \mu,\nu=0,1
\ee

 $\bullet$ Einstein equation along $S^2$ is equivalent to

\be
R_{ ij}+ 8 g_{ ij}  h_z h_{\bar z} - {1\over 8} g_{ ij} \big( g^{(1)}_z \bar g^{(1)}_{\bar z}   +g^{(1)}_{\bar z} \bar g^{(1)}_{ z}\big)-{3\over 8} g_{ ij} \big( g^{(2)}_z \bar g^{(2)}_{\bar z} +  g^{(2)}_{\bar z} \bar g^{(2)}_{ z}\big)=0, \quad i,j=2,3
\ee

$\bullet$ Einstein equation along $K_3$ is equivalent to

\be
R_{ ab}- 8 g_{ab}  h_z h_{\bar z} - {1\over 8} g_{ab} \big( g^{(1)}_z \bar g^{(1)}_{\bar z}   +g^{(1)}_{\bar z} \bar g^{(1)}_{ z}\big)+{1\over 8} g_{ab} \big( g^{(2)}_z \bar g^{(2)}_{\bar z} +  g^{(2)}_{\bar z} \bar g^{(2)}_{ z}\big)=0, \quad a,b=4,\cdots,7
\ee

$\bullet$ The Einstein equation along the $\Sigma_2$ directions has several components
\be
R_{z\bar z} + {1\over 8} g_{z\bar z} \big( g^{(1)}_z \bar g^{(1)}_{\bar z}   +g^{(1)}_{\bar z} \bar g^{(1)}_{ z}\big)-{1\over 8}g_{z\bar z} \big( g^{(2)}_z \bar g^{(2)}_{\bar z} +  g^{(2)}_{\bar z} \bar g^{(2)}_{ z}\big)- g_{z\bar z}  \big(P_z \bar P_{\bar z} +\bar P_z  P_{\bar z} \big)=0
\ee
and
\be
R_{zz}-2 \rho^2 P_z \bar P_z + 8 \rho^2 h_z h_z-{1\over 2}\rho^2\big( -g^{(1)}_z \bar g^{(1)}_{ z}+g^{(2)}_z \bar g^{(2)}_{ z} \big)=0
\ee
together with the complex conjugate equations.

$\bullet$ The equation of motion for the complex scalar takes the form
\be
{1\over H^2 \rho^2}\Big(\partial_z (\rho H^2 P_{\bar z})+\partial_{\bar z} (\rho H^2 P_{ z})\Big)-2i (q_z P_{\bar z}+q_{\bar z} P_z) -{1\over 2 f_1^2 f_2^2} \big(( g^{(1)}_z  g^{(1)}_{\bar z} -g^{(2)}_z  g^{(2)}_{\bar z} \big)=0
\ee
as well as the complex conjugate equation.

$\bullet$ The equation of motion for the anti-symmetric tensor fields  takes the form
\bea
&&{1\over H^2 \rho^2}\Big(\partial_z  (\rho f_2^2 f_3^4  g^{(1)}_{\bar z} ) + \partial_{\bar z}  (\rho f_2^2 f_3^4  g^{(1)}_{ z} ) \Big) -{i\over f_1^2}\big( q_z g^{(1)}_{\bar z} +q_{\bar z} g^{(1)}_{ z}\big) - {1\over f_1^2}\big( P_z \bar g^{(1)}_{\bar z} +P_{\bar z} \bar g^{(1)}_{ z}\big)\no \\
&&+{4i\over f_1^2}\big( h_z g^{(2)}_{\bar z} + h_{\bar z} g^{(2)}_{ z}\big)=0
\eea
 and
 \bea
&&{1\over H^2 \rho^2}\Big(\partial_z  (\rho f_1^2 f_3^4  g^{(2)}_{\bar z} ) + \partial_{\bar z}  (\rho f_1^2 f_3^4  g^{(2)}_{ z} ) \Big) -{i\over f_2^2}\big( q_z g^{(2)}_{\bar z} +q_{\bar z} g^{(2)}_{ z}\big) - {1\over f_2^2}\big( P_z \bar g^{(2)}_{\bar z} +P_{\bar z} \bar g^{(2)}_{ z}\big)\no \\
&&-{4i\over f_2^2}\big( h_z g^{(1)}_{\bar z} + h_{\bar z} g^{(1)}_{ z}\big)=0
\eea

The strategy in proving that these equations are automatically satisfied is to replace all fields in terms of the harmonic 
and holomorphic functions and their derivatives, using the relations given in section \ref{totsolution} and in appendix \ref{bianchi3app}. Since the harmonic and holomorphic functions are independent it can be checked that the equation of motion is indeed satisfied term by term (using Mathematica).

\end{document}